\documentclass[11pt]{article}
\usepackage{jheppub} 
\usepackage{amsfonts,amsmath,amssymb,epsf}
\usepackage{amsmath,braket}
\usepackage{amsthm}
\usepackage{physics}
\usepackage{caption}
\usepackage{subcaption}
\usepackage{graphicx,hyperref,color}
\usepackage[dvipsnames]{xcolor}
\usepackage{float}
\hypersetup{
    unicode=false,          
    pdftoolbar=true,        
    pdfmenubar=true,        
    linktocpage=true,       
    pdffitwindow=false,     
    pdfstartview={FitH},    
    pdfnewwindow=true,      
    colorlinks=true,       
    linkcolor=blue,          
    citecolor=blue,        
    filecolor=blue,      
    urlcolor=blue           
}

\numberwithin{equation}{section}									

\newcommand{\be}{\begin{equation}}
\newcommand{\ba}{\begin{eqnarray}}
\newcommand{\ea}{\end{eqnarray}}
\newcommand{\ee}{\end{equation}}

\newcommand{\s}{\sqrt}
\newcommand{\vp}{\varphi}

\newcommand{\ti}{\tilde}
\newcommand{\ap}{\alpha}

\newcommand{\ddd}{\cdot\cdot\cdot}
\newcommand{\no}{\nonumber \\}
\newcommand{\la}{\langle}
\newcommand{\lb}{\rangle}
\newcommand{\bea}{\begin{eqnarray}}
\newcommand{\eea}{\end{eqnarray}}
\newcommand{\bes}{\begin{equation*}}
\newcommand{\beas}{\begin{eqnarray*}}
\newcommand{\eeas}{\end{eqnarray*}}
\newcommand{\bas}{\begin{array*}}
\newcommand{\eas}{\end{array*}}
\newcommand{\ees}{\end{equation*}}

\newcommand{\ep}{\epsilon}

   \let\vp=\varphi

\newcommand{\tm}[2]{\tau^{#1|#2}}
\newcommand{\ttm}[2]{\tilde{\tau}^{#1|#2}}
\newcommand{\smat}[2]{\mathcal{S}_{#1}^{~#2}}

\newcommand{\matr}{\mathbb{M}}
\newcommand{\dens}{\mathbb{D}}
\newcommand{\herm}{\mathbb{H}}
\newcommand{\SVD}{\mathrm{SVD}}
\newcommand{\vN}{\mathrm{vN}}
\newcommand{\JSK}{\mathrm{JSK}}
\newcommand{\FP}{\mathrm{FP}}
\newcommand{\ABB}{\mathrm{ABB}}

\usepackage{tikz}
\usetikzlibrary{arrows,arrows.meta,intersections, calc,positioning,decorations.pathreplacing,decorations.pathmorphing,shapes}
\usetikzlibrary{patterns}
\usetikzlibrary{decorations.markings}
\usetikzlibrary{knots}

\arxivnumber{2307.06531}


\title{\boldmath SVD Entanglement Entropy}

\author{Arthur J.\ Parzygnat$^{a,b}$,}
\affiliation{$^{a}$Graduate School of Informatics, \\ Nagoya University, Chikusa-ku, 464-8601 Nagoya, Japan}
\affiliation{$^{b}$Department of Mathematics,\\ Massachusetts Institute of Technology, Cambridge, Massachusetts 02139, USA}

\author{Tadashi Takayanagi$^{c,d,e}$,}
\affiliation{$^{c}$Center for Gravitational Physics and Quantum Information,\\
Yukawa Institute for Theoretical Physics, Kyoto University, \\
Kitashirakawa Oiwakecho, Sakyo-ku, Kyoto 606-8502, Japan}
\affiliation{$^{d}$Inamori Research Institute for Science,\\
620 Suiginya-cho, Shimogyo-ku,
Kyoto 600-8411 Japan}
\affiliation{$^{e}$Kavli Institute for the Physics and Mathematics
 of the Universe (WPI),\\
University of Tokyo, Kashiwa, Chiba 277-8582, Japan}

\author{Yusuke Taki$^{c}$,}

\author{Zixia Wei$^{c,f}$}
\affiliation{$^{f}$Interdisciplinary Theoretical and Mathematical Sciences Program (iTHEMS), \\RIKEN, Wako 351-0198, Japan}

\emailAdd{arthurjp@mit.edu}
\emailAdd{takayana@yukawa.kyoto-u.ac.jp}
\emailAdd{yusuke.taki@yukawa.kyoto-u.ac.jp}
\emailAdd{zixia.wei@yukawa.kyoto-u.ac.jp}
\emailAdd{zixiawei@fas.harvard.edu}

\abstract{
In this paper, we introduce a new quantity called SVD entanglement entropy. This is a generalization of entanglement entropy in that it depends on two different states, as in pre- and post-selection processes. This SVD entanglement entropy takes non-negative real values and is bounded by the logarithm of the Hilbert space dimensions. The SVD entanglement entropy can be interpreted as the average number of Bell pairs distillable from intermediates states. We observe that the SVD entanglement entropy gets enhanced when the two states are in the different quantum phases in an explicit example of the transverse-field Ising model. Moreover, we calculate the R{\'e}nyi SVD entropy in various field theories and examine holographic calculations using the AdS/CFT correspondence. }

\begin{document}


\vspace*{-2cm}
\begin{flushright}
YITP-23-87 \\
RIKEN-iTHEMS-Report-23
\end{flushright}

\bigskip

\maketitle
\flushbottom








\newpage
\section{Introduction}

Postselection, the process of conditioning on a particular type of outcome, is an important and often-used procedure not only in quantum information science~\cite{Aharonov:1988PRL,Ramos:2020nature,Arvidsson:2020NatCom,Lupu:2022PRL}, but also in condensed matter physics, high energy physics, and quantum gravity (however, see also~\cite{Zhu:2011PRA,Ferrie:2014PRL,Combes:2014PRA} for some subtleties). It is also employed to study the black hole information problem~\cite{Horowitz:2003he,Gottesman:2003up}. Thinking of the usefulness of entanglement entropy in the studies of quantum information-theoretic properties of quantum many-body systems, quantum field theories, and quantum gravity, it is intriguing to ask for a generalization of entanglement entropy to quantum systems while implementing postselection. Since the entanglement entropy is defined for a single quantum state, we need to extend it to a quantity that depends on not only the initial (preselected) state $|\psi_1\lb$ but also the final (postselected/projected) state $|\psi_2\lb$.

One interesting candidate of such a quantity is the pseudo entropy, introduced in \cite{Nakata:2021ubr}. 
Refer to earlier work \cite{2014PhRvA..90b2116S} for a similar quantity in the context of conditional entropy of post-selected states (see also~\cite{FuPa23,Tu:2021xje}). Consider two pure states $|\psi_1\lb$ and $|\psi_2\lb$ in a Hilbert space $\mathcal{H}$, which we can regard as the initial state and final state, respectively, thinking of a post-selection process. Then, introduce the transition matrix $\tau^{1|2}$ as
\ba
\label{eq:transitionmatrix}
\tau^{1|2}=\frac{|\psi_1\lb\la\psi_2|}{\la \psi_2|\psi_1\lb}
\ea
(generalizations of this to mixed states are discussed in~\cite{Parzygnat:2022pax,Guo:2022jzs}, for example).
We decompose the total Hilbert space into $A$ and $B$ as the direct product
\ba
{\cal H}={\cal H}_A\otimes {\cal H}_B
\ea
and we define the reduced transition matrix as
\ba
\tau^{1|2}_A=\mbox{Tr}_B\left[\frac{|\psi_1\lb\la\psi_2|}{\la \psi_2|\psi_1\lb}\right].\label{rdtr}
\ea
The pseudo entropy \cite{Nakata:2021ubr} is defined by its complex analytically extended `von~Neumann entropy'
\ba
S(\tau^{1|2}_A)=-\mbox{Tr}\left[\tau^{1|2}_A\log \tau^{1|2}_A \right].\label{PEd}
\ea
Notice that when $|\psi_1\lb=|\psi_2\lb(\equiv |\psi\lb)$, this quantity reduces to the standard entanglement entropy, 
\ba
S(\rho_A)=-\mbox{Tr}[\rho_A\log \rho_A],\label{EEf}
\ea
where $\rho_A\equiv\mbox{Tr}_B[|\psi\lb\la\psi|]$ is the reduced density matrix for the subsystem $A$.

The pseudo entropy  (\ref{PEd}) has the advantage that its gravity dual is simply given by the minimal area in a Euclidean time-dependent asymptotically anti-de Sitter (AdS) background as an extension of holographic entanglement entropy \cite{Ryu:2006bv,Ryu:2006ef,Hubeny:2007xt}, which was the original motivation why this quantity was introduced in \cite{Nakata:2021ubr}. However, the pseudo entropy in general takes complex values because the transition matrix is not necessarily hermitian \cite{Nakata:2021ubr} and it can even take values larger than the logarithm of the Hilbert space dimensions \cite{Ishiyama:2022odv}. Thus, its quantum information-theoretic interpretation is not straightforward. Nevertheless, for a special class of initial and final states, it was shown that we can interpret the pseudo entropy as the number of Bell pairs distilled from the intermediate states \cite{Nakata:2021ubr}. Moreover, it has been pointed out that the real part of pseudo entropy can be used as a quantum order parameter that distinguishes different quantum phases \cite{Mollabashi:2020yie,Mollabashi:2021xsd,Nishioka:2021cxe,Akal:2022qei}. The real part of pseudo entropy was also recently introduced in~\cite{FuPa23} in order to extend many theorems of standard entropies to a dynamical context~\cite{Parzygnat:2022pax}, including a quantum analogue of the standard classical entropic Bayes' theorem~\cite{FuPa23,FuPa21}. For further progress on pseudo entropy, refer to 
\cite{Akal:2020wfl,Camilo:2021dtt,Goto:2021kln,Miyaji:2021lcq,Hikida:2021ese,Murciano:2021dga,Guijosa:2022jdo,Berkooz:2022fso,Hikida:2022ltr,Akal:2022qei,Mori:2022xec,Mukherjee:2022jac,Guo:2022sfl,Miyaji:2022dna,Bhattacharya:2022wlp,Guo:2022jzs,Doi:2022iyj,Narayan:2022afv,Li:2022tsv,He:2023eap,Kanda:2023zse,Doi:2023zaf,Chen:2023gnh,Narayan:2023ebn,Jiang:2023loq,Chu:2023zah,Chandra:2023rhx,He:2023wko}. In spite of these developments, we still do not have a clear physical or quantum information-theoretic understanding of the imaginary part of pseudo entropy.

In addition to developing further ideas on pseudo entropy, it is intriguing at the same time to explore other entropic quantities that can be applicable to post-selection processes and that take non-negative real values. In this paper, as such a candidate, we would like to propose a new quantity based on a singular value decomposition (SVD) of the transition matrix $\tm{1}{2}_A$, which we call SVD entanglement entropy or SVD entropy for short. This quantity always takes real non-negative values and is bounded by the logarithm of the Hilbert space dimensions. We will provide a quantum information-theoretic interpretation of this quantity in terms of the number of Bell pairs in the intermediate states for any choice of initial and final states. After we analyze this quantity in qubit systems, we will examine explicit examples in intergrable conformal field theories (CFTs), holographic CFTs via the AdS/CFT correspondence, Chern-Simons topological field theory, and quantum spin systems. We will also work out some mathematical and statistical properties of the SVD entanglement entropy. 

Quantities analogous to our SVD entropy have appeared in earlier works. For example, an un-normalized version of the SVD entropy has recently been introduced in the context of quantifying temporal correlations in quantum systems~\cite{JSK23}. As another example, a Schatten 2-norm normalized version of the SVD entropy has been used extensively in the field of genetics to measure the complexity of genome expression data~\cite{Alter:2000PNAS}. In addition, the SVD entropy as defined here has been used in the analysis of stock market indices~\cite{Nakaji:2022PRR} and as a complexity measure for ecological networks~\cite{Strydom:2021FEE}.  However, our introduction of the SVD \emph{entanglement} entropy, as well as its connection to field theories, appears to be new to our knowledge. 

This paper is organized as follows. In Section~\ref{sec:def}, we explain the definition and basic properties of SVD entanglement entropy, including an analogue of the Page curve. In Section~\ref{sec:two-qubit}, we study examples of calculations of SVD entanglement entropy in two-qubit systems. We also provide a quantum information-theoretic interpretation of SVD entanglement entropy in terms of the average number of distillable Bell pairs in the intermediate states. In Section~\ref{sec:SVDentint2dcfts}, we compute the SVD entanglement entropy in integrable two-dimensional CFTs in the presence of local operator excitations. In Section~\ref{sec:hol_CFT}, we examine a holographic calculation of SVD entanglement entropy. In Section~\ref{sec:CS_theory}, we calculate the SVD entanglement entropy  in a Chern-Simons gauge theory. In Section~\ref{sec:spin_systems}, we analyze the SVD entanglement entropy under the quantum phase transition in the transverse-field Ising spin chain. In Section~\ref{sec:conclusions}, we summarize our conclusions and discuss future problems.

\section{Definition and basic properties of SVD entanglement entropy}\label{sec:def}

We propose a new quantity that provides a real and non-negative entropic quantity, which we call SVD entanglement entropy, aimed at post-selection setups as we explain below.

\subsection{Definition of SVD entanglement entropy}

First, we introduce the following density matrix constructed from the reduced transition matrix (\ref{rdtr}):
\ba\label{rho12}
\rho^{1|2}_A=\frac{\s{(\tau^{1|2}_A)^\dagger\tau^{1|2}_A}}{\mbox{Tr}\left[\s{(\tau^{1|2}_A)^\dagger\tau^{1|2}_A}\right]}.
\ea
From this, we define a new quantity, called the SVD entanglement entropy of $\tau_{A}^{1|2}$, as the von~Neumann entropy of $\rho^{1|2}_A$:
\ba
\label{eqn:SVDEEformula}
S(\rho^{1|2}_A)=-\mbox{Tr}\left[\rho^{1|2}_A\log \rho^{1|2}_A \right].
\ea
As we will see later, this can be interpreted as the Shannon entropy for the normalized eigenvalues of a singular value decomposition (SVD) of the reduced transition matrix $\tau^{1|2}_A$.

Since $\rho^{1|2}_A$ is in fact a density matrix, the SVD entropy is always non-negative and is bounded as 
\ba
0\leq S(\rho^{1|2}_A)\leq \log d_A,
\ea
where $d_A$ is the dimension of $\mathcal{H}_A$. It is also immediately clear that when the two states are identical $|\psi_1\lb=|\psi_2\lb$, the SVD entropy coincides with the ordinary entanglement entropy (\ref{EEf}). This shows that this new quantity is a natural generalization of entanglement entropy.

Here we note that the factor $\langle\psi_2|\psi_1\rangle$ cancels in \eqref{rho12}. Therefore, SVD entropy may be defined even in cases where $\ket{\psi_1}$ and $\ket{\psi_2}$ are orthogonal, as long as the (unnormalized) 
reduced transition matrix 
\ba
\tilde{\tau}^{1|2}_A=\Tr_B\ket{\psi_1}\bra{\psi_2}
\ea
is nonzero (cf.\ Example 1 in Section~\ref{sec:two-qubit} and Section~\ref{sec:VPS} for explicit examples). 

For later purposes and practical calculations in field theories, it is also useful to introduce a  R\'enyi-type generalization.
We first generalize $\rho^{1|2}_A$ as follows
\ba
\rho^{(m)1|2}_A 
=\frac{\left((\tau^{1|2}_A)^\dagger\tau^{1|2}_A\right)^{\frac{m}{2}}}{
\mbox{Tr}\left[\left((\tau^{1|2}_A)^\dagger\tau^{1|2}_A\right)^{\frac{m}{2}}\right]},  \label{rma12}
\ea
so that $\rho^{(1)1|2}_A =\rho^{1|2}_A$. Based on this, we define a R\'enyi version $S^{(n,m)}_A$ of SVD entropy by
\ba
S^{(n,m)}_A=S^{(n)}(\rho^{(m)1|2}_A)=\frac{1}{1-n}\log\mbox{Tr}\left[(\rho^{(m)1|2}_A)^n\right],\label{RNPE}
\ea
where $S^{(n)}(\rho)$ is the $n$-th R{\'e}nyi entropy defined by $S^{(n)}(\rho)=\frac{1}{1-n}\log\Tr[\rho^n]$. 
Notice that the original SVD entropy is given by the limit
\ba
S^{(1,1)}_A=\lim_{n\to 1}S^{(n)}\left(\rho^{(1)1|2}_A\right)=S(\rho^{1|2}_A).
\ea

When the two states are identical $|\psi_1\lb=|\psi_2\lb (\equiv|\psi\lb)$, the R\'enyi  SVD entropy is expressed in terms of the R\'enyi entanglement entropy (EE):
\ba
S^{(n,m)}_A\Bigl|_{|\psi_1\lb=|\psi_2\lb}=S^{(n)}(\rho^{(m)1|2}_A)\Bigl|_{|\psi_1\lb=|\psi_2\lb}=\frac{1}{1-n}\left[(1-mn)S^{(mn)}_A+n(m-1)S^{(m)}_A\right],\label{fomnm}
\ea
where $S^{(n)}_A$ is the $n$-th  R\'enyi  entanglement entropy defined by $S^{(n)}_A=\frac{1}{1-n}\log\mbox{Tr}[\rho_A^n]$, where $\rho_A$ is the standard reduced density matrix $\rho_A=\mbox{Tr}_B[|\psi\lb\la\psi|]$.

\subsection{Basic properties}\label{subsec:properties}

First of all, we can easily see that this new entropy $S(\rho^{1|2}_A)$ vanishes when $A$ is the total system, or more precisely when $\tau_A^{1|2}$ is pure. 
Moreover, if either $|\psi_1\lb$ or $|\psi_2\lb$ has no entanglement, i.e. if each can be written as a direct product $|\vp\lb_A|\vp'\lb_B$ of $A$ and $B$, then the SVD entropy vanishes, $S(\rho^{1|2}_A)=0$.

\begin{figure}
\centering
\begin{subfigure}[b]{0.45\textwidth}
\includegraphics[width=1.0\textwidth]{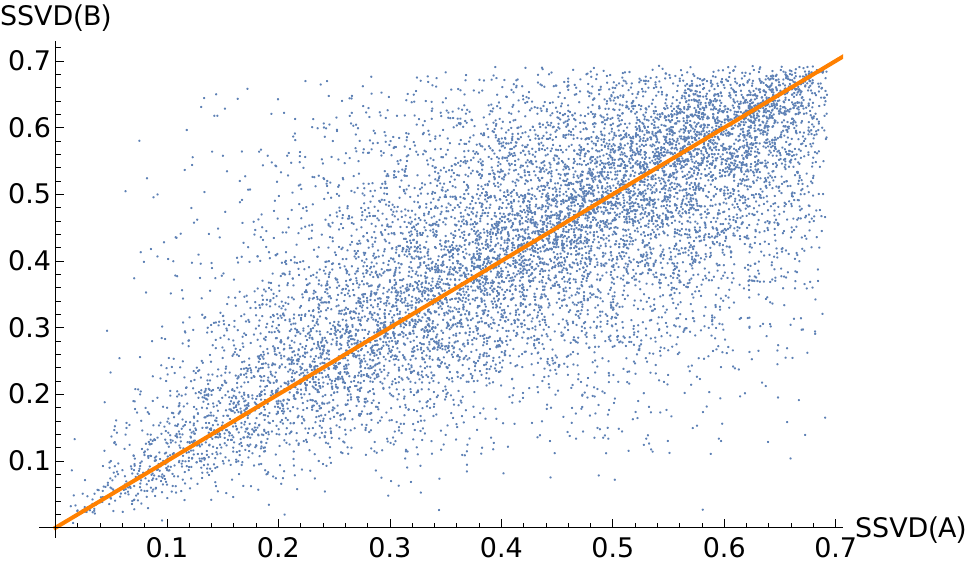}
\subcaption{$d_{A}=2=d_{B}$}
\end{subfigure}
\qquad
\begin{subfigure}[b]{0.45\textwidth}
\includegraphics[width=1.0\textwidth]{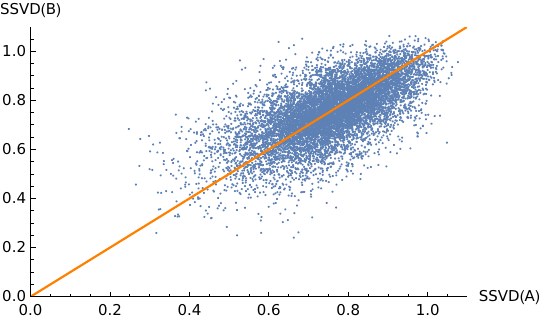}
\subcaption{$d_{A}=3=d_{B}$}
\end{subfigure}
\caption{Unlike what happens for ordinary entanglement entropy or pseudo entropy, the SVD entanglement entropy values $S(\rho^{1|2}_{A})$ and $S(\rho^{1|2}_{B})$ are not equal in general. These scatter plots show the values of $(S(\rho^{1|2}_{A}),S(\rho^{1|2}_{B}))$ for Haar-random samples of $N=10000$ pairs of unit vectors $|\psi_{1}\rangle,|\psi_{2}\rangle\in\mathcal{H}_{A}\otimes\mathcal{H}_{B}$ for two different dimensions $d_{A}=d_{B}$. If the two SVD entanglement entropies were equal, they would all lie along the diagonal line. Nevertheless, the distribution seems symmetric about the diagonal line suggesting that the subregion SVD entanglement entropies might be equal on average. This property also seems to hold for higher values of $d_{A}=d_{B}$. As the dimensions increase, the average SVD entanglement entropy increases (see also Section~\ref{sec:EVSVDEE}).}
\label{fig:NewPEAvsB}
\end{figure}

However, we in general find $S(\rho^{1|2}_A)\neq S(\rho^{1|2}_B)$ as opposed to the entanglement entropy and the pseudo entropy. To support this claim, Figure~\ref{fig:NewPEAvsB} provides numerical illustrations for how the values of the two entropies are distributed when sampling $N=10000$ Haar random pure states $\ket{\psi_{1}}$ and $\ket{\psi_{2}}$. 
In fact, one can find examples where
\ba
\mbox{Tr}\left[\left((\tau^{1|2}_A)^\dagger\tau^{1|2}_A\right)^n\right]\neq \mbox{Tr}\left[\left((\tau^{1|2}_B)^\dagger\tau^{1|2}_B\right)^n\right]
\ea
(see Section~\ref{sec:SVDentint2dcfts} and Figure~\ref{fig:SABreplica} for example).
This is in sharp contrast with the property
\ba
\mbox{Tr}[(\tau^{1|2}_A)^n]= \mbox{Tr}[(\tau^{1|2}_B)^n].
\ea 
It is also helpful to note that we can diagonalize the reduced transition matrix $\tau^{1|2}_A$ via the singular value decomposition (SVD) as follows
\ba
\tau^{1|2}_A=U\cdot \Lambda\cdot  V^{\dagger},
\label{unitary}
\ea
where $U$ and $V$ are unitary matrices and $\Lambda$ is a diagonal matrix with non-negative eigenvalues $\lambda_i\geq 0\ \ (i=1,2,\ddd, d_{A})$:
\ba
\Lambda=\mbox{diag}(\lambda_1,\lambda_2,\dots,\lambda_{d_{A}}).
\ea
However, note that since $\sum_i\lambda_{i}$ is not necessarily normalized to be $1$, we introduce the normalized eigenvalues 
\ba
\hat{\lambda}_i=\frac{\lambda_i}{\sum_i\lambda_i},
\ea
so that $\sum_i\hat{\lambda}_i=1$. Since the $\lambda_{i}$ are the singular values of $\tau^{1|2}_{A}$, while the $\hat{\lambda}_{i}$ are the eigenvalues of $\rho^{1|2}_{A}$,
the SVD entropy can be expressed as follows
\ba
S(\rho^{1|2}_A)=-\sum_{i}\hat{\lambda}_i\log\hat{\lambda}_i,
\ea
which is clearly real and non-negative.
This analysis also shows that we get the same entropy even if we replace $\rho^{1|2}_A$ with the conjugated one
by
\ba
{\rho'}^{1|2}_A=\frac{\s{\tau^{1|2}_A(\tau^{1|2}_A)^\dagger}}{\mbox{Tr}\left[\s{\tau^{1|2}_A(\tau^{1|2}_A)^\dagger}\right]}.
\ea

SVD entropy and its R\'enyi versions can be rewritten by using the Schatten norms, which are defined as follows. For an operator $X$, which is not assumed to be Hermitian, the Schatten $p$-norm for $p\in [1,\infty)$ is defined as
\begin{align}\label{norm}
    \norm{X}_p:=\left(\sum_i\lambda_i^p\right)^{\frac{1}{p}},
\end{align}
where $\lambda_i\ge0\, (i=1,2,\ldots,\dim{\mathcal{H}})$ are the eigenvalues of $\sqrt{X^\dagger X}$, i.e.\ the singular values of $X$. 
Note that $\norm{X}_1$ is identical to the trace norm of $X$. 
By using these Schatten norms, 
\begin{align}
    \Tr\left[\left(\rho_A^{1|2}\right)^n\right]=\frac{\sum_{i}\lambda_i^n}{\left(\sum_i\lambda_i\right)^n}=\left(\frac{\norm{\tm{1}{2}_A}_n}{\norm{\tm{1}{2}_A}_1}\right)^n.
\end{align}
Therefore, the R\'enyi version of SVD entropy can be expressed as 
\ba
S_{A}^{(n)}(\rho^{1|2}_{A})=\frac{n}{1-n}\log\left(\frac{\norm{\tm{1}{2}_A}_n}{\norm{\tm{1}{2}_A}_1}\right).
\ea
Furthermore, we can also express the generalized versions $S^{(n,m)}_A$ in a similar way. Namely, since 
\begin{align}
    \Tr\left[\left(\rho_A^{(m)1|2}\right)^n\right]=\frac{\sum_{i}\lambda_i^{mn}}{\left(\sum_i\lambda_i^{m}\right)^n}=\left(\frac{\norm{\tau_A^{1|2}}_{mn}}{\norm{\tau^{1|2}_A}_m}\right)^{mn},
\end{align}
we have 
\begin{align}
    S_A^{(n,m)}=\frac{mn}{1-n}\log\left(\frac{\norm{\tau_A^{1|2}}_n^{m}}{\norm{\tau^{1|2}_A}_1^{m}}\right).
\end{align}

\subsection{Local unitary and SVD entanglement entropy}\label{subsec:localunitary}
Let us consider a quantum system decomposed into two parts, called $A$ and $B$, respectively. 
Let $\ket{\psi_1},\ket{\psi_2}\in\mathcal{H}_{A}\otimes\mathcal{H}_B$ be two states related by a local unitary operator on $A$:
\begin{align}
    \ket{\psi_1}=U\otimes I\ket{\psi_2}.
\end{align}
In this situation, the reduced transition matrices for subsystems $A$ and $B$ are 
\begin{align}
    \ttm{1}{2}_A&=U\Tr_B[|\psi_2\rangle\langle \psi_2|],\\
    \ttm{1}{2}_B&=\Tr_A[U\otimes I|\psi_2\rangle\langle\psi_2|].
\end{align}
Since the overall $U$ in $\ttm{1}{2}_A$ vanishes inside $(\tm{1}{2}_A)^\dagger\tm{1}{2}_A$, the SVD entropy for $A$ is unchanged from the standard entanglement entropy for both states:
\begin{align}
    S(\rho_A^{1|2}) = S(\rho_A^{1}) = S(\rho_A^{2}). 
\end{align}
From another point of view, when considering the singular values of the reduced transition matrix on $A$, one can always perform a unitary transformation to erase the effect of $U$. On the other hand, the SVD entropy on subsystem $B$ changes because the operator $U$ does not vanish.

Thus, we have found that one characteristic property of SVD entropy for a subsystem is that the value changes when an operation is performed on the complement subsystem. We will later observe an analogous behavior in field theory calculations when we include excitations.

\subsection{Comparing SVD entropy to other extensions of entropy}
\label{sec:comparingSVD}

Many extensions of the von~Neumann entropy have been proposed in the literature, such as the recent ones in~\cite{2014PhRvA..90b2116S,Nakata:2021ubr,JSK23,FuPa23,Tu:2021xje}. We will next examine their similarities and differences for finite quantum systems in order to compare these alternative proposals more precisely (for brevity, we will leave out pseudo entropy in this section since pseudo entropy will be, and has been, analyzed in other sections). Note that we will examine these properties for the entropy itself, not necessarily the entanglement entropy, the latter of which is the entropy applied after isolating a subsystem of a larger system via a partial trace. 

The von~Neumann entropy is a functional that is defined on the set of states of a physical system. Given the matrix algebra $\matr_{m}$ of $m\times m$ matrices, the von~Neumann entropy is the function $S_{\vN}:\dens_{m}\to[0,\infty)$ given on a density matrix $\rho\in\dens_{m}$ by 
\begin{equation}
S_{\vN}(\rho)
=-\Tr[\rho\log\rho]
=-\sum_{i=1}^{m}p_{i}\log p_{i},
\end{equation}
where $p_{i}$ are the eigenvalues of $\rho$ (including repetitions if multiplicity is greater than $1$). 
Here, $\dens_{m}$ denotes the convex space of density matrices in $\matr_{m}$.
We use the subscript ``$\vN$'' because we will consider several other entropies in this section, and it will be helpful to distinguish them. Meanwhile, the SVD entropy introduced above in this work is an extension of the von~Neumann entropy that is valid not only on the domain of density matrices, $\dens_{m}$, but also on the domain of all matrices, $\matr_{m}$. In this section, we will study several properties of the SVD entropy, isolating which properties are common to it and the von~Neumann entropy. In addition, we will compare it to alternative proposals for extensions of entropies to beyond just the convex space of density matrices. In what follows, we also set $\herm_{m}$ to denote the convex space of $m\times m$ hermitian matrices. 

The general definition of the SVD entropy $S_{\SVD}:\matr_{m}\to[0,\infty)$ that we focus on in this work is given by 
\ba
\label{eqn:SVDEgeneral}
S_{\SVD}(A)=S_{\vN}\left(\frac{|A|}{\lVert A\rVert_{1}}\right)\equiv-\sum_{i=1}^{m}\frac{\sigma_{i}}{\lVert A\rVert_{1}}\log\left(\frac{\sigma_{i}}{\lVert A\rVert_{1}}\right),
\ea
where $|A|:=\sqrt{A^{\dag}A}$ denotes the operator absolute value matrix, $\lVert A\rVert_{1}=\tr|A|=\sum_{i=1}^{m}\sigma_{i}$ is the trace norm of $A$ (the Schatten 1-norm), and the $\sigma_{i}$ are the singular values of $A$ (including repetitions if multiplicity is greater than $1$). Note that one might more appropriately call this the (trace-norm) normalized SVD entropy, since the matrix $|A|$ is normalized by its trace norm so that $|A|/\lVert A\rVert_{1}$ is a density matrix as long as $A\ne0$. If $A=0$, then the SVD entropy is set equal to $0$ (note that this implies the SVD entropy is discontinuous at $0$). This definition of SVD entropy specializes to the SVD entanglement entropy formula~\eqref{eqn:SVDEEformula} via the relation 
\begin{equation}
S_{\SVD}(\tau^{1|2}_{A})=S_{\vN}(\rho^{1|2}_{A}).
\end{equation}
Interestingly, the formula~\eqref{eqn:SVDEgeneral} for SVD entropy has been used in the analysis of stock market index data~\cite{Nakaji:2022PRR}, as a measure of the complexity of ecological networks~\cite{Strydom:2021FEE}, and as a measure of brain-computer interfacing data complexity~\cite{Roberts:1999MBEC}. 

The fact that we normalize the matrix before applying the usual von~Neumann entropy formula $S_{\vN}(A)=-\tr(A\log A)$ (which is a well-defined function for arbitrary positive matrices $A$) means that the SVD entropy in~\eqref{eqn:SVDEgeneral} is slightly different from the one considered in~\cite{JSK23}, the latter of which extends the von~Neumann entropy to trace $1$ Hermitian matrices using the standard SVD without normalization.  
We will denote the entropy of~\cite{JSK23} by $S_{\JSK}$, and its value on a Hermitian matrix $A\in\herm_{m}$ is given by 
\ba
S_{\JSK}(A)=S_{\vN}\big(|A|\big)
\equiv-\sum_{i=1}^{m}\sigma_{i}\log\sigma_{i}.
\ea
The entropy function $S_{\JSK}$ was introduced in~\cite{JSK23} to study the entropy of pseudo-density operators (unrelated to pseudo entropy), which are extensions of density matrices to include temporal correlations of states evolving in time and subject to measurements~\cite{FJV15}. The JSK and SVD entropies are related to each other via 
\ba
S_{\SVD}(A)=\frac{S_{\JSK}(A)}{\lVert A\rVert_{1}}+\log\big(\lVert A\rVert_{1}\big),
\ea
which provides a better comparison to the entropy from~\cite{JSK23}. 

\begin{figure}
\centering
\includegraphics[width=9cm]{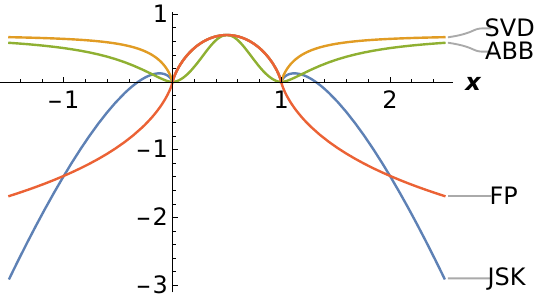}
\caption{Given the matrix 
$A=\mathrm{diag}(x,1-x)$,
with $x\in\mathbb{R}$, the  different entropy formulas, $S_{\SVD}(A)$, $S_{\ABB}(A)$, $S_{\JSK}(A)$, and $S_{\FP}(A)$ are compared as functions of $x$. Note that they all agree for $x\in[0,1]$ except for $S_{\ABB}(A)$, the latter of which is therefore not an extension of the standard entropy.}
\label{fig:threeents}
\end{figure}

Our normalization for $S_{\SVD}$ also differs from the Schatten 2-norm normalized SVD entropy used in the analysis of genetic expression data~\cite{Alter:2000PNAS} and as a truncation error for quantum lattice systems~\cite{weinstein2012doing}. This version of SVD entropy is given by
\ba
S_{\ABB}(A)=S_{\vN}\left(\frac{A^{\dag}A}{\lVert A\rVert_{2}^2}\right)
\equiv-\sum_{i=1}^{m}\frac{\sigma_{i}^{2}}{\lVert A\rVert_{2}^2}\log\left(\frac{\sigma_{i}^{2}}{\lVert A\rVert_{2}^2}\right).
\ea
This latter entropy $S_{\ABB}$ is used in~\cite{Alter:2000PNAS} to measure the complexity of genome expression data. The relationship to quantum mechanics only comes from the fact that the data are organized using matrices, but the data are all classical. In particular, $S_{\ABB}$ is not an extension of the von~Neumann entropy (cf.\ Figure~\ref{fig:threeents}). 
Nevertheless, this entropy is also a special case of the R{\'e}nyi version of our entropy. In detail, setting
\ba
S_{\SVD}^{(n,m)}(A):=\frac{1}{1-n}\log\Tr\left[\left(\frac{|A|^{mn}}{\lVert|A|^{m}\rVert_{1}^{n}}\right)\right],
\ea
we find that 
\ba
S_{\ABB}(A)
=\lim_{n\to1}S_{\SVD}^{(n,2)}(A).
\ea

In Figure~\ref{fig:threeents}, we show the differences between the SVD entropy introduced here and the entropies from~\cite{JSK23,FuPa23,Alter:2000PNAS} in the special case of the hermitian matrix $A=\left[\begin{smallmatrix}x&0\\0&1-x\end{smallmatrix}\right]$, with $x\in\mathbb{R}$. They are given by 
\begin{align}
S_{\JSK}(A)&=-|x|\log|x|-|1-x|\log|1-x| \nonumber \\
S_{\SVD}(A)&=\frac{-|x|\log|x|-|1-x|\log|1-x|}{|x|+|1-x|}+\log\big(|x|+|1-x|\big) \nonumber \\
S_{\ABB}(A)&=\frac{-x^2\log(x^2)-(1-x)^2\log\big((1-x)^2\big)}{x^2+(1-x)^2}+\log\big(x^2+(1-x)^2\big) \nonumber \\
S_{\FP}(A)&=-x\log|x|-(1-x)\log|1-x|,
\end{align}
where 
\ba
S_{\FP}(A)=-\Tr\big[A\log|A|\big]
\equiv-\sum_{i=1}^{m}\lambda_{i}\log|\lambda_{i}|
\ea
is the entropy from~\cite{FuPa23}, which is defined in terms of the eigenvalues $\lambda_{i}$ (as opposed to just the singular values) of the Hermitian matrix $A\in\herm_{m}$. The entropy functional $S_{\FP}$ from~\cite{FuPa23} was built to extend the standard von~Neumann entropy to dynamical entities~\cite{AJP22,FuPa21}, namely states over time, which are quantum extensions of joint probability distributions~\cite{HHPBS17,FuPa22,Parzygnat:2022pax}. A transition matrix, as in~\eqref{eq:transitionmatrix}, is an example of a state over time (see~\cite{Parzygnat:2022pax} for details). Therefore, $S_{\FP}$ also specializes to the entropy formula used for pre- and post-selected states in~\cite{2014PhRvA..90b2116S}. In the paper \cite{Tu:2021xje}, a quantity in the same form  $-\mbox{Tr}[\rho_A \log |\rho_A|]$ was used to an alternative definition of entanglement entropy (\ref{EEf}) in non-hermitian quantum systems.

\subsection{Invariance, convexity, and additivity properties of the SVD entropy}

Here, we summarize and prove some properties of SVD entropy as compared to the usual von~Neumann entropy. These properties go beyond those discussed earlier in Section~\ref{subsec:properties}. We continue to implement the notations $S_{\vN},S_{\SVD},S_{\JSK},$ and $S_{\FP}$ from Section~\ref{sec:comparingSVD}. 

First, the SVD entropy is invariant under unitary transformations in the sense that 
\ba
S_{\SVD}(UAU^{\dag})=S_{\SVD}(A)
\ea
for all $A\in\matr_{m}$ and unitaries $U\in\matr_{m}$. In fact, the SVD entropy is two-sided invariant under \emph{independent} unitaries in the sense that 
\ba
S_{\SVD}(UAV^{\dag})=S_{\SVD}(A)
\ea
for all $A\in\matr_{m}$ and unitaries $U,V\in\matr_{m}$. This is because the singular values of a matrix satisfy such an invariance property, namely, if $A=U'\Sigma V'^{\dag}$ is an SVD of $A$, with $U',V'\in\matr_{m}$ unitary, then $UAV^{\dag}=(UU')\Sigma(VV')^{\dag}$ shows explicitly that the singular value matrix $\Sigma$ is unaffected by such an operation. 

Second, the SVD entropy is additive in the sense that 
\ba
S_{\SVD}(A\otimes B)=S_{\SVD}(A)+S_{\SVD}(B).
\ea
This follows from 
\begin{align}
S_{\SVD}(A\otimes B)
&=S_{\vN}\left(\frac{|A\otimes B|}{\lVert A\otimes B\rVert_{1}}\right)
=S_{\vN}\left(\frac{|A|}{\lVert A\rVert_{1}}\otimes\frac{|B|}{\lVert B\rVert_{1}}\right) \nonumber \\
&=S_{\vN}\left(\frac{|A|}{\lVert A\rVert_{1}}\right)+S_{\vN}\left(\frac{|B|}{\lVert B\rVert_{1}}\right)
=S_{\SVD}(A)+S_{\SVD}(B)
\end{align}
using the fact that the ordinary von~Neumann entropy is additive (note that this additivity property fails for the entropy $S_{\JSK}$ from~\cite{JSK23}, but it still holds for the entropy $S_{\FP}$ defined in~\cite{FuPa23}). 

Third, the SVD entropy satisfies a weak form of concavity~\cite{JSK23}, namely
\begin{equation}
\label{eq:weakconcav}
\lambda S_{\SVD}(A)+(1-\lambda)S_{\SVD}(B)\le S_{\SVD}\left(\lambda\frac{|A|}{\lVert A\rVert_{1}}+(1-\lambda)\frac{|B|}{\lVert B\rVert_{1}}\right)
\end{equation}
for all $A,B\in\matr_{m}$ and $\lambda\in[0,1]$. 
This follows from the concavity of the usual von~Neumann entropy. In more detail, 
\begin{align}
S_{\SVD}\left(\lambda\frac{|A|}{\lVert A\rVert_{1}}+(1-\lambda)\frac{|B|}{\lVert B\rVert_{1}}\right)&= S_{\vN}\left(\lambda\frac{|A|}{\lVert A\rVert_{1}}+(1-\lambda)\frac{|B|}{\lVert B\rVert_{1}}\right) \nonumber\\
&\ge \lambda S_{\vN}\left(\frac{|A|}{\lVert A\rVert_{1}}\right)+(1-\lambda)S_{\vN}\left(\frac{|B|}{\lVert B\rVert_{1}}\right) \nonumber\\
&=\lambda S_{\SVD}(A)+(1-\lambda)S_{\SVD}(B), 
\end{align}
where the concavity was used in the step with the inequality. The first equality holds because the argument of $S_{\SVD}$ is a density matrix and because $S_{\SVD}$ is an extension of the von~Neumann entropy. 

\begin{figure}
\centering
\begin{subfigure}[b]{0.45\textwidth}
\includegraphics[width=1.0\textwidth]{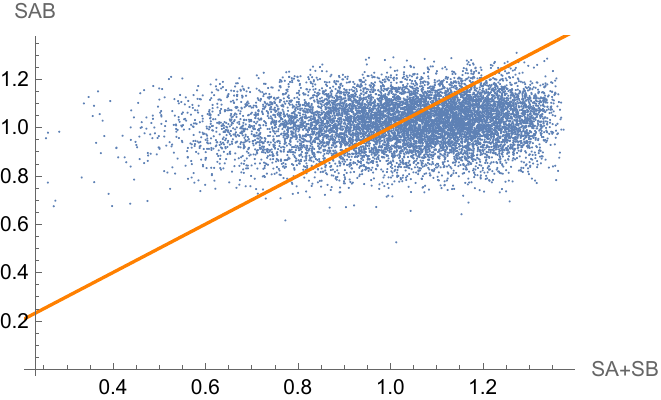}
\subcaption{A scatter plot of points whose coordinates are given by $(S(\rho^{1|2}_{A})+S(\rho^{1|2}_{B}),S(\rho^{1|2}_{AB}))$. Points below the diagonal line correspond to states that satisfy subadditivity. The plot illustrates that subadditivity fails.}
\label{fig:NewPEsubadd}
\end{subfigure}
\qquad
\begin{subfigure}[b]{0.45\textwidth}
\includegraphics[width=1.0\textwidth]{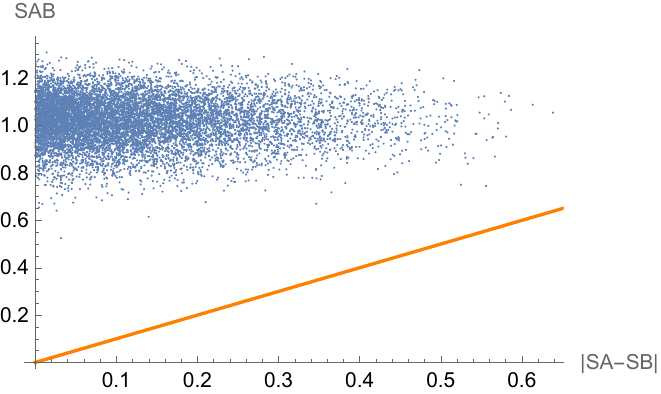}
\subcaption{A scatter plot of points whose coordinates are given by $(|S(\rho^{1|2}_{A})-S(\rho^{1|2}_{B})|,S(\rho^{1|2}_{AB}))$. Points above the diagonal line correspond to states that satisfy the Araki-Lieb (AL) inequality, suggesting that the AL inequality might hold generically.}
\label{fig:NewPEArakiLieb}
\end{subfigure}
\caption{Two scatter plots of 10000 Haar random states $|\psi_{1}\rangle,|\psi_{2}\rangle\in\mathcal{H}_{A}\otimes\mathcal{H}_{B}\otimes\mathcal{H}_{E}$, where the dimensions are given by $d_{A}=d_{B}=2$ and $d_{E}=4$, are used to illustrate the failure of subadditivity of SVD entropy and the plausibility of the Araki-Lieb (AL) inequality, respectively. However, the AL inequality does not hold for all states.}
\label{fig:subaddArLi}
\end{figure}

Certain refined upper and lower bounds, monotonicity, and properties like subadditivity and strong subadditivity do not necessarily hold. We will come back to these properties in detail for qubit systems and lattice systems later. Here, we make this more precise in the case of subadditivity. For this, we begin with two vector states $|\psi_{1}\rangle,|\psi_{2}\rangle\in\mathcal{H}_{A}\otimes\mathcal{H}_{B}\otimes\mathcal{H}_{E}$, where the additional Hilbert space $\mathcal{H}_{E}$ is used as an environment to generate mixed states in a way analogous to purification~\cite{Sc36,Hughston:1993PLA} (equivalently, the GNS construction~\cite{Segal:1947BAMS,Parzygnat:2016ACS}). From these data, we define 
\[
\tilde{\tau}^{1|2}:=|\psi_{1}\rangle\langle\psi_{2}|, \qquad
\tilde{\tau}^{1|2}_{AB}:=\Tr_{E}\left[\tilde{\tau}^{1|2}\right],
\qquad
\tilde{\tau}^{1|2}_{A}:=\Tr_{BE}\left[\tilde{\tau}^{1|2}\right],
\qquad
\tilde{\tau}^{1|2}_{B}:=\Tr_{AE}\left[\tilde{\tau}^{1|2}\right],
\]
\[
\rho^{1|2}_{AB}:=\frac{\sqrt{(\tilde{\tau}^{1|2}_{AB})^{\dag}\tilde{\tau}^{1|2}_{AB}}}{\Tr[\sqrt{(\tilde{\tau}^{1|2}_{AB})^{\dag}\tilde{\tau}^{1|2}_{AB}}]},
\qquad
\rho^{1|2}_{A}:=\frac{\sqrt{(\tilde{\tau}^{1|2}_{A})^{\dag}\tilde{\tau}^{1|2}_{A}}}{\Tr[\sqrt{(\tilde{\tau}^{1|2}_{A})^{\dag}\tilde{\tau}^{1|2}_{A}}]},
\qquad
\rho^{1|2}_{B}:=\frac{\sqrt{(\tilde{\tau}^{1|2}_{B})^{\dag}\tilde{\tau}^{1|2}_{B}}}{\Tr[\sqrt{(\tilde{\tau}^{1|2}_{B})^{\dag}\tilde{\tau}^{1|2}_{B}}]}.
\]
Subadditivity of SVD entropy would say that $S_{\SVD}(\tilde{\tau}^{1|2}_{AB})\le S_{\SVD}(\tilde{\tau}^{1|2}_{A})+S_{\SVD}(\tilde{\tau}^{1|2}_{B})$, i.e., $S(\rho^{1|2}_{AB})\le S(\rho^{1|2}_{A})+S(\rho^{1|2}_{B})$. However, we find that subadditivity fails, even for qubit systems (see Figure~\ref{fig:subaddArLi}). In particular, strong subadditivity cannot hold for SVD entropy. Nevertheless, numerical evidence suggests that the Araki-Lieb (AL) inequality $S(\rho^{1|2}_{AB})\ge|S(\rho^{1|2}_{A})-S(\rho^{1|2}_{B})|$ might hold generically (see Figure~\ref{fig:subaddArLi})~\cite{Araki:1970cmp}, which would give a non-trivial lower bound for the joint SVD entropy. However, the AL inequality cannot hold in full generality, and a one-parameter family of counterexamples for two qubits is given by setting the environment Hilbert space to be trivial, i.e. $\mathcal{H}_{E}=\mathbb{C}$, and setting 
\begin{align}
\ket{\psi_{1}}&=c\ket{10}+s\ket{01}\\
\ket{\psi_{2}}&=c\ket{00}+s\ket{11}
\end{align}
where $c=\cos\theta$, $s=\sin\theta$, and $\theta\in[0,2\pi)$. Then 
\begin{align}
\tilde{\tau}^{1|2}_{A}&=c^2\ket{1}\bra{0}+s^2\ket{0}\bra{1}\\
\tilde{\tau}^{1|2}_{B}&=cs\ket{0}\bra{1}+cs\ket{1}\bra{0}
\end{align}
and
\begin{align}
\rho^{1|2}_{A}&=c^2\ket{0}\bra{0}+s^2\ket{1}\bra{1}\\
\rho^{1|2}_{B}&=\frac{1}{2}\ket{0}\bra{0}+\frac{1}{2}\ket{1}\bra{1}
\end{align}
so that 
\begin{equation}
S(\rho^{1|2}_{A})=-c^2\log(c^2)-s^2\log(s^2),
\quad
S(\rho^{1|2}_{B})=\log(2),
\quad
S(\rho^{1|2}_{AB})=0,
\end{equation}
thereby illustrating the failure of the AL inequality for most values of $\theta$.

\subsection{The expected value of SVD entanglement entropy}
\label{sec:EVSVDEE}

Given a Hilbert space decomposition $\mathcal{H}=\mathcal{H}_{A}\otimes \mathcal{H}_{B}$, most states are entangled with respect to such a decomposition whenever the dimensions $d_{A}:=\mathrm{dim}(\mathcal{H}_{A})$ and $d_{B}:=\mathrm{dim}(\mathcal{H}_{A})$ are large and the average entanglement entropy tends to increase as the dimensions $d_{A}$ and $d_{B}$ increase~\cite{Page:1993ae,Page:1993wv,Sen:1996,HLW06}. We would now like to explore a similar property for the SVD entanglement entropy. We do this numerically using Haar-random pure states~\cite{HLW06}. 

\begin{figure}
\centering
\begin{subfigure}[b]{0.45\textwidth}
\begin{tikzpicture}
\node at (0,0) 
   {\includegraphics[width=1.0\textwidth]{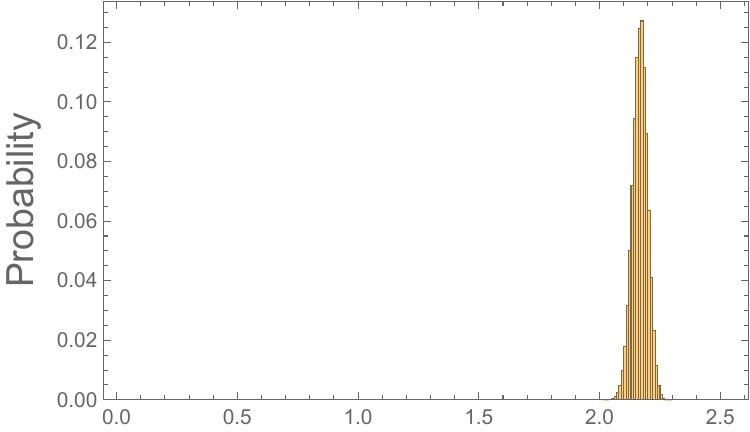}};
\node at (-0.25,0.45) 
   {\includegraphics[width=0.6\textwidth]{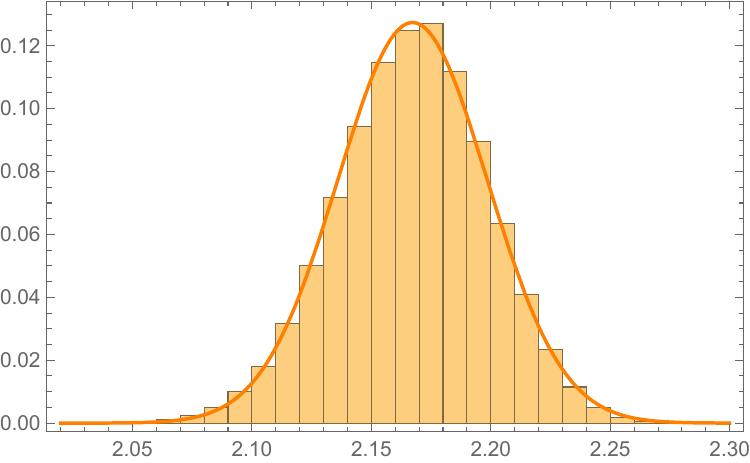}};
\end{tikzpicture}
\end{subfigure}
\qquad
\begin{subfigure}[b]{0.45\textwidth}
\begin{tikzpicture}
\node at (0,0) 
   {\includegraphics[width=1.0\textwidth]{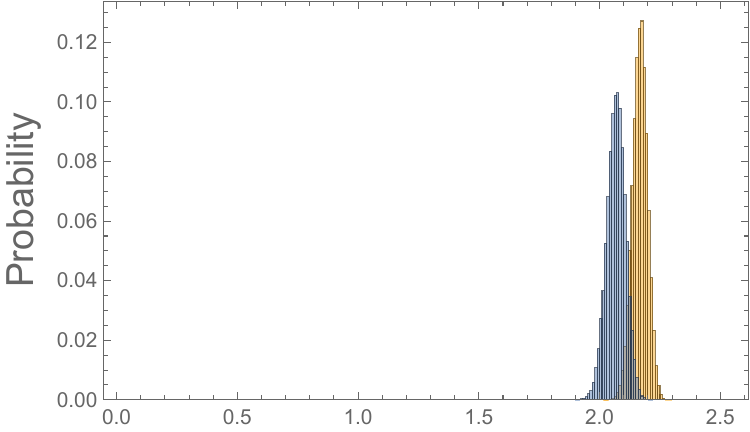}};
\node at (-0.25,0.45) 
   {\includegraphics[width=0.6\textwidth]{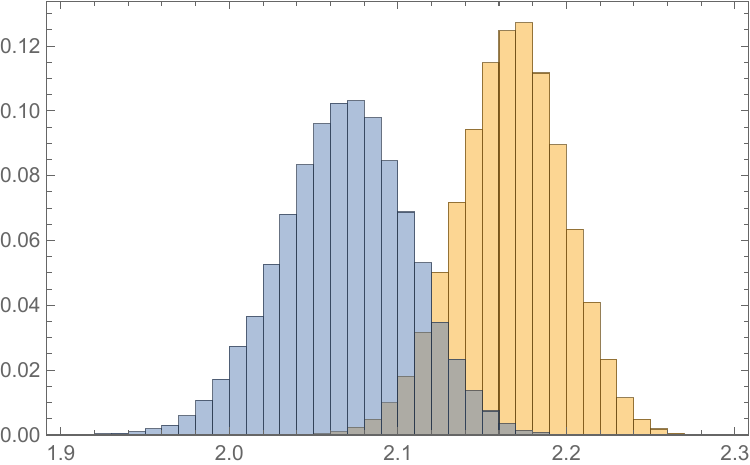}};
\end{tikzpicture}
\end{subfigure}
\caption{The figure on the left illustrates a histogram of values for the values of SVD entropy $S(\rho^{1|2}_{A})$ (in orange). The curve is well-approximated by a Gaussian distribution (the curve is drawn in orange on the zoomed-in figure). The figure on the right compares this to the standard entanglement entropy $S(\rho^{1}_{A})$ (in blue). In particular, the standard entanglement entropy histogram is shifted slightly to the left of the histogram for the SVD entanglement entropy, indicating a lower average value of entropy as compared to the SVD entanglement entropy.}
\label{fig:histogHr}
\end{figure}

In Figure~\ref{fig:histogHr}, we take a sample of $N=65536$ Haar-random pure states $|\psi_{1}\rangle,|\psi_{2}\rangle$ in $\mathcal{H}_{A}\otimes \mathcal{H}_{B}$, where the dimensions $d_{A}$ and $d_{B}$ of the Hilbert spaces are both equal to $d=13$ (the dimension $d$ is chosen large enough to illustrate that the average SVD entanglement entropy is close to the maximal value $\log(d)$, but small enough to easily provide the numerical data used to construct the histogram). The standard entanglement entropy is computed for each  $|\psi_{1}\rangle$  using $\rho_{A}^{1}:=\Tr_{B}[|\psi_{1}\rangle\langle\psi_{1}|]$ and it is compared with the SVD entropy of $\tilde{\tau}^{1|2}_{A}$, which corresponds to the von~Neumann entropy of $\rho^{1|2}_{A}:=\big|\tilde{\tau}^{1|2}_{A}\big|/\big\lVert\tilde{\tau}^{1|2}_{A}\big\rVert_{1}$, 
where $\tilde{\tau}^{1|2}_{A}=\Tr_{B}[\tilde{\tau}^{1|2}]$ and $\tilde{\tau}^{1|2}:=|\psi_{1}\rangle\langle\psi_{2}|$ (normalizing factors are accounted for in the definition of $\rho^{1|2}_{A}$). 

Several interesting observations can be made from these data. First, the expected/mean value for the SVD entropy is higher than the expected value for the standard entanglement entropy. Second, the distribution of SVD entropy values  closely follows a Gaussian curve, unlike that of the pseudo entropy~\cite{Nakata:2021ubr}. Third, the standard deviation for the distribution of SVD entropy values is smaller than that of the standard entanglement entropy, so that the distribution is more sharply peaked at its mean value. Rigorous proofs of these observations, provided that they hold in full generality, are still lacking. 

\subsection{Analogue of the Page curve}

\begin{figure}
\centering
\includegraphics[width=9cm]{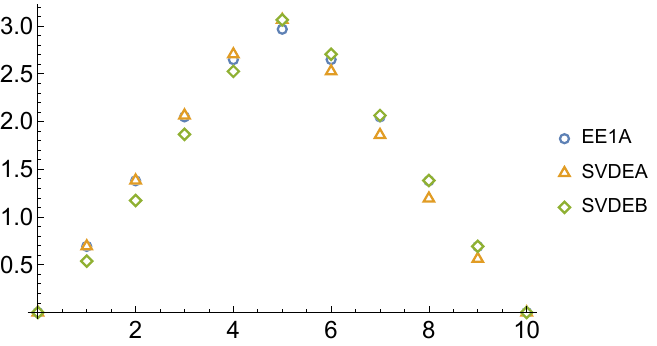}
\caption{In this plot, $M=5$, so that the horizontal axis variable $t$ takes values in $\{0,1,\dots,10\}$. A total of 100 states $|\psi_{1}\rangle$ and $|\psi_{2}\rangle$ were chosen randomly according to the Haar measure on $\mathcal{H}=\mathbb{C}^{2^{10}}=\mathbb{C}^{1024}$. Three types of averages are computed. These are EE1A, SVDEA, and SVDEB denote the usual entanglement entropy for state $|\psi_{1}\rangle$ on region A, the SVD entropy of the transition matrix on region A, and the SVD entropy of the transition matrix on region B, respectively.}
\label{fig:pagec}
\end{figure}

For simplicity, fix a Hilbert space $\mathcal{H}$ of dimension $d=2^{2M}$ for some $M\in\mathbb{N}$. For each $t\in\{0,1,\dots,2M\}$, set 
\[
\mathcal{H}_{A(t)}:=\mathbb{C}^{2^{t}}
\quad\text{ and }\quad
\mathcal{H}_{B(t)}:=\mathbb{C}^{2^{2M-t}}
\]
so that $\mathcal{H}\cong\mathcal{H}_{A(t)}\otimes\mathcal{H}_{B(t)}$ for each $t$. In~\cite{Page:1993ae}, Page computed the average entanglement entropy for subsystems given a Haar random initial pure state on $\mathcal{H}$ (see also~\cite{Sen:1996}). Then, in~\cite{Page:1993wv}, Page showed that the average entanglement entropy increases roughly linearly as a function of $t$ until $t=M$ and then decreases roughly linearly as a function of $t$ until $t=2M$ (see Figure~\ref{fig:pagec}). This curve is called the \emph{Page curve}.

In the case of SVD entropy, one starts with two Haar-random pure states $|\psi_{1}\rangle,|\psi_{2}\rangle$ in $\mathcal{H}$. Then, in analogy to Page's computation, one can compute the SVD entropy on subregions $A(t)$ and $B(t)$. Note that since $S(\rho^{1|2}_{A})\ne S(\rho^{1|2}_{B})$, it is important to specify which of the two regions one is computing the SVD entropy for. Although we do not compute these average entropies analytically, in Figure~\ref{fig:pagec}, we illustrate that one obtains a curve quite similar to, but not exactly the same as, the Page curve. Further analysis needs to be done to see if this is an effect due to the small sample sizes taken or if this persists for the true analytical averages. Interestingly, a Page curve was also recently computed for the alternative entanglement entropy formula $-\Tr[\rho_{A}\log|\rho_{A}|]$ in~\cite{PhysRevLett.130.010401}.

\section{SVD entanglement entropy in two qubit systems}
\label{sec:two-qubit}

Here we start our study of concrete examples of calculating SVD entropy by examining two qubit systems to explore its properties. This analysis will also reveal a quantum information-theoretic meaning of SVD entanglement entropy. In all examples in this section, systems $A$ and $B$ refer to the first and second qubit, respectively. 
In addition, the pre- and post-selected two-qubit states will be written in the generic form
\begin{align}
|\psi_1\lb & =a_1|00\lb+b_1|10\lb+c_1|01\lb+d_1|11\lb, 
\\
|\psi_2\lb & =a_2|00\lb+b_2|10\lb+c_2|01\lb+d_2|11\lb.
\end{align}
in most parts of this section. 

\subsection{Entanglement distillation and SVD entropy}

It is well-known that entanglement entropy provides a quantitative measure of quantum entanglement for pure states. For this we can explicitly show that the von~Neumann entropy (\ref{EEf}) coincides with the averaged number of Bell pairs which can be distilled from a given bipartite pure state via local operations and classical communication (LOCC) 
\cite{Nielsen:1998sz,Bennett:1995tk}. Below we would like to extend this argument to the SVD entropy.

We decompose the reduced transition matrix $\tau^{1|2}_A$ as follows
\ba
U^\dagger \cdot \tau^{1|2}_A\cdot V=\lambda_1|0\lb_A\la 0|+\lambda_2|1\lb_A\la 1|.
\ea
Since $\lambda_1+\lambda_2$ might not equal $1$, we introduce $\hat{\lambda}_1=\frac{\lambda_1}{\lambda_1+\lambda_2}$ and $\hat{\lambda}_2=\frac{\lambda_2}{\lambda_1+\lambda_2}$.

Now we take the asymptotic limit by taking $M\to\infty$ copies of the two-qubit system as usual in the interpretation of entanglement entropy as the number of distillable Bell pairs via LOCC. In this setup, the reduced transition matrix can be decomposed as 
\ba
(U^{\dagger})^{\otimes M}\cdot \left(\tau^{1|2}_A\right)^{\otimes M}\cdot V^{\otimes M}=\left(\lambda_1|0\lb_A\la 0|+\lambda_2|1\lb_A\la 1|\right)^{\otimes M}.
\ea

We introduce the projections onto Bell pairs on $M$ copies of two qubits, which act on the subsystem $A$, which has $M$ qubits (as in (3.58) and (3.59) of \cite{Nakata:2021ubr}):
\ba
\Pi_k=\sum_{i=1}^{{}_MC_k}|P^{(k)}_i\lb\la P^{(k)}_i|, 
\ea
so that the projections sum to the identity, i.e.
\ba
\sum_{k=0}^{M} \Pi_k=I.
\ea
In the above, we set ${}_MC_k=\frac{M!}{N!(M-N)!}$.

The main difference between this calculation and the one from~\cite{Nakata:2021ubr} is that here we perform the $V$ and $U^\dagger$ transformation before and after the projection such that the probability to measure the sector of $\Pi_k$ is given by 
\ba
p_k&=&\frac{\mbox{Tr}\left[\Pi_k\cdot (U^{\dagger})^{\otimes M}\cdot \left(\tau^{1|2}_A\right)^{\otimes M}\cdot V^{\otimes M} \right]}{\sum_{k=0}^M
\mbox{Tr}\left[\Pi_k\cdot (U^{\dagger})^{\otimes M}\cdot \left(\tau^{1|2}_A\right)^{\otimes M}\cdot V^{\otimes M} \right]}\no
&=& \left(\hat{\lambda}_1\right)^{k} \left(\hat{\lambda}_2\right)^{M-k}{}_MC_k.
\ea
Thus, the average number of the distillable Bell pairs in this post-selection process is estimated as 
\ba
\sum_{k=0}^M p_k\cdot\log {}_MC_k\simeq M\cdot S(\rho^{1|2}_A),
\ea
where 
\ba
S(\rho^{1|2}_A)=-\hat{\lambda}_1\log\hat{\lambda}_1-\hat{\lambda}_2\log\hat{\lambda}_2.
\ea
This shows that the distillable entanglement in this process is equal to the new entropy we introduced.

\subsection{Examples}
\label{sec:qubitsexamples}
In this subsection, we investigate SVD entropy for several examples of two-qubit system in order to illustrate characteristic behaviors of the quantity.  

\paragraph{Example 1}

Let us first consider the two states
\begin{align}
    \ket{\psi_1}&=\frac{1}{\sqrt{2}}(\ket{00}+e^{i\theta}\ket{11}),\\
    \ket{\psi_2}&=\frac{1}{\sqrt{2}}(\ket{00}+\ket{11}).
\end{align}
The reduced transition matrix is 
\begin{align}
    \tm{1}{2}_A=\frac{1}{1+e^{i\theta}}(\ket{0}\bra{0}+e^{i\theta}\ket{1}\bra{1}).
\end{align}
Therefore, the reduced transition matrix is not necessarily hermitian and its pseudo R\'enyi entropy
\begin{align}
    S^{(n)}\left(\tm{1}{2}_A\right)=\frac{1}{1-n}\log \left[\frac{\cos\frac{n\theta}{2}}{2^{n-1}\cos^n\left(\frac{\theta}{2}\right)}\right]
\end{align}
becomes complex valued for generic $\theta$. By taking the limit $n\to 1$, we have 
\begin{align}
    S\left(\tm{1}{2}_A\right)=\log 2+\log\left[\cos\frac{\theta}{2}\right]+\frac{\theta}{2}\tan\frac{\theta}{2}.
\end{align}
This behavior may look confusing because the periodicity of $\theta$, which the original setup has, is lost in the $n\to 1$ limit in this analytic form. 

Alternatively, we would like to consider the SVD entropy. We can easily check that 
\begin{align}
    \rho_A^{1|2}=\frac{1}{2}(\ket{0}\bra{0}+\ket{1}\bra{1}).
\end{align}
Therefore, we simply have 
\begin{align}
    S(\rho_A^{1|2})=\log 2,
\end{align}
which is equal to the entanglement entropy $S(\rho^1_A)=S(\rho^2_A)$. 
This result does not depend on $\theta$. In particular, $\theta=\pi$ gives an example of SVD entanglement entropy for two orthogonal states. The reason for the trivial dependence on $\theta$ is clear from the definition of SVD entropy. Namely, we can eliminate the phase factor $e^{i\theta}$ by a local unitary transformation on $|\psi_1\lb$.

\paragraph{Example 2}

Next, we consider the two states 
\begin{align}
    \ket{\psi_1}&=\frac{1}{2}\left(\ket{00}+\ket{11}+e^{i\theta}(\ket{01}+\ket{10})\right),\\
    \ket{\psi_2}&=\frac{1}{\sqrt{2}}(\ket{00}+\ket{11}).
\end{align}
The entanglement entropies for $\ket{\psi_1}$ and $\ket{\psi_2}$ are 
\begin{align}
    S(\rho^1_A)&=-\frac{1+\cos\theta}{2}\log\frac{1+\cos\theta}{2}-\frac{1-\cos\theta}{2}\log\frac{1-\cos\theta}{2}
    ,\\
    S(\rho^2_A)&=\log 2.
\end{align}
The reduced transition matrix is evaluated as
\begin{align}
    \tm{1}{2}_A=\frac{1}{2}
    \begin{pmatrix}
        1 & e^{i\theta}\\
        e^{i\theta} & 1
    \end{pmatrix}.
\end{align}
Since the eigenvalues of $\tm{1}{2}_A$ are $(1\pm e^{i\theta})/2$, the pseudo entropy is
\begin{align}
    S(\tm{1}{2}_A)=-\frac{1+e^{i\theta}}{2}\log\frac{1+e^{i\theta}}{2}-\frac{1-e^{i\theta}}{2}\log\frac{1-e^{i\theta}}{2},
\end{align}
which takes complex values. 

\begin{figure}[t]
    \centering
    \includegraphics[width=0.5\textwidth]{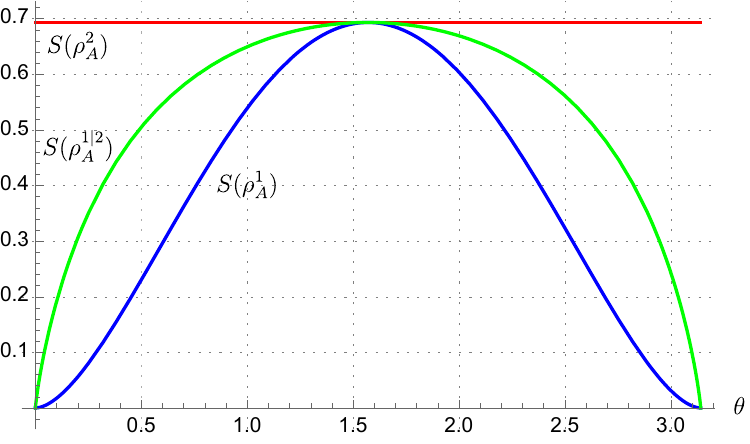}
    \caption{Plots of SVD entropy $S(\rho^{1|2}_A)$ and the ordinary entanglement entropies $S(\rho^1_A),S(\rho^2_A)$. We observe that SVD entropy in this case has upper and lower bound given by $\max\{S(\rho^1_A),S(\rho^2_A)\}$ and $\min\{S(\rho^1_A),S(\rho^2_A)\}$, respectively, though this is not a general property of SVD entropy (as will be shown in Figure~\ref{fig:qubit01} and later examples).}
    \label{fig:qubit}
\end{figure}

Next, let us calculate the SVD entropy. The new reduced density matrix $\rho^{1|2}_A$ is 
\begin{align}
    \rho^{1|2}_A=\frac{1}{2}
    \begin{pmatrix}
        1&\cos\theta \\
        \cos\theta &1
    \end{pmatrix}.
\end{align}
Therefore, the SVD entropy is given by
\begin{align}
    S(\rho^{1|2}_A)=\log\left(\sqrt{\frac{1+\cos\theta}{2}}+\sqrt{\frac{1-\cos\theta}{2}}\right)-\frac{\sqrt{\frac{1+\cos\theta}{2}}\log\sqrt{\frac{1+\cos\theta}{2}}+\sqrt{\frac{1-\cos\theta}{2}}\log\sqrt{\frac{1-\cos\theta}{2}}}{\sqrt{\frac{1+\cos\theta}{2}}+\sqrt{\frac{1-\cos\theta}{2}}}.
\end{align}
As depicted in Figure \ref{fig:qubit}, $S(\rho^{1|2}_A)$ takes values between $S(\rho^1_A)$ and $S(\rho^2_A)$ for every $\theta$. However, as we will see in the next example, the upper bound can fail in general, even for two-qubit systems. 

\paragraph{Example 3}

\begin{figure}[hhh]
    \centering
    \includegraphics[width=0.4\textwidth,trim={0cm 1cm 0cm 1.0cm},clip]{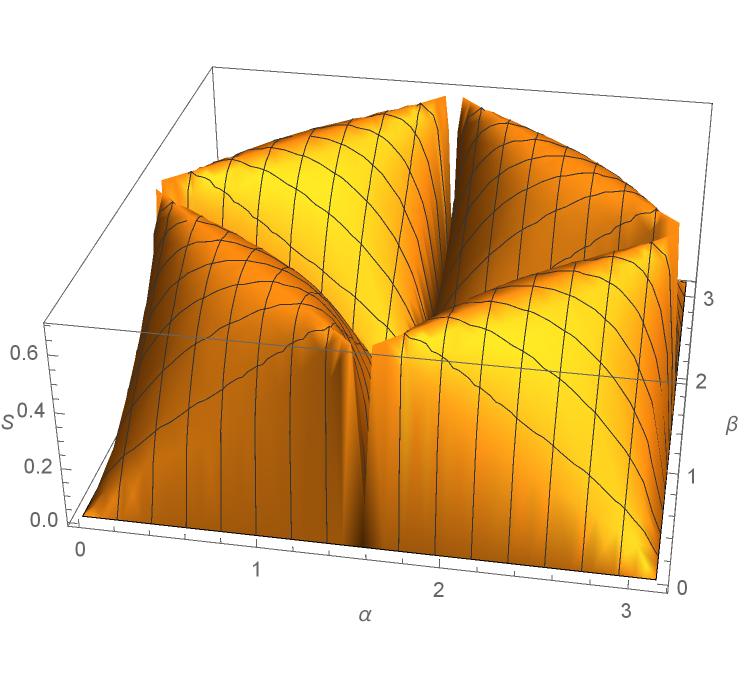}
    \hspace{15mm}
    \includegraphics[width=0.4\textwidth]{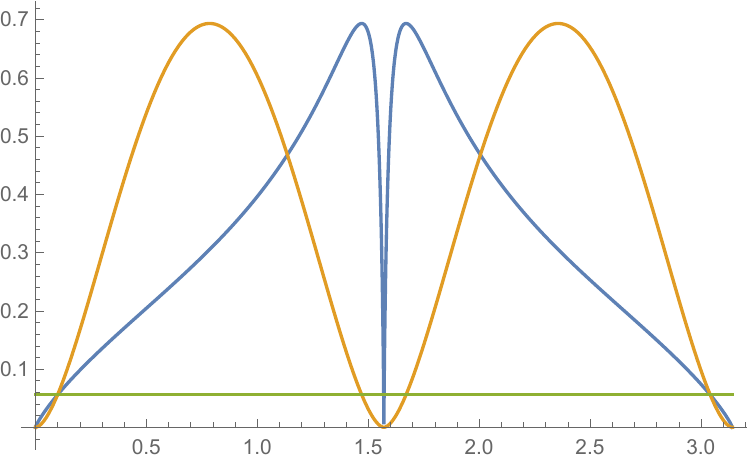}
    \caption{Plots of SVD entropy for the two qubit states in example 3 (\ref{x01}). The left panel shows $S(\rho^{1|2}_A)$ as a function of $\ap$ and $\beta$. The right panel shows plots of $S(\rho^{1|2}_A)$ (blue), $S(\rho^{1}_{A})$ (orange), and $S(\rho^{2}_{A})$ (green) as functions of $\ap$ in (\ref{x01}) and where $\beta=0.1$.} 
    \label{fig:qubit01}
\end{figure}

We choose
\ba
|\psi_1\lb=\cos\ap |00\lb+\sin\ap |11\lb,\no
|\psi_2\lb=\cos\beta |00\lb+\sin\beta |11\lb.\label{x01}
\ea
Then we find (we introduce $s_1=\sin\ap$ and $s_2=\sin\beta$  etc.)
\ba
\rho^{1|2}_A=\frac{1}{|c_1c_2|+|s_1s_2|}\Big(|c_1c_2||0\lb\la 0|+
|s_1s_2||1\lb\la 1|\Big).
\ea
Thus, we obtain the final expression
\ba
S(\rho^{1|2}_A)=-\frac{|c_1c_2|}{|c_1c_2|+|s_1s_2|}
\log\left[\frac{|c_1c_2|}{|c_1c_2|+|s_1s_2|}\right]-\frac{|s_1s_2|}{|c_1c_2|+|s_1s_2|}
\log\left[\frac{|s_1s_2|}{|c_1c_2|+|s_1s_2|}\right].
\ea
This is plotted in Figure~\ref{fig:qubit01}.
We find $S=\log 2$ for $\ap\pm\beta=\frac{\pi}{2}(2m+1)$. Also $S(\rho^{1|2}_A)=0$ when $\ap=\frac{\pi}{2}m$ or 
$\beta=\frac{\pi}{2}m$ for an integer $m$.

\subsection{Upper and lower bounds for SVD entanglement entropy}

\begin{figure}
\centering
\begin{tabular}{ccc}
\includegraphics[width=5cm]{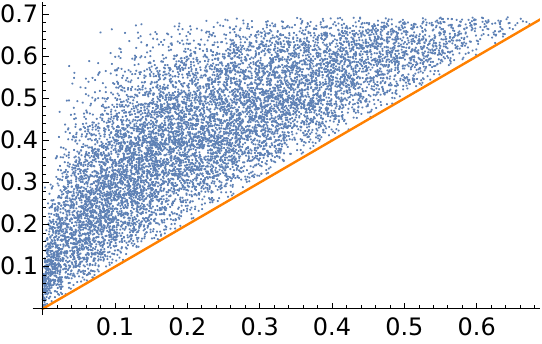}&\includegraphics[width=5cm]{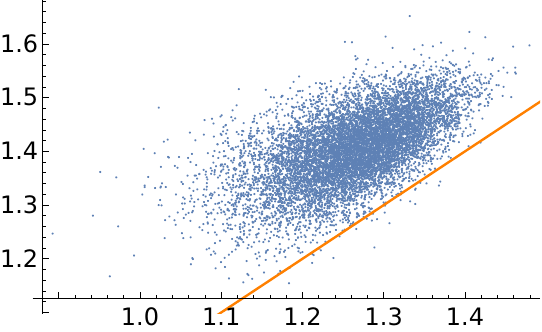}&\includegraphics[width=5cm]{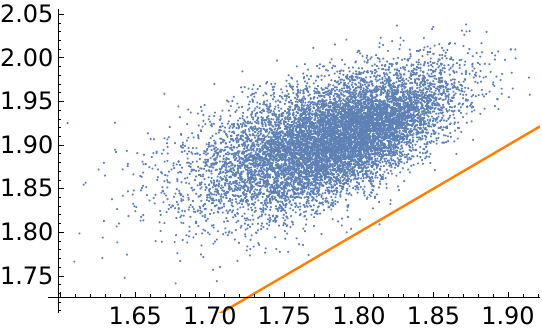}\\
\end{tabular}
\caption{These scatter plots are obtained by randomly sampling $|\psi_{1}\rangle,|\psi_{2}\rangle\in\mathcal{H}_{A}\otimes\mathcal{H}_{B}$ according to the Haar measure. 
From the left, the dimension $d_{A}=d_{B}$ takes values in $2$, $6$, and $10$, respectively. 
The $(x,y)$ components of the data points are given by $(\min\{S(\rho^{1}_{A}),S(\rho^{2}_{A})\},S(\rho^{1|2}_{A}))$. Therefore, any point above the orange diagonal line indicates that $\min\{S(\rho^{1}_{A}),S(\rho^{2}_{A})\}\le S(\rho^{1|2}_{A})$ is satisfied, while points below the orange diagonal indicate violations of this inequality.  Above, we illustrate that this inequality seems valid for $d_{A}=2$ but fails for $d_{A}\ge3$ (the case $d_{A}=3$ is not shown above, but several violations were found). Nevertheless, violations of this bound seem to be less likely as $d_{A}$ increases, suggesting the possibility that the probability of violation may tend to zero as the dimension increases.} 
\label{fig:mindAinc}
\end{figure}

One might wonder if there are any reasonable inequalities that hold for SVD entanglement entropy. Since SVD entropy is computed in terms of the standard von~Neumann entropy of an associated density matrix, it must be the case that it is bounded from below by $0$ and bounded from above by $\log(d_{A})$, i.e.\
\begin{equation}
0 \le S(\rho^{1|2}_{A}) \le \log(d_{A}),
\end{equation}
which illustrates a departure from the pseudo entropy of $\tau^{1|2}_{A}$. 
In the examples of Section~\ref{sec:qubitsexamples}, we explored the possibilities of improving this upper bound of $\log(d_{A})$ and possibly finding an improved lower bound in terms of the minimum of the marginal SVD entropies. 
In Figure~\ref{fig:mindAinc}, we provide numerical evidence suggesting that a non-trivial lower bound of $\min\{S(\rho_{A}^{1}),S(\rho_{A}^{2})\}$ might hold for qubits, fails in general for qudits, but seems to have a higher probability of holding as $d_{A}$ increases. 
Rigorous proofs of these observations for qubits and asymptotically large systems, provided that they hold, are still lacking.

\subsection{Monotonicity of SVD entanglement entropy under local unitaries}

We can rigorously prove, however, a restricted version of the lower bound observed above, which we will call monotonicity of SVD entanglement entropy under local unitaries. 

Let us consider a two-qubit system where the two qubits are called $A$ and $B$, respectively. Let $\ket{\psi_1}$ be 
\begin{align}
    \ket{\psi_1} = c\ket{00} + s\ket{11}, 
\end{align}
and let 
\begin{align}
    \ket{\psi_2} = c\ket{\alpha 0} + s \ket{\beta 1},
\end{align}
be a state that is obtained by performing a local unitary on $A$. In other words, there is a unitary $U$ such that 
\begin{align}
    \ket{\alpha} = U \ket{0},~
    \ket{\beta} = U \ket{1},
\end{align}
so that $\ket{\psi_2}=(U\otimes I)\ket{\psi_1}$.
Let us consider the un-normalized transition matrix made from these two states: 
\begin{align}
    |\psi_2\rangle \langle\psi_1| = c^2|00\rangle\langle00| + cs |\alpha 0\rangle\langle11| + cs |\beta1\rangle\langle00| + s^2 |\beta 1\rangle\langle 11|. 
\end{align}
We have already seen in Section~\ref{subsec:localunitary} that the SVD entropy of subsystem $A$ is unchanged from the standard entanglement entropy for both states regardless of the unitary $U$: 
\begin{align}
    S(\rho_A^{1|2}) = S(\rho_A^{1}) = S(\rho_A^{2}). 
\end{align}
On the other hand, the SVD entropy on subsystem $B$ in general changes as $U$ varies. 
The reduced transition matrix on $B$ looks like
\begin{align}
    \tau_B \propto c^2 U_{00}|0\rangle\langle0| 
    + cs U_{10} |0\rangle\langle1| 
    + cs U_{01} |1\rangle\langle0| 
    + s^2 U_{11}|1\rangle\langle1|.
\end{align}
Let us quickly see two examples. 
When $U = X$, where $X$ is the Pauli matrix, 
\begin{align}
    S(\rho_B^{1|2}) = \log 2, 
\end{align}
no matter what $c$ and $s$ are. 
The second example is when $c=s=1/\sqrt{2}$, in which case
\begin{align}
    S(\rho_B^{1|2}) = \log 2, 
\end{align}
no matter what $U$ is. These observations motivate us to conjecture that 
\begin{align}\label{eq:monotonicity}
    S(\rho_B^{1|2}) \geq S(\rho_B^{1}) = S(\rho_B^{2}). 
\end{align}
This is indeed the case in two-qubit systems. In fact, we can also find that the equality holds if and only if $c=1/\sqrt{2}$ (i.e. $\ket{\psi_1}$ is maximally entangled), or $c=1$ (i.e. $\ket{\psi_1}$ is separable), or $U = {\rm diag}(e^{i\theta_1},e^{i\theta_2})$. 
The proof is given below. Note that the conclusion here is a restricted version of 
\begin{align}
    S(\rho^{1|2}_{B}) \stackrel{?}{\geq} \min\{S(\rho^{1}_{B}), S(\rho^{2}_{B})\},
\end{align}
to the cases where $S(\rho^{1}_{B}) = S(\rho^{2}_{B})$.

\begin{proof}
    A general unitary $U$ can be written as 
    \begin{align}
        U = 
        e^{i\theta}\begin{pmatrix}
            p & -q^* \\
            q & p^* \\
        \end{pmatrix},
    \end{align}
    where $|p|^2 + |q|^2 = 1$. With this expression, 
    \begin{align}\label{eq:squared_matrix}
        \left(\rho^{1|2}_B\right)^2 \propto  
        \begin{pmatrix}
            c^4|p|^2 + c^2s^2|q|^2 & (c^3s-cs^3)p^*q \\
            (c^3s-cs^3)pq^* & s^4|p|^2 + c^2s^2|q|^2 \\
        \end{pmatrix}.
    \end{align}
    Let us denote the two eigenvalues of $\rho^{1|2}_B$ as $\lambda_+$ and $\lambda_-$. Then $\lambda_+$ and $\lambda_-$ are related as $\lambda_+ + \lambda_- = 1$. Without loss of generality, let us set $s\leq c$ and $\lambda_-\leq \lambda_+$ below. 
    To show \eqref{eq:monotonicity}, it is necessary and sufficient to show 
    \begin{align}\label{eq:pre_squared_inequality}
     0 \leq s^2 \leq \lambda_- \leq \lambda_+ \leq c^2 \leq 1. 
    \end{align}
    Furthermore, this is equivalent to showing
    \begin{align}\label{eq:squared_inequality}
     0 \leq \frac{s^4}{c^4+s^4} \leq \frac{\lambda_-^2}{\lambda_+^2+\lambda_-^2} \leq \frac{\lambda_+^2}{\lambda_+^2+\lambda_-^2} \leq 
     \frac{c^4}{c^4+s^4}
     \leq 1. 
    \end{align}
    The equivalence between \eqref{eq:pre_squared_inequality} and \eqref{eq:squared_inequality} follows directly from the fact that $f(x) = \frac{x^2}{x^2 + (1-x)^2}$ is a monotonic increasing function of $x$ when $x\in[0,1]$. 

    In the following, we will aim at showing \eqref{eq:squared_inequality}. From the expression \eqref{eq:squared_matrix}, one can find that $\frac{\lambda_-^2}{\lambda_+^2+\lambda_-^2}$ and $\frac{\lambda_+^2}{\lambda_+^2+\lambda_-^2}$ are two solutions of 
    \begin{align}
        g(x) \equiv x^2 - x + \frac{c^4 s^4}{\left((c^4+s^4)|p|^2 + 2c^2s^2 |q|^2\right)^2} = 0. 
    \end{align}
    Here, since $c^4+s^4 \geq 2c^2s^2$ and the equality holds only when $c = 1/\sqrt{2}$, 
    \begin{align}
        g(x) \geq x^2 -x + \frac{c^4s^4}{(c^4 + s^4)^2},
    \end{align}
    and the equality holds if and only if $c = 1/\sqrt{2}$ or $c=1$ or $|p|=1$.

    If we substitute $x = \frac{c^4}{c^4 + s^4}$ into $g(x)$, we find 
    \begin{align}
        g\left(\frac{c^4}{c^4 + s^4}\right) \geq \left(\frac{c^4}{c^4 + s^4}\right)^2 - \frac{c^4}{c^4 + s^4} + \frac{c^4s^4}{(c^4 + s^4)^2} = 0. 
    \end{align}
    Again, the equality holds if and only if $c = 1/\sqrt{2}$ or $c=1$ or $|p|=1$. 

    Therefore, \eqref{eq:squared_inequality} is shown and the fourth $\leq$ in \eqref{eq:squared_inequality} turns into $=$ if and only if $c = 1/\sqrt{2}$ or $c=1$ or $|p|=1$. As a result, \eqref{eq:monotonicity} is shown and the equality holds if and only if $c = 1/\sqrt{2}$ or $c=1$ or $|p|=1$.
\end{proof}

This monotonicity of SVD entanglement entropy under local unitaries does not hold in general dimensions. A simple counterexample can be easily constructed in a two-qutrit system. 

\paragraph{Example 4}
Consider a two-qutrit system and let
\begin{align}
    \ket{\psi_1} = \frac{1}{\sqrt{2}} (\ket{00} + \ket{11}), \\
    \ket{\psi_2} = \frac{1}{\sqrt{2}} (\ket{00} + \ket{21}).
\end{align}
Then, $\ket{\psi_2} = (U\otimes I) \ket{\psi_1}$, 
where 
\begin{align}
    U = |0\rangle\langle0| + |1\rangle\langle2| + |2\rangle\langle1|.
\end{align}
In this case, it is straightforward to show that $0 = S(\rho_B^{1|2}) < S(\rho_B^{1}) = S(\rho_B^{2}) = \log 2$, so that monotonicity does not hold.

\section{SVD entanglement entropy in integrable 2d CFTs}
\label{sec:SVDentint2dcfts}

Here we would like to calculate the SVD entropy in two dimensional conformal field theories (CFTs). However, the SVD entropy itself is difficult to calculate because it involves the square root of the 
reduced transition matrix, which is not clear how to implement by a path-integral. Instead we would like to compute its R\'enyi versions (\ref{RNPE}). We choose the subsystem $A$ to be an interval with length $L$ at a specific time. Note that if $|\psi_1\lb$ and  $|\psi_2\lb$ are both ground states in two dimensional CFTs, the standard results of (R{\'e}nyi) entangelement entropy in two dimensional CFTs \cite{Holzhey:1994we,Calabrese:2004eu} dictate
\be
S^{(n)}_{A(\text{vac})}=\frac{c}{6}\left(1+\frac{1}{n}\right)\log\frac{L}{\ep}.
\label{sneee}
\ee
Hence, we find
\ba
S^{(n,m)}_{A(\text{vac})}=\frac{c}{6}\cdot \frac{n+1}{mn}\cdot \log \frac{L}{\ep},
\label{vacnm}
\ea
where $\ep$ is an infinitesimally small constant associated with the UV cutoff. This can be found by plugging (\ref{sneee}) into the formula (\ref{fomnm}). Note that we can calculate $S^{(n,m)}_{A}$ using the replica method by adding the twist operators at the boundary of the subsystem $A$ when $m/2$ and $n$ are positive integers. When $m$ is odd, we need the analytical continuation of $m$ from even integers because $\rho^{(m)1|2}_A$ involves 
fractional powers of transition matrices as can be seen from (\ref{rma12}).

\begin{figure}[hhhh]
    \centering
    \includegraphics[width=.4\textwidth]{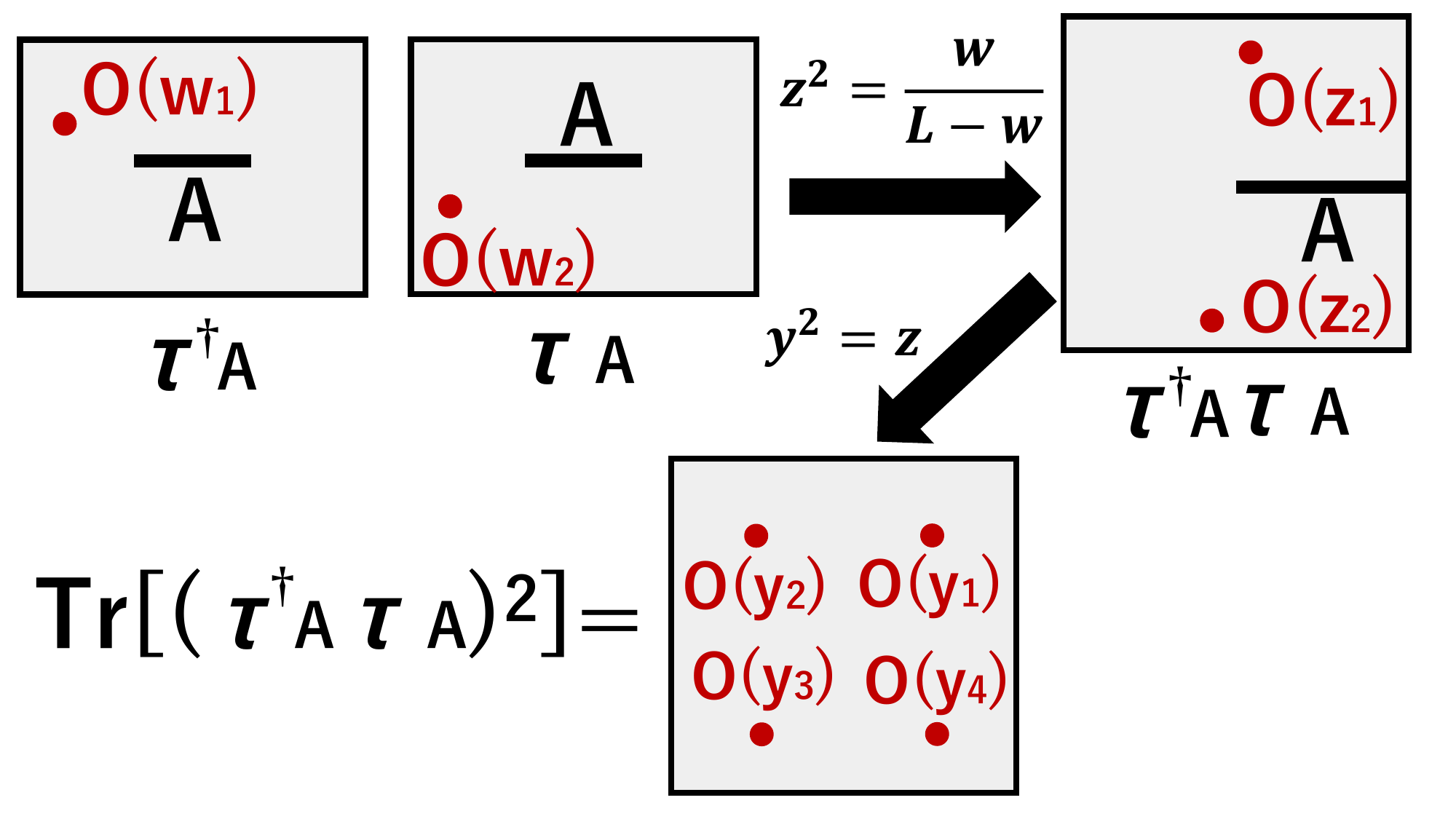}
    \hspace{10mm}
      \includegraphics[width=.4\textwidth]{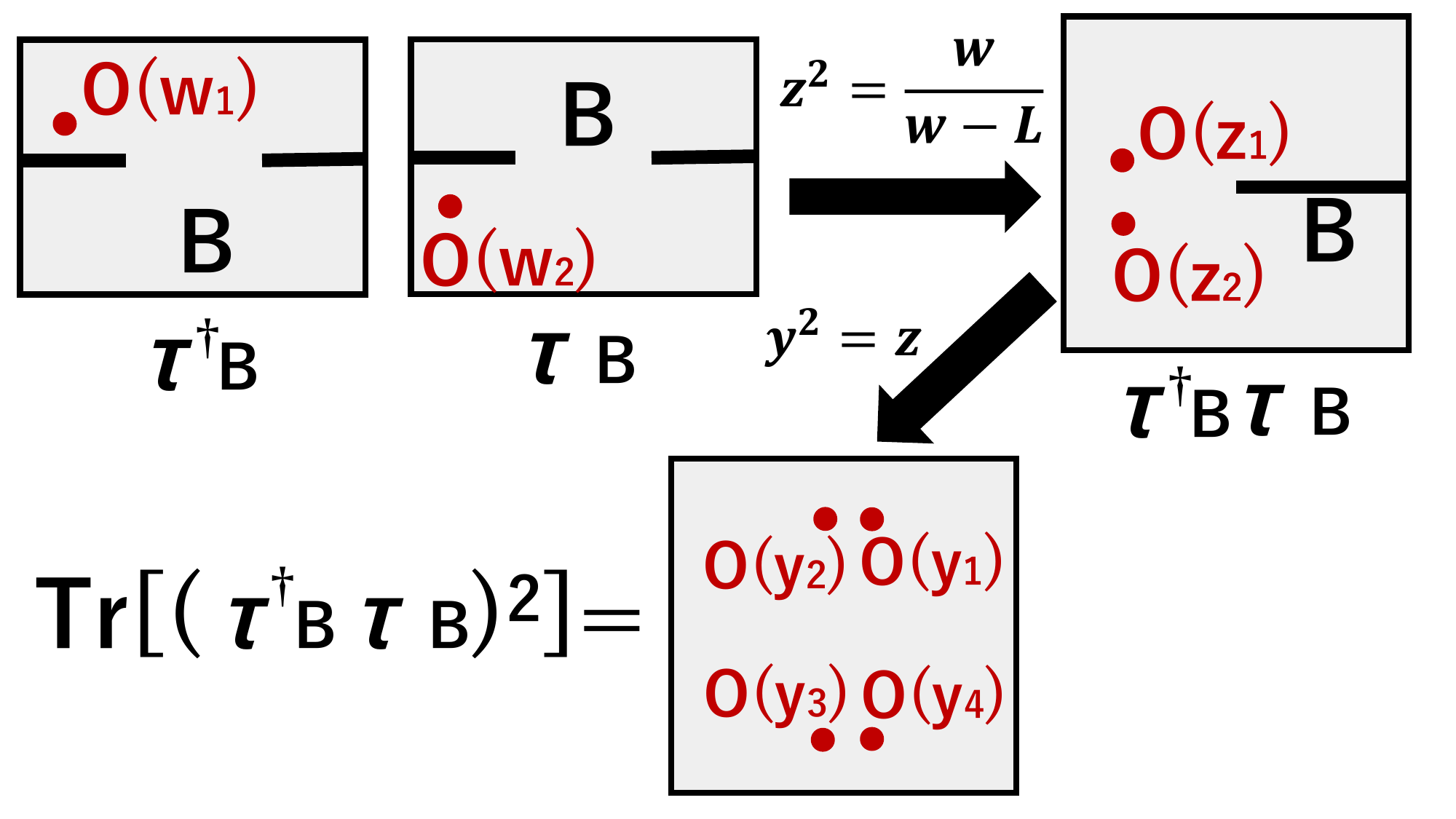}
    \caption{The replica calculation of $S^{(2,2)}_A$ (left) and $S^{(2,2)}_B$ (right). This illustrates that $S^{(2,2)}_A\neq S^{(2,2)}_B$ in general. In this figure, $\tau_{A}=\tau^{1|2}_{A}$ and $\tau_{B}=\tau^{1|2}_{B}$.}
    \label{fig:SABreplica}
\end{figure}

We would like to focus on the difference between the SVD entropy for excited states created by local primary operators and that for the vacuum states (\ref{vacnm}):  
\ba
\Delta S^{(n,m)}_A=S^{(n,m)}_A-S^{(n,m)}_{A(\text{vac})}.
\ea
More specifically, we focus on the $n=m=2$ case, i.e. $\Delta S^{(2,2)}_A$ and  $\Delta S^{(2,2)}_B$.
In the replica calculation, this can be computed as in Figure~\ref{fig:SABreplica}. Since this replica method calculation for local excitations is a straightforward generalization of that in the entanglement entropy \cite{Nozaki:2014hna,Nozaki:2014uaa,He:2014mwa} (see also \cite{Alcaraz:2011tn} for an earlier approach) and that in the pseudo entropy \cite{Nakata:2021ubr}, our description below will be minimal.

\subsection{The vacuum and a primary state}
\label{sec:VPS}

We consider the setup of a two dimensional CFT on a plane, whose complex coordinates are $(w,\bar{w})$. We choose the subsystem $A$ to be the interval $0\leq \mbox{Re}[w]\leq L$ at Im$[w]=0$.
We assume that $|\psi_1\lb$ is the locally excited state and 
 $|\psi_2\lb$ is the vacuum state, i.e.
\ba
|\psi_1\lb=e^{-bH_{\text{CFT}}}\mathcal{O}(a)|0\lb,\ \ \ \ \ |\psi_2\lb=|0\lb,
\label{localOP}
\ea
where the operator is inserted at $w=a-ib$. The parameter $a$ describes the spatial location and $b$ provides the UV regularization. Note that in this case we have $\la\psi_1|\psi_2\lb=0$.

We employ the conformal map
\ba
z^2=\frac{w}{L-w},  \label{cofa}
\ea
to calculate $S_A^{(n,2)}$ such that the cut along the subsystem $A$ is mapped into the positive real axis of the $z$ plane.
On the other hand, for $S_B^{(n,2)}$ we choose the map
\ba
z^2=\frac{w}{w-L}.  \label{cofb}
\ea
After these transformations, the matrices $\tau^\dagger_A\tau_A$ and $\tau^\dagger_B\tau_B$, where $\tau_{A}=\tau^{1|2}_{A}$ and $\tau_{B}=\tau^{1|2}_{B}$, are written as a path-integral on the $z$-plane with the cut along the positive real axis.

In the replica calculation, we choose the two locations for the operator insertions to be  
\ba
w_1=a+ib,\ \ \ \ w_2=a-ib,
\ea
as depicted in Figure~\ref{fig:SABreplica}. The parameter $a$ describes the location of the operator insertion, while the parameter $b$ is used to control the UV regularization or smearing of the local operator $\mathcal{O}$ because otherwise the two point function of local operators inserted at the same location is divergent.

For the subsystem $A$ we find 
\ba
&& z_1=\left(\frac{a+ib}{L-a-ib}\right)^{\frac{1}{2}}
=e^{\frac{\pi}{2}i}\left(\frac{-a-ib}{L-a-ib}\right)^{\frac{1}{2}},\no
&& z_2=-\left(\frac{a-ib}{L-a+ib}\right)^{\frac{1}{2}}
=e^{\frac{3\pi}{2}i}\left(\frac{-a+ib}{L-a+ib}\right)^{\frac{1}{2}}.\label{z1}
\ea

For the subsbstem $B$ we obtain
\ba
&& z_1=\left(\frac{a+ib}{a+ib-L}\right)^{\frac{1}{2}}
=e^{\pi i}\left(\frac{-a-ib}{L-a-ib}\right)^{\frac{1}{2}},\no
&& z_2=-\left(\frac{a-ib}{a-ib-L}\right)^{\frac{1}{2}}
=e^{\pi i}\left(\frac{-a+ib}{L-a+ib}\right)^{\frac{1}{2}}.
\label{z2}
\ea

To calculate Tr$[(\tau^\dagger_A\tau_A)^n]$, we glue 
the sheet with the cut whose path-integral represents the matrix $\tau^\dagger_A\tau_A$ successively $n$ times. 
This can be done by the conformal map
\ba
y^n=z.
\ea
Below we focus on the $n=2$ case.

For the subsystem $A$ we find
\ba
&& y_1=\left(\frac{a+ib}{L-a-ib}\right)^{\frac{1}{4}}
=e^{\frac{\pi}{4}i}\left(\frac{-a-ib}{L-a-ib}\right)^{\frac{1}{4}}=-y_3,\no
&& y_2=i\left(\frac{a-ib}{L-a+ib}\right)^{\frac{1}{4}}
=e^{\frac{3\pi}{4}i}\left(\frac{-a+ib}{L-a+ib}\right)^{\frac{1}{4}}=-y_4.
\ea
For the subsbstem $B$ we obtain
\ba
&& y_1=\left(\frac{a+ib}{a+ib-L}\right)^{\frac{1}{4}}
=e^{\frac{\pi}{2}i}\left(\frac{-a-ib}{L-a-ib}\right)^{\frac{1}{4}}=-y_3,\no
&& y_2=i\left(\frac{a-ib}{a-ib-L}\right)^{\frac{1}{4}}
=e^{\frac{\pi}{2}i}\left(\frac{-a+ib}{L-a+ib}\right)^{\frac{1}{4}}=-y_4.
\ea

For $n=2$, the vacuum-subtracted R\'enyi entropy is calculated by 
\ba
\Delta S^{(2,2)}_A=-\log F_2,
\ea
where $F_2$ is expressed by correlation functions of the CFT on the $n$-replicated manifolds $\Sigma_n$:
\ba
F_2=\frac{\mbox{Tr}[\tau^\dagger_A\tau_A\tau^\dagger_A\tau_A]}{\left(\mbox{Tr}[\tau^\dagger_A\tau_A]\right)^2}
=\frac{\la \mathcal{O}(w_1)\mathcal{O}(w_2)\mathcal{O}(w_3)\mathcal{O}(w_4)\lb_{\Sigma_4}}
{\left(\la \mathcal{O}(w_1)\mathcal{O}(w_2)\lb_{\Sigma_2}\right)^2}.
\ea 
By using the above conformal maps and well-known facts in two dimensional conformal field theories, we find that $F_2$ can take the form (refer to \cite{He:2014mwa} for more details)
\begin{align}
    F_2=|\eta|^{4h}|1-\eta|^{4h}G(\eta,\bar{\eta}).
    \label{fptwhhr}
\end{align}
Here, $h$ is the conformal dimension and $\eta$ is the cross ratio, which is explicitly given by 
\ba
\eta\equiv\frac{(y_1-y_2)(y_3-y_4)}{(y_1-y_3)(y_2-y_4)}=-\frac{(y_2-y_1)^2}{4y_1y_2}.
\ea
The function $G$ in (\ref{fptwhhr}) is the essential part of the four point function, which is model dependent and is known to depend only on the cross ratio $\eta$, owing to the conformal invariance. 

In $c=1$ free scalar CFT, for $\mathcal{O}=e^{i\frac{\phi}{2}}+e^{-i\frac{\phi}{2}}$ with the conformal dimension $h=\frac{1}{8}$, we find
\ba
F_2=\frac{1}{2}\left(1+|\eta|+|1-\eta|\right).
\ea
This gives $\Delta S_{A}^{(2,2)}=0$ because $F_2(\eta)=1$ for all values of $\eta$ in the range $0<\eta<1$, which holds in the present case.

\begin{figure}[hhhh]
    \centering
    \includegraphics[width=.4\textwidth]{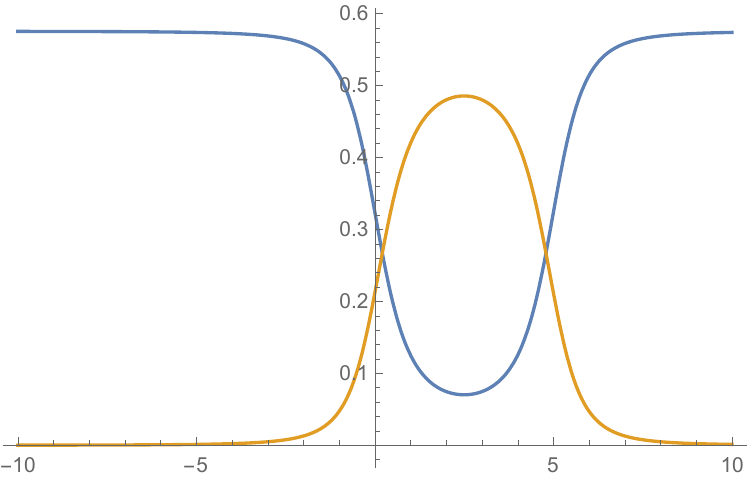}
    \hspace{10mm}
      \includegraphics[width=.4\textwidth]{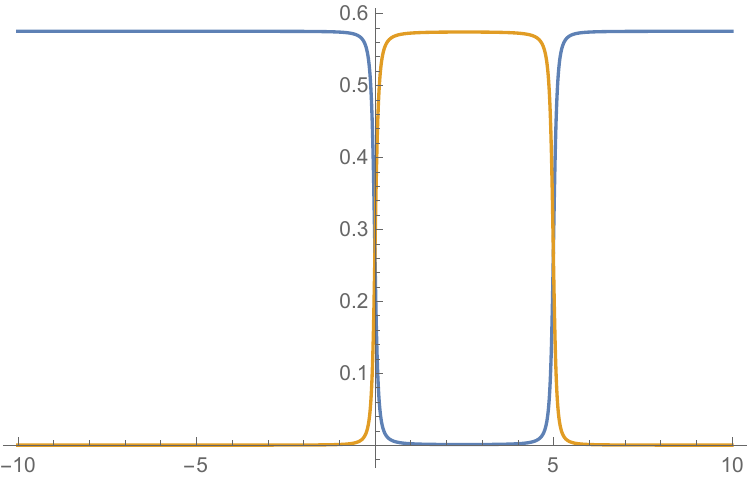}
    \caption{We plot $\Delta S^{(2,2)}_A$ (blue) and $\Delta S^{(2,2)}_B$ (orange) as functions of the location of the operator insertion $a$ for $L=5$ in the Ising CFT when $\mathcal{O}$ is the energy operator $\psi_L\psi_R$. We set $b=1$ and $b=0.1$ in the left and right panel, respectively.}
    \label{fig:Ising}
\end{figure}

In the Ising $c=\frac{1}{2}$ CFT, for the energy operator 
$\mathcal{O}=\psi_L\psi_R$, where $\psi_L$ and $\psi_R$ are the massless Majorana fermion in the left and right moving sectors, with the conformal dimension $h=\frac{1}{2}$, we obtain
\ba
F_2=|1-\eta+\eta^2|^2.
\ea
In this case $S^{(2)}_A$ and $S^{(2)}_B$ are non-zero,
as shown in Figure~\ref{fig:Ising}. The peak value in the limit $b\to 0$ is 
\ba
\Delta S^{(2,2)}_A=\log\frac{16}{9}.
\ea

Here we find $S^{(2,2)}_A$ does not increase when the local operator acts on the subsystem $A$. This can be heuristically understood by noting that a unitary transformation $U_A$ acting on $A$, which describes the local excitation on $A$, does not change the SVD entropy since we can absorb such a unitary transformation as in 
(\ref{unitary}). On the other hand, $S^{(2,2)}_B$ can change its value due to a local excitation on $A$, as explained in Section~\ref{subsec:localunitary}.

\subsection{Time Evolution in free scalar CFT}

Next we would like to analyze the time evolution of $\Delta S^{(2,2)}_A$
and  $\Delta S^{(2,2)}_B$ for the locally excited states (\ref{localOP}) in 
$c=1$ free scalar CFT. 
To analyze the time evolution we set
\ba\begin{aligned}
& w_1=-l+i(b-it),\ \ \ & \bar{w}_1&=-l-i(b-it), \\
& w_2=-l+i(-b-it),\ \ \ & \bar{w}_2&=-l-i(-b-it). 
\end{aligned}\ea
The subsystem is now taken to be the length $L$ interval $[0,L]$ and $b$ is the UV regularization parameter of the local excitation.

First let us compute $\Delta S^{(2,2)}_A$. Using the map (\ref{cofa}) for subsystem $A$, we find

\ba\begin{aligned}
&y_1=e^{\frac{\pi}{4}i}\left(\frac{l-t-ib}{L-t+l-ib}\right)^{\frac{1}{4}}=-y_3,\\
&y_2=e^{\frac{3\pi}{4}i}\left(\frac{l-t+ib}{L-t+l+ib}\right)^{\frac{1}{4}}=-y_4.
\end{aligned}\ea

For $0<t<l$, in the $b\to 0 $ limit, we find
\ba\begin{aligned}
&y_1=e^{\frac{\pi}{4}i}\left(\frac{l-t}{L-t+l}\right)^{\frac{1}{4}}=-y_3,\\
&y_2=e^{\frac{3\pi}{4}i}\left(\frac{l-t}{L-t+l}\right)^{\frac{1}{4}}=-y_4.
\end{aligned}\ea
Therefore, we obtain $\eta=\bar{\eta}=\frac{1}{2}$, which leads to 
$\Delta S^{(2,2)}_A=0$. This result is expected since the entangled pair created at $x=-l$ does not reach the subsystem $A$ at this early time.

For $l<t<l+L$,  in the $b\to 0 $ limit, we find
\ba\begin{aligned}
&y_1=\left(\frac{t-l}{L-t+l}\right)^{\frac{1}{4}}=-y_3,\\
&y_2=-\left(\frac{t-l}{L-t+l}\right)^{\frac{1}{4}}=-y_4.
\end{aligned}\ea
Therefore, we obtain $\eta=1$ and $\bar{\eta}=\frac{1}{2}$, which leads to 
\ba
\Delta S^{(2,2)}_A=\log(4-2\s{2}). \label{SVDva}
\ea
Indeed, in this later time, one of the entangled pairs created at $x=-l$ is propagated to $A$.
\\

Next we calculate $S^{(2,2)}_B$. By using the map (\ref{cofb}) for subsystem $A$, which is the interval $[0,L]$, we find

\ba\begin{aligned}
&y_1=e^{\frac{\pi}{2}i}\left(\frac{l-t-ib}{L-t+l-ib}\right)^{\frac{1}{4}}=-y_3,\\
&y_2=e^{\frac{3\pi}{2}i}\left(\frac{l-t+ib}{L-t+l+ib}\right)^{\frac{1}{4}}=-y_4.
\end{aligned}\ea

For $0<t<l$, in the $b\to 0 $ limit, we find
\ba\begin{aligned}
&y_1=e^{\frac{\pi}{2}i}\left(\frac{l-t}{L-t+l}\right)^{\frac{1}{4}}=-y_3,\\
&y_2=e^{\frac{3\pi}{2}i}\left(\frac{l-t}{L-t+l}\right)^{\frac{1}{4}}=-y_4.
\end{aligned}\ea
Therefore, we obtain $\eta=\bar{\eta}=0$, which leads to 
$\Delta S^{(2,2)}_B=0$. 

For $l<t<l+L$,  in the $b\to 0 $ limit, we find
\ba\begin{aligned}
&y_1=e^{\frac{\pi}{4}i}\left(\frac{t-l}{L-t+l}\right)^{\frac{1}{4}}=-y_3,\\
&y_2=e^{\frac{3\pi}{4}i}\left(\frac{t-l}{L-t+l}\right)^{\frac{1}{4}}=-y_4.
\end{aligned}\ea
Therefore, we obtain $\eta=\frac{1}{2}$ and $\bar{\eta}=0$, which leads to 
$\Delta S^{(2,2)}_B=\log(4-2\s{2})$.  In this way, the result of $\Delta S^{(2,2)}_B$ is the same as $\Delta S^{(2,2)}_A$. As opposed to the $t=0$ case studied in the previous section, the SVD entropy after time evolution becomes non-vanishing even for the free scalar CFT. It would be an interesting future problem
to give a physical or quantum information-theoretic interpretation of the value
(\ref{SVDva}).

\subsection{Two primary states}
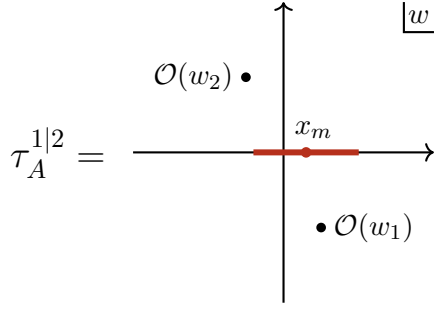
\begin{figure}
    \centering
    \begin{tikzpicture}[thick]
        \begin{scope}
            \draw[->] (0,-2)--(0,2);
            \draw[->] (-2,0)--(2,0);
            \draw[BrickRed,line width=0.80mm] (-0.4,0)--(1,0);
            \draw[draw=none,fill=BrickRed] (0.3,0) circle (0.07);
            \node at (0.4,0.3) {$x_m$};
            \draw (1.6,2)--(1.6,1.6)--(2,1.6);
            \node at (1.8,1.8) {$w$};
            \node at (1.2,-1) {$\mathcal{O}(w_1)$};
            \draw[fill] (0.5,-1) circle (0.05);
            \draw[fill] (-0.5,1) circle (0.05);
            \node at (-1.2,1) {$\mathcal{O}(w_2)$};
            \node at (-3,0) {\Large $\tm{1}{2}_A=$};
        \end{scope}
    \end{tikzpicture}
    \caption{Path integral representation of the transition matrix for two primary states $\ket{\psi_1}=\mathcal{O}(w_1,\bar{w}_1)\ket{0},\,\ket{\psi_2}=\mathcal{O}(\bar{w}_2,w_2)\ket{0}$ with $w_1=a_1-ib_1,w_2=a_2+ib_2$. We denote the center of subsystem $A$ by $x_m$.} 
    \label{fig:2ndRenyi}
\end{figure}
It is also interesting to consider two locally excited states given by 
\begin{align}\begin{aligned}
    \ket{\psi_1}&=\mathcal{O}(w_1,\bar{w}_1)\ket{0},\\
    \ket{\psi_2}&=\mathcal{O}(\bar{w}_2,w_2)\ket{0},
\end{aligned}\end{align}
where $\mathcal{O}$ is the same operator as the previous subsection:
\begin{align}
    \mathcal{O}=e^{\frac{i}{2}\phi}+e^{-\frac{i}{2}\phi},
\end{align}
whose conformal dimension is $h=\bar{h}=1/8$.
We parameterize the real and imaginary parts of $w_1$ and $w_2$ as 
\begin{align}
    w_1=a_1-ib_1,\quad w_2=a_2+ib_2.
\end{align}
The unnormalized transition matrix 
\begin{align}
    \ttm{1}{2}_A=\ket{\psi_1}\bra{\psi_2}
\end{align}
is given by the path integral with two operator insertions at $w=w_1,w_2$, keeping the subsystem $A$ open (see Figure~\ref{fig:2ndRenyi}).
We take the subsystem $A$ to be an interval with length $L$ centered at $x=x_m$, where the subscript ``$m$'' stands for ``midpoint.'' 
The 2nd R\'{e}nyi entropy for the density matrix 
\begin{align}
    \rho^{(2)1|2}_A=\frac{(\tm{1}{2}_A)^\dagger\tm{1}{2}_A}{\Tr[(\tm{1}{2}_A)^\dagger\tm{1}{2}_A]}
\end{align}
is given by 
\begin{align}\label{2Renyipp}
    \Delta S^{(2,2)}_A=-\log\frac{\langle\mathcal{O}_{1}\mathcal{O}^\dagger_{2}\mathcal{O}^\dagger_{3}\mathcal{O}_{4}\mathcal{O}_{5}\mathcal{O}^\dagger_{6}\mathcal{O}^\dagger_{7}\mathcal{O}_{8}\rangle_{\Sigma_4}}{\left(\langle\mathcal{O}_{1}\mathcal{O}^\dagger_{2}\mathcal{O}^\dagger_{3}\mathcal{O}_{4}\rangle_{\Sigma_2}\right)^2},
\end{align}
where we have introduced a shorthand notation $\mathcal{O}_i\equiv \mathcal{O}(w_i,\bar{w}_i)$ and the operators on the replicated manifolds $\Sigma_4,\Sigma_2$ are fixed at
\begin{align}\begin{aligned}
    w_3&=\bar{w}_1,\quad w_4=\bar{w}_2,\\
    w_{i+4}&=-w_i,\quad (i=1,2,3,4).
\end{aligned}\end{align}
Here we assume that $\bar{w}_1,\bar{w}_2$ are the complex conjugates of $w_1,w_2$, so that we do not consider any analytic continuations such as the real time evolution. The correlation function in \eqref{2Renyipp} is calculated in the same way as the previous subsection. Here we consider the following two cases, which have the same setup as in Sections 4.1 and 4.2 of \cite{Nakata:2021ubr}. 

\begin{figure}
    \centering
    \includegraphics[width=0.49\textwidth]{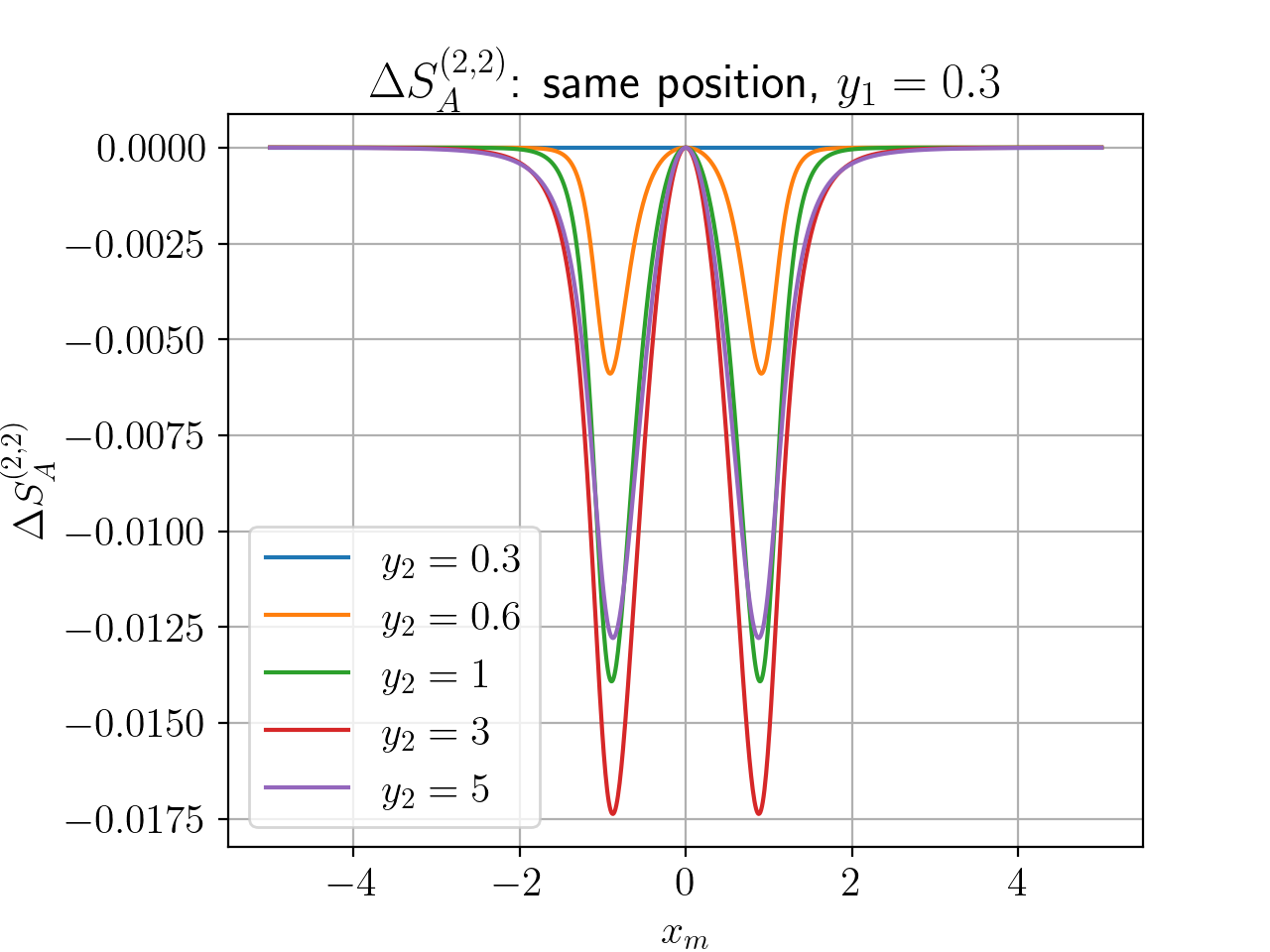}
    \includegraphics[width=0.49\textwidth]{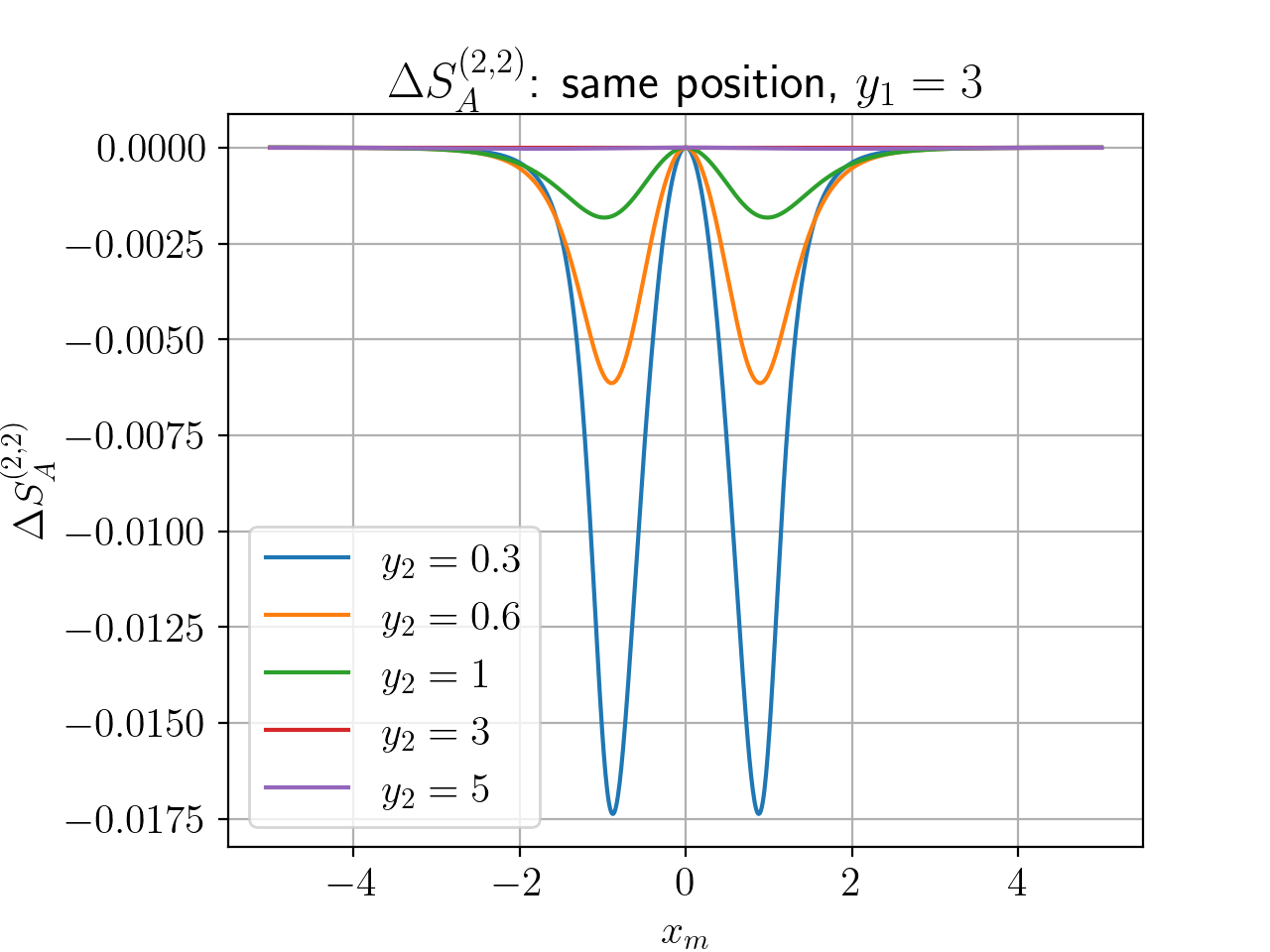}
    \includegraphics[width=0.49\textwidth]{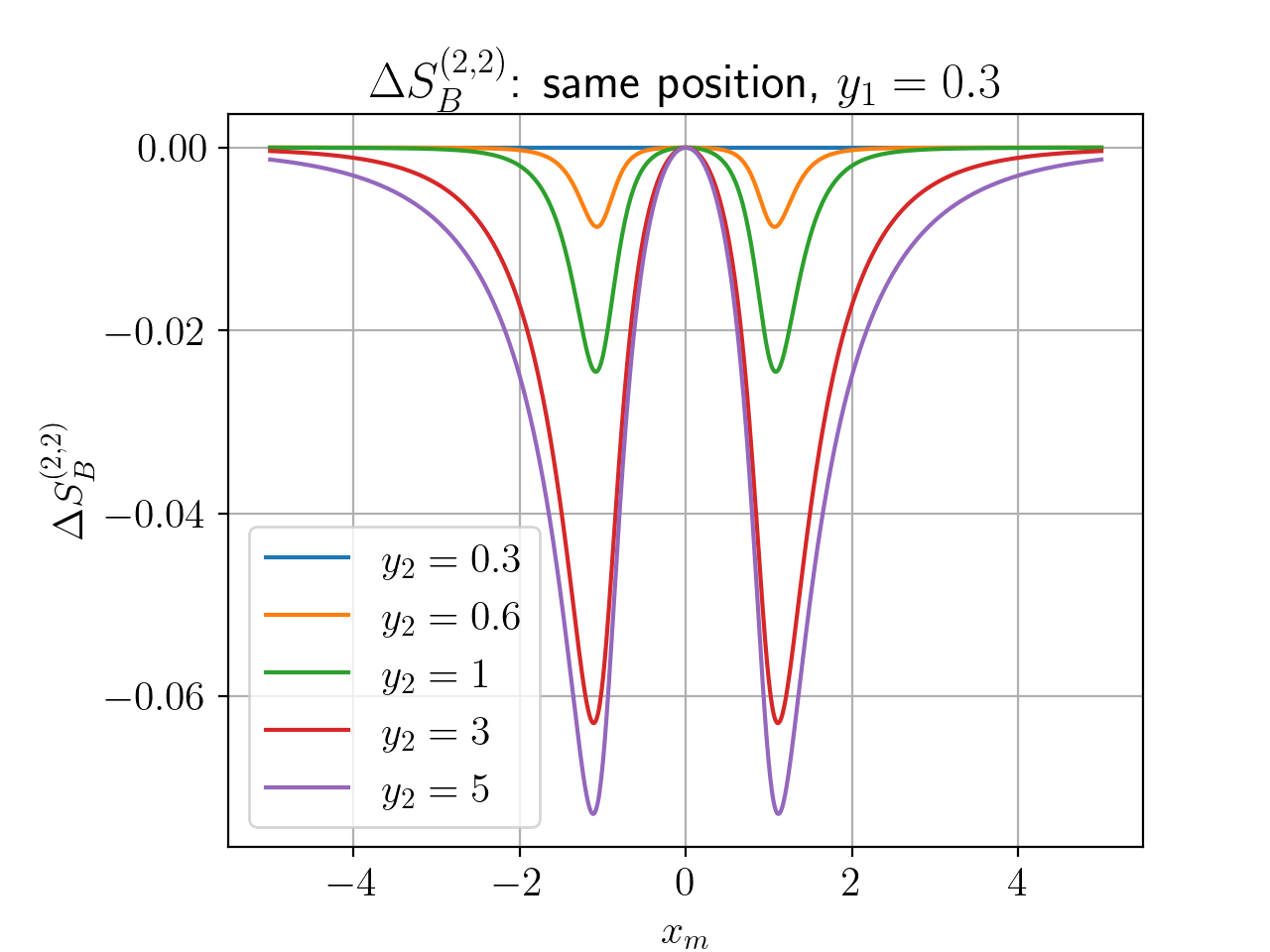}
    \includegraphics[width=0.49\textwidth]{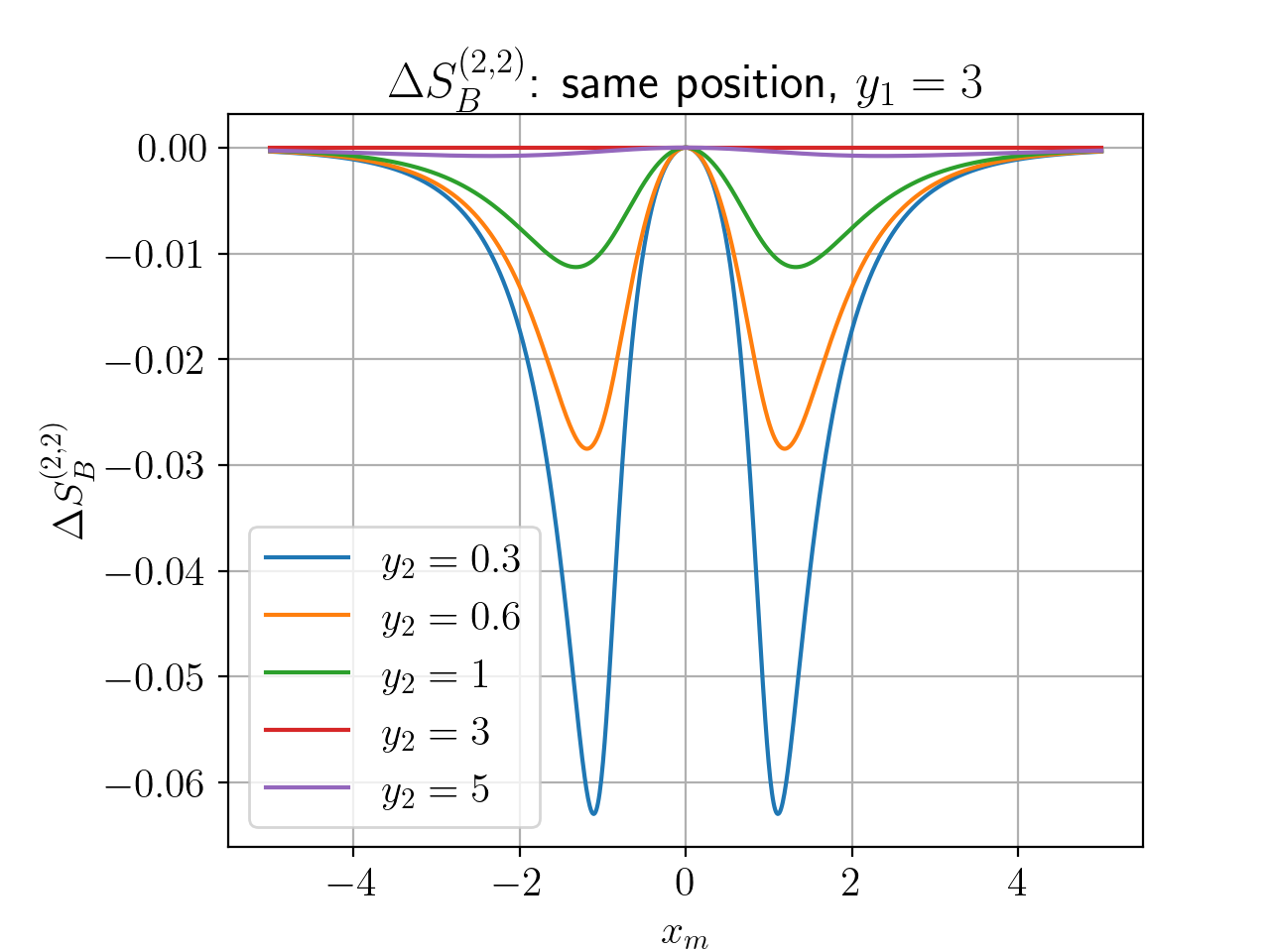}
    \caption{We set the length of subsystem $A$ as $L=2$ and the locations of the two excitations as $a_1=a_2=0$. The upper panels are plots of the second R\'{e}nyi entropy $\Delta S^{(2,2)}_A$ for subsystem $A$ as a function of $x_m$, the center of subsystem $A$, when $b_1=0.3$ and $b_1=3$, respectively. The blue, orange, green, red, and purple curves correspond to $b_2=0.3,0.6,1,3,$ and $5$, respectively. The lower panels are plots of the second R\'{e}nyi entropy $\Delta S^{(2,2)}_B$ for subsystem $B$ as a function of $x_m$. The parameters are the same as the upper ones.}
    \label{fig:twoprimaries}
\end{figure}

\paragraph{Exciting at the same position with different cutoffs}
In Figure~\ref{fig:twoprimaries}, the SVD entropies $\Delta S^{(2,2)}_A$ and $\Delta S^{(2,2)}_B$ are plotted in the case where two operators are excited at the same position $a_1=a_2=0$ and the cutoffs $b_1,b_2$ are different. First, a universal feature seen in both $\Delta S^{(2,2)}_A$ and $\Delta S^{(2,2)}_B$ is that the absolute values are large when $b_1$ and $b_2$ are different and sufficiently small. This behavior is similar to the original pseudo entropy as seen in Sections 4.1 and 4.2 of \cite{Nakata:2021ubr}, although the original pseudo entropies for subsystems $A$ and $B$ are identical. Second, we can see that the absolute value of $\Delta S^{(2,2)}_B$ is larger than that of $\Delta S^{(2,2)}_A$ using the same configurations of operator excitations. This behavior is analogous to a general property described in Section~\ref{subsec:localunitary}. Namely, a unitary operation acting on $\mathcal{H}_A$ ($\mathcal{H}_B$) affects the SVD entropy $S_B$ ($S_A$) defined on the complement of the subsystem.

\begin{figure}
    \centering
    \includegraphics[width=0.49\textwidth]{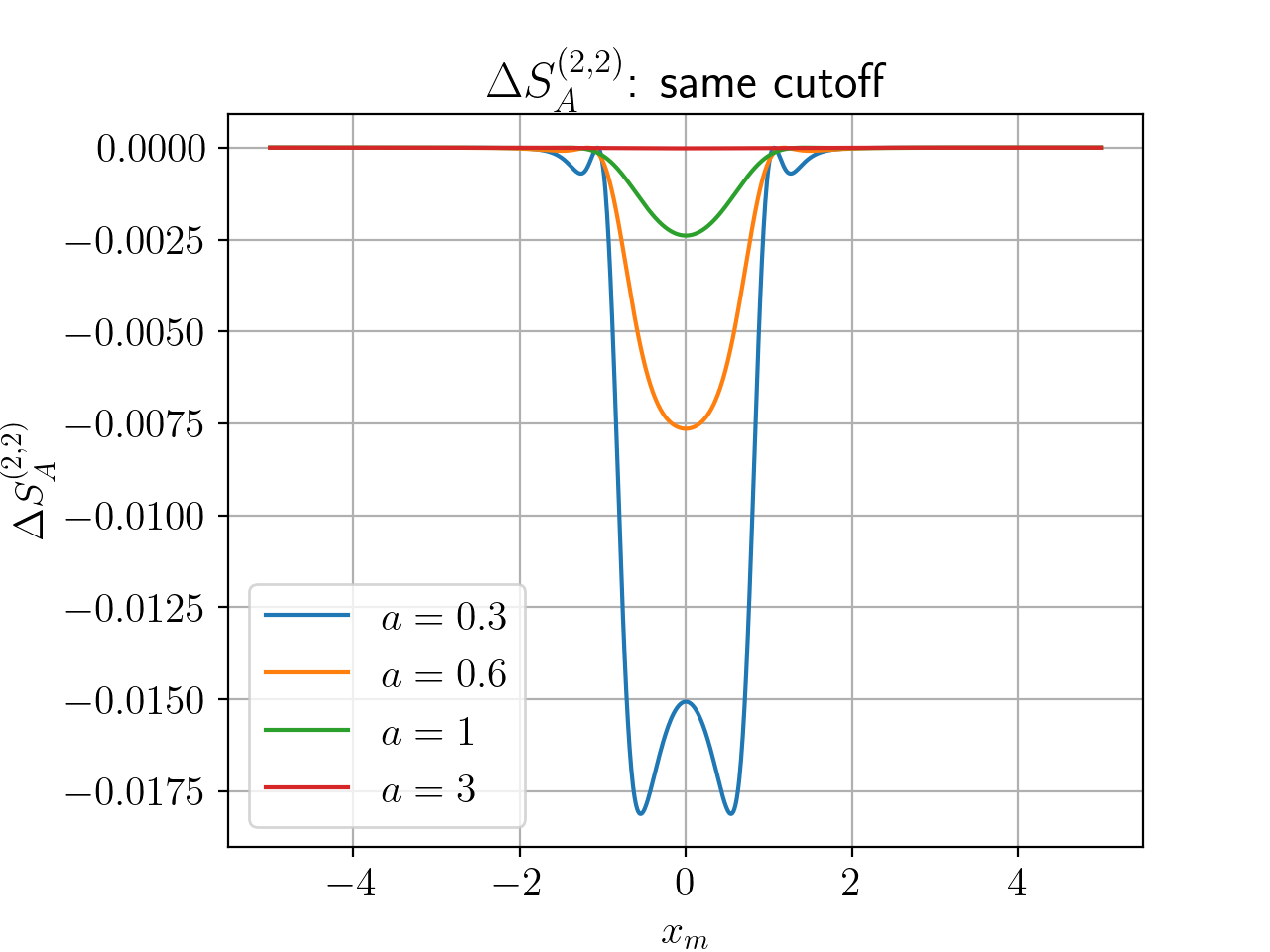}
    \includegraphics[width=0.49\textwidth]{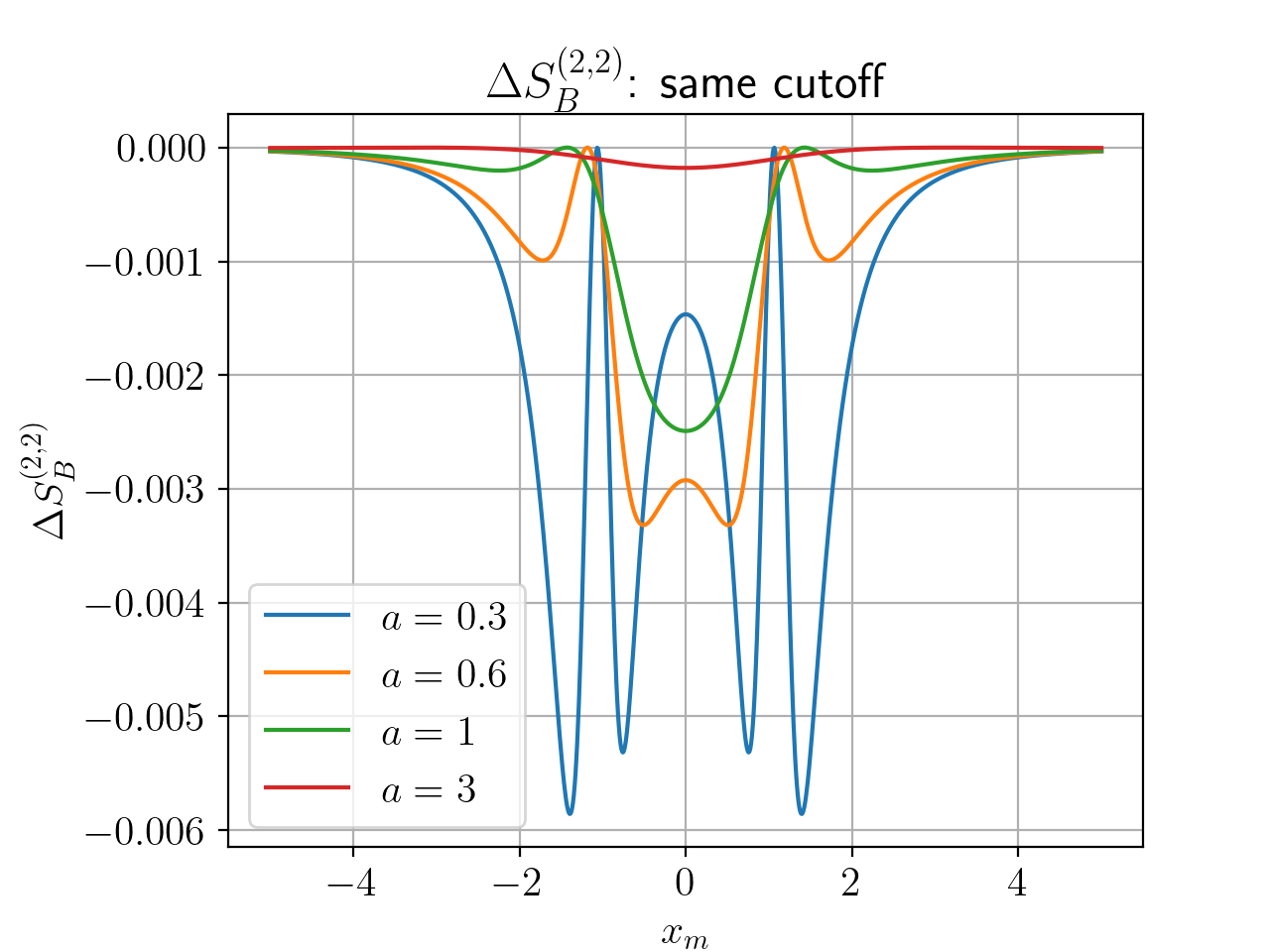}
    \caption{Two excitation points are $a_1=0.2$ and $a_2=-0.2$, while the cutoffs are the same $b_1=b_2=b$. The left and right panels show $\Delta S^{(2,2)}_A$ and $\Delta S^{(2,2)}_B$ as functions of the center $x_m$ of subsystem $A$. In both figures, the length of subsystem is fixed to be $L=2$ and the blue, orange, green, and red curves represent $b=0.3,0.6,1,$ and $3$, respectively.}
    \label{fig:samecutoff}
\end{figure}

\paragraph{Exciting at different positions with same cutoff} 
In Figure~\ref{fig:samecutoff}, the SVD entropies $\Delta S^{(2,2)}_A,\Delta S^{(2,2)}_B$ are plotted in the case where two operators are excited at different positions fixed at $a_1=-a_2=0.2$, while the cutoffs are identical $b_1=b_2=b$. These behave as expected from the above case: the absolute values of $\Delta S^{(2,2)}_A,\Delta S^{(2,2)}_B$ get larger as the edges of the subsystem approach the operator insertions and the cutoff $b$ is small. Around the two peaks where the right edge is close to $a_2=-0.2$ and the left edge is close to $a_1=0.2$, $\left|\Delta S^{(2,2)}_B\right|$ is larger than $\left|\Delta S^{(2,2)}_A\right|$ as we expected from the above cases. However, a clearer understanding is missing for the middles peaks due to the complicated contributions from two excited operators.  

\begin{figure}
    \centering
    \includegraphics[width=0.6\textwidth]{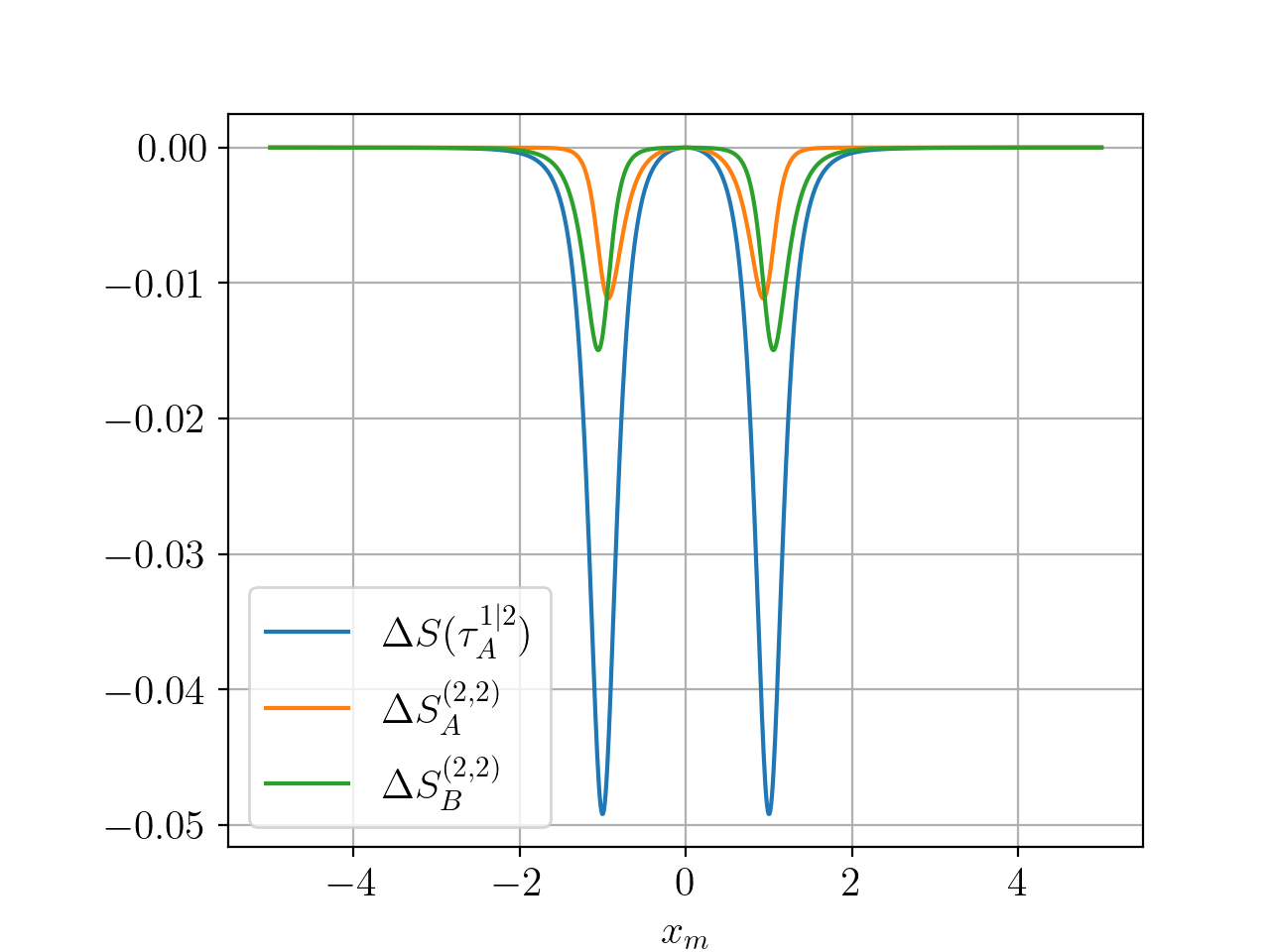}
    \caption{Comparison among pseudo entropy (blue) and SVD entropy for subsystems $A$ (orange) and $B$ (green) with the values $a_1=a_2=0,b_1=0.5,$ and $b_2=0.2$.}
    \label{fig:PEandSVD}
\end{figure}

\paragraph{Comparison with pseudo entropy}
As we have observed, the behavior of SVD entropy is similar to pseudo entropy, with one difference being that $\Delta S^{(2,2)}_A\neq \Delta S^{(2,2)}_B$ for the SVD entropy. Another difference is that the absolute value of the difference $\Delta S^{(2,2)}$ of SVD entropy from the vacuum value is suppressed compared with that of the original pseudo entropy, as shown in Figure~\ref{fig:PEandSVD}.

\section{SVD entanglement entropy in Holographic CFTs} \label{sec:hol_CFT}

Another interesting direction is to calculate SVD entropy for holographic CFTs, which correspond to CFTs in the strong coupling limit and in the large $N$ (or large $c$) limit. Holographic CFTs have gravity dual descriptions in light of the AdS/CFT correspondence \cite{Maldacena:1997re}.
In AdS/CFT, the entanglement entropy (\ref{EEf}) can be computed geometrically by the area of minimal surface $\Gamma_A$ in a geometry dual to the original quantum state $|\psi\lb$ \cite{Ryu:2006bv,Ryu:2006ef,Hubeny:2007xt} as
\ba
S(\rho_A)=\frac{A(\Gamma_A)}{4G_N},\label{RTF}
\ea
where $\Gamma_A$ is the minimal area surface whose boundary coincides with the boundary of subsystem $A$.

Thus, if we can find a gravity dual of $\rho^{1|2}_A$, we can calculate the SVD entropy of the associated reduced transition matrix. However, for given states $|\psi_1\lb$ and 
 $|\psi_2\lb$ in a CFT, this is difficult in general because we need to take the square root in (\ref{rho12}) in a path-integral formalism of a CFT. To circumvent this issue, here we instead consider a holographic calculation of a related functional, namely 
\ba
S^{(1,2)}_A=S(\rho^{(2)1|2}_A)=-\mbox{Tr}\left[\rho^{(2)1|2}_A\log \rho^{(2)1|2}_A\right] ,
\ea
where $\rho^{(2)1|2}_A$ is defined by (\ref{rma12}), more explicitly, 
\ba
\rho^{(2)1|2}_A 
=\frac{\tau^{1|2\dagger}_A \tau^{1|2}_A}{
\mbox{Tr}\left[\tau^{1|2\dagger}_A\tau^{1|2}_A\right]}.\label{RSVDF}
\ea
For this, if we are able to prepare the gravity dual of the state $\rho^{(2)1|2}_A$, which should be more easily achievable than for $\rho^{1|2}_A$ because there is no square root in $\rho^{(2)1|2}_A$, then we can calculate the R{\'e}nyi version of the SVD entropy (\ref{RSVDF}) using the metric of the dual geometry by applying the geometric formula (\ref{RTF}). This is what we accomplish in this section. 

Before we go on to the evaluation of $S^{(1,2)}_A$ in an explicit example, we would like to mention that we can formally calculate the genuine SVD entropy $S^{(1,1)}_A$ by computing the area of the holographic entanglement entropy in the gravity dual of $U_A^\dagger\cdot \tau^{1|2}_A\cdot V_A$ (see Eq.~(\ref{unitary})). Similarly, we can calculate  $S^{(1,1)}_B$ from the dual geometry of $U_B^\dagger\cdot \tau^{1|2}_B\cdot V_B$. This strongly suggests  $S^{(1,1)}_A\neq S^{(1,1)}_B$ in general, as opposed to the pseudo entropy.

\subsection{Holographic SVD entropy from Janus Solution}

Consider the setup in 2d CFT where $|\psi_1\lb$ is defined by the vacuum state of a CFT defined by an exactly imaginary perturbation 
\ba
\lambda\int d^2w\,\mathcal{O}(w,\bar{w}),
\ea
of an original CFT, while $|\psi_2\lb$ is given by the vacuum $|0\lb$ of the original CFT without the perturbation.

\begin{figure}[hhhh]
    \centering
    \includegraphics[width=.6\textwidth]{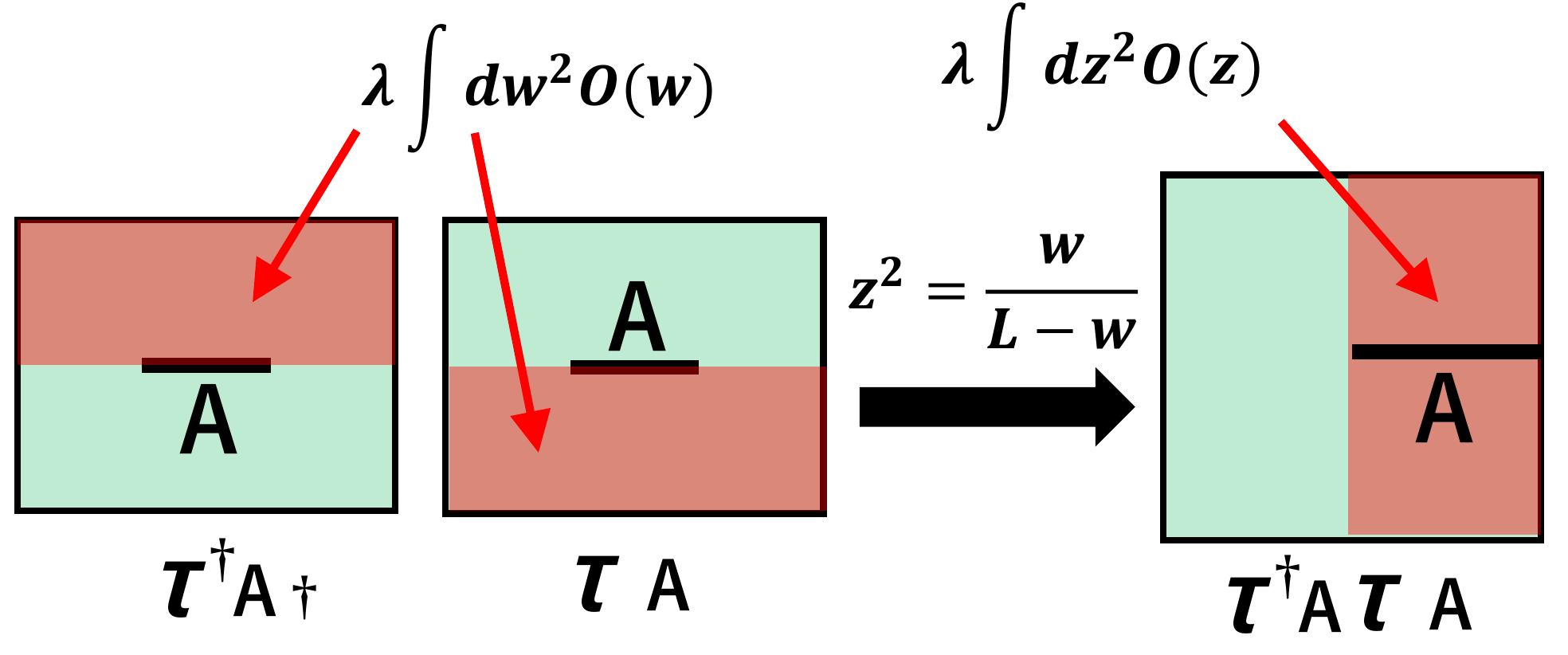}
    \caption{Holographic Calculation of $S_A$ using the Janus Solution.}
    \label{fig:Janus}
\end{figure}

By applying the conformal map (\ref{cofa}), we can calculate $S^{(1,2)}_A$ as the holographic entanglement entropy in the geometry dual to $\rho^{(2)1|2}_A$, as depicted in Figure~\ref{fig:Janus}. Note that the conformal factor via the transformation  (\ref{cofa}) does not show up as we consider the exactly marginal perturbation.
Thus, the SVD entropy $S^{(1,2)}_A$ can be computed by the formula (\ref{RTF}),
where $\Gamma_A$ is the minimal surface that connects the two endpoints of $A$, which are chosen to be $w=0$ and $w=L$, equivalently
$z=0$ and $z=\infty$, respectively. To regulate the result, we further perform the conformal map
\ba
\ti{z}=\frac{z}{z+1},
\ea
so that the end points of $A$ are now $\ti{z}=0$ and $\ti{z}=1$. Now we would like to estimate the UV cutoff $\delta$ in this $\ti{z}$ coordinate. By considering the region near $\ti{z}=0$, i.e. $w=0$, we find that the UV cutoff $\ep$ in the $w$ coordinate is mapped into that in the $\ti{z}$ coordinate as 
\ba
\delta =\s{\frac{\ep}{L}}.
\ea
If both $|\psi_1\lb$ and $|\psi_2\lb$ are both the vacuum $|0\lb$, 
$S_A$ is given by 
\ba
S^{(1,2)}_A=\frac{c}{3}\log\frac{1}{\delta}=\frac{c}{6}\log\frac{L}{\ep},\label{EEvac}
\ea
which agrees with what we expect, i.e.~(\ref{vacnm}). 

In the presence of the exactly marginal perturbation, we can employ the Janus solution \cite{Bak:2007jm} as the gravity dual to $\rho^{(2)1|2}_A$.  
It is an asymptotically AdS$_3$ solution in the Einstein-scalar theory. The metric $ds^2$ and massless scalar field $\phi$ of the Janus solution read
\ba
&& ds^2=dy^2+\left(\frac{1+\s{1-2\gamma^2}\cosh 2y}{2}\right)ds^2_{AdS2},\no
&& \phi=\phi_0+\frac{1}{\s{2}}\log\left[\frac{1+\s{1-2\gamma^2}+\s{2}\gamma\tanh y}{1+\s{1-2\gamma^2}-\s{2}\gamma\tanh y}\right].\label{soljan}
\ea
The difference of the value of the massless scalar field in the $y\to \infty$ and $y\to -\infty$ limit is dual to the exact marginal perturbation. Thus, the Janus deformation parameter $\gamma$ is related to the parameter $\lambda$ such that $\lambda\propto \gamma$ when $\gamma$ is small. When $\gamma=0$, the scalar field becomes $\phi_{0}$ and the metric is the one of pure AdS$_3$.

The Janus metric (\ref{soljan})  allows us to determine the coefficient in front of $\log\frac{b}{\delta}$ in (\ref{EEvac}) by considering the geodesic length in that background. This leads to the following estimation of SVD entropy:
\ba
S^{(1,2)}_A=\frac{c}{6}\s{\frac{1+\s{1-2\gamma^2}}{2}}\log \frac{L}{\ep},
\ea
where $\gamma$ is linearly related to $\lambda$.
Note that this shows that the SVD entropy is decreasing as the exactly marginal deformation grows.

\subsection{SVD entanglement entropy for locally excited states in Holographic CFTs}

In the holographic CFTs with the large $c$ limit, we can apply the Heavy-Heavy-Light-Light conformal block approximation \cite{Fitzpatrick:2014vua}.
This leads to the following result 
\ba
\Delta S^{(1,2)}_A=\frac{c}{6}\log\left|\frac{\eta^{\frac{1-\ap}{2}}(1-\eta^\ap)}{\ap(1-\eta)}\right|^2, \label{PEH}
\ea
where $\ap=\s{1-24h/c}$. Even though this expression was originally derived for the calculation of entanglement entropy in holographic CFTs \cite{Asplund:2014coa,Nozaki:2013wia}, the same formula can be applied to our calculation as both are described by the replica method with the same Riemann surface.

The cross ratio $\eta$ in (\ref{PEH}) is given by 
\ba
\eta=\frac{z_{12}z_{34}}{z_{13}z_{24}}=\frac{z_1}{z_4}.
\ea
Here $z_1$ and $z_4$ are identified with $z_1$ and $z_2$ in (\ref{z1}), respectively.  For $\Delta S^{(1,2)}_B$ we can use (\ref{z2}) to calculate the cross ratio.

\begin{figure}[hhhh]
    \centering
    \includegraphics[width=.3\textwidth]{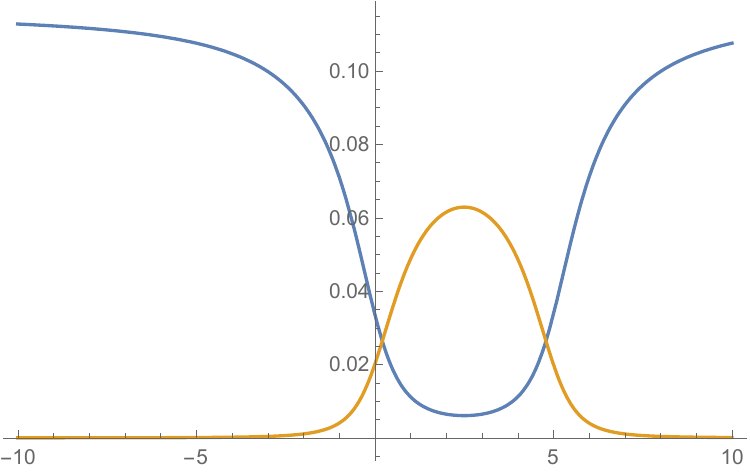}
      \includegraphics[width=.3\textwidth]{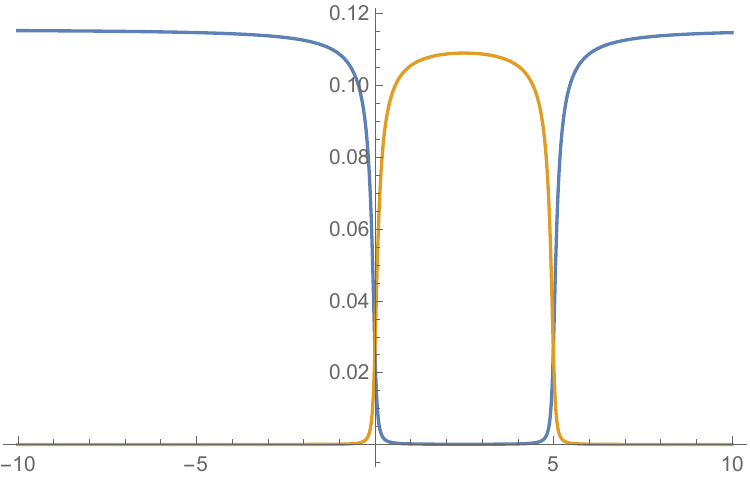}
      \includegraphics[width=.3\textwidth]{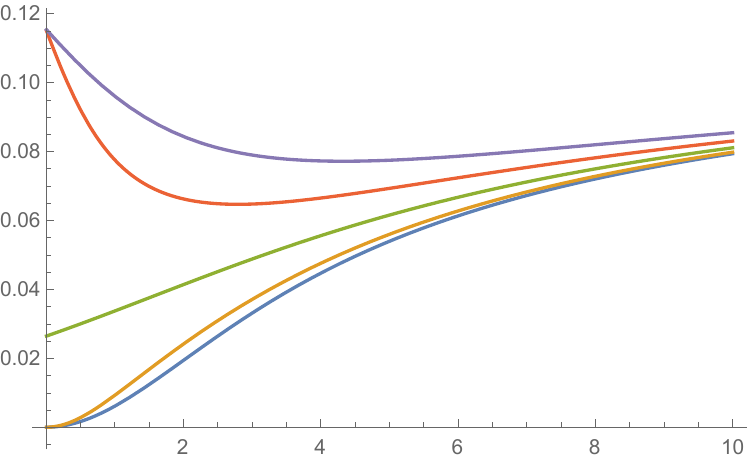}
    \caption{In the left and middle panel, we show $\Delta S^{(1,2)}_A$ (blue) and $\Delta S^{(1,2)}_B$ (orange) as a function of $a$ for $L=5$ and $\ap=1/2$. We set $b=1$ in the left panel and $b=0.1$ in the middle panel. In the right panel, we plot $\Delta S^{(1,2)}_A$ as a function of $b$ for fixed values of $a$. The blue, orange, green, red and purple curves (which appear in this order from bottom to top) correspond to $a=2.5, 3.75, 5, 6.25,$ and $7.5$, respectively.}
    \label{fig:HHLL}
\end{figure}

The result is plotted in Figure~\ref{fig:HHLL}. We observe that when the local operator is acted on the subsystem $A$, $\Delta S^{(2,2)}_A$ is suppressed, while $\Delta S^{(2,2)}_B$ is enhanced, and vice versa. This is similar to what we found in the integral CFTs and can be explained in the same way.

\section{SVD entanglement entropy in Chern-Simons theory}\label{sec:CS_theory}
We investigate SVD entropy in three-dimensional $SU(N)$ Chern-Simons theory with level $k$. In the case of pseudo entropy, we can see non-trivial behaviors when we consider states with four quasi-particles excitation on a slice $S^2$ \cite{Nishioka:2021cxe}. In the perspective of Chern-Simons theory, quasi-particles are realized as the end points of Wilson lines going through the interior of the ball $B^3$ with boundary $S^2$. In other words, the states are prepared by the path integral on $B^3$ by inserting Wilson lines ending on the boundary. Four-excited states are interesting to study pseudo entropy with because we can prepare different but overlapped states by changing the linking number of Wilson lines. 

As in Section 2.5.2 of \cite{Nishioka:2021cxe}, we define the states $\ket{\psi_a}$ on a slice $S^2=A\cup B$ as the path integral by inserting two Wilson lines that have $a$ crossings. Pictorially, 
\begin{align}\label{eq:psia}
  \ket{\psi_a}=
    \begin{tikzpicture}[thick,scale=1.5,baseline={([yshift=-.5ex]current bounding box.center)}]
        \begin{scope}[decoration={markings, mark=at position 0.7 with {\arrow{>}}}]
        \draw[fill=lightgray!20!white] (0,0) ellipse (1.5 and 1);
        \draw[dotted] (0,1) arc (90:-90:0.2 and 1);
        \node at (-1.2,0.9) {$A$};
        \node at (1.2,0.9) {$B$};
        \begin{scope}[shift={(-1.1,0.4)}] 
          \draw[BrickRed] (-0.05,-0.05)--(0.05,0.05);
          \draw[BrickRed] (-0.05,0.05)--(0.05,-0.05);
        \end{scope}
        \begin{scope}[shift={(-1.1,-0.4)}] 
          \draw[BrickRed] (-0.05,-0.05)--(0.05,0.05);
          \draw[BrickRed] (-0.05,0.05)--(0.05,-0.05);
        \end{scope}
        \begin{scope}[shift={(1.1,0.4)}] 
          \draw[BrickRed] (-0.05,-0.05)--(0.05,0.05);
          \draw[BrickRed] (-0.05,0.05)--(0.05,-0.05);
        \end{scope}
        \begin{scope}[shift={(1.1,-0.4)}] 
          \draw[BrickRed] (-0.05,-0.05)--(0.05,0.05);
          \draw[BrickRed] (-0.05,0.05)--(0.05,-0.05);
        \end{scope}
        \begin{knot}[background color=lightgray!20!white]
            \strand[BrickRed] (1.1,0.4) .. controls (0.9,-0.45) and (0.7,-0.45) .. (0.52,-0.1);
            \strand[BrickRed] (1.1,-0.4) .. controls (0.9,0.45) and (0.7,0.45) .. (0.52,0.1);
        \end{knot}
        \begin{knot}[background color=lightgray!20!white,flip crossing=1]
            \strand[BrickRed] (-1.1,0.4) .. controls (-0.8,-0.5) and (-0.6,-0.5) .. (-0.38,0) .. controls (-0.2,0.42) and (0,0.42) ..(0.15,0.1);
            \strand[BrickRed] (-1.1,-0.4) .. controls (-0.8,0.5) and (-0.6,0.5) .. (-0.38,0) .. controls (-0.2,-0.42) and (0,-0.42) ..(0.15,-0.1);
        \end{knot}
        \node[BrickRed] at (-0.95,0.6) {\footnotesize$j$} ;
        \node[BrickRed] at (-0.95,-0.6) {\footnotesize$j$} ;
        \node[BrickRed] at (0.95,0.6) {\footnotesize$\bar{j}$} ;
        \node[BrickRed] at (0.95,-0.6) {\footnotesize$\bar{j}$} ;
        \node[BrickRed] at (0.4, 0) {\large $\dots$};
        \draw (0,1) arc (90:270:0.2 and 1);
        \node[fill=lightgray!20!white] at (0,-0.6) {\footnotesize $|a|$ crossings};
      \end{scope}
    \end{tikzpicture}
  \ .
\end{align}
Here we also allow $a$ to take negative values, which denote twisting in the opposite direction. We fix the representation of Wilson line to be the fundamental representation $j$ of $SU(N)$, then the quasi-particles emerging at the end points have the fundamental representation $j$ and the anti-fundamental representation $\bar{j}$. The unnormalized transition matrix $\ttm{a}{b}_A:=\tr_B\ket{\psi_a}\bra{\psi_b}$ is expressed as 
\begin{align}\label{eq:tmab}
  \ttm{a}{b}_A=
    \begin{tikzpicture}[thick,scale=1.5,baseline={([yshift=-.5ex]current bounding box.center)}]
        \begin{scope}[decoration={markings, mark=at position 0.7 with {\arrow{>}}}]
        \draw[fill=lightgray!20!white] (0,0) ellipse (1.5 and 1);
        \draw[dotted] (0,1) arc (90:-90:0.2 and 1);
        \node at (-1.2,0.9) {$A$};
        \node at (1.2,0.9) {$\bar{A}$};
        \begin{scope}[shift={(-1.1,0.4)}] 
          \draw[BrickRed] (-0.05,-0.05)--(0.05,0.05);
          \draw[BrickRed] (-0.05,0.05)--(0.05,-0.05);
        \end{scope}
        \begin{scope}[shift={(-1.1,-0.4)}] 
          \draw[BrickRed] (-0.05,-0.05)--(0.05,0.05);
          \draw[BrickRed] (-0.05,0.05)--(0.05,-0.05);
        \end{scope}
        \begin{scope}[shift={(1.1,0.4)}] 
          \draw[BrickRed] (-0.05,-0.05)--(0.05,0.05);
          \draw[BrickRed] (-0.05,0.05)--(0.05,-0.05);
        \end{scope}
        \begin{scope}[shift={(1.1,-0.4)}] 
          \draw[BrickRed] (-0.05,-0.05)--(0.05,0.05);
          \draw[BrickRed] (-0.05,0.05)--(0.05,-0.05);
        \end{scope}
        \begin{knot}[background color=lightgray!20!white]
            \strand[BrickRed] (1.1,0.4) .. controls (0.9,-0.45) and (0.7,-0.45) .. (0.52,-0.1);
            \strand[BrickRed] (1.1,-0.4) .. controls (0.9,0.45) and (0.7,0.45) .. (0.52,0.1);
        \end{knot}
        \begin{knot}[background color=lightgray!20!white,flip crossing=1]
            \strand[BrickRed] (-1.1,0.4) .. controls (-0.8,-0.5) and (-0.6,-0.5) .. (-0.38,0) .. controls (-0.2,0.42) and (0,0.42) ..(0.15,0.1);
            \strand[BrickRed] (-1.1,-0.4) .. controls (-0.8,0.5) and (-0.6,0.5) .. (-0.38,0) .. controls (-0.2,-0.42) and (0,-0.42) ..(0.15,-0.1);
        \end{knot}
        \node[BrickRed] at (-0.95,0.6) {\footnotesize$j$} ;
        \node[BrickRed] at (-0.95,-0.6) {\footnotesize$j$} ;
        \node[BrickRed] at (0.95,0.6) {\footnotesize$\bar{j}$} ;
        \node[BrickRed] at (0.95,-0.6) {\footnotesize$\bar{j}$} ;
        \node[BrickRed] at (0.4, 0) {\large $\dots$};
        \draw (0,1) arc (90:270:0.2 and 1);
        \node[fill=lightgray!20!white] at (0,-0.6) {\footnotesize $|a-b|$ crossings};
      \end{scope}
    \end{tikzpicture}
  \ ,
\end{align}
by gluing the subregions $B$ of $\ket{\psi_a}$ and $\bra{\psi_b}$. 

To evaluate the SVD entropy, we have to compute
\begin{align}\label{trttn}
    \Tr\left[\left(\left(\ttm{a}{b}_A\right)^\dagger\ttm{a}{b}_A\right)^n\right].
\end{align}
Since the hermitian conjugate is performed by reversing the orientation of the manifold, $\left(\ttm{a}{b}_A\right)^\dagger$ has the same number of crossings as $\ttm{a}{b}_A$, but with flipping the direction. Therefore $\left(\ttm{a}{b}_A\right)^\dagger \ttm{a}{b}_A$ has no crossings, and \eqref{trttn} is just equal to the partition function on $S^3$ with two unlinked Wilson loops inserted. Such a partition function can be calculated by Witten's surgery method \cite{Witten:1988hf}. Then 
\begin{align}\label{toptr}
    \Tr\left[\left(\left(\ttm{a}{b}_A\right)^\dagger\ttm{a}{b}_A\right)^n\right]=\frac{(\smat{j}{0})^2}{\smat{0}{0}}=\frac{d_j^2}{\mathcal{D}},
\end{align}
where $d_j:=\smat{j}{0}/\smat{0}{0}$ is called the quantum dimension and $\mathcal{D}:=(\smat{0}{0})^{-1}$ is called the total dimension. 
By analytically continuing $n$ to $n/2$, we find
\begin{align}
    \Tr[\left(\rho^{a|b}_A\right)^n]=\frac{ \Tr\left[\left(\left(\ttm{a}{b}_A\right)^\dagger\ttm{a}{b}_A\right)^{\frac{n}{2}}\right]}{\left(\Tr\sqrt{\left(\ttm{a}{b}_A\right)^\dagger\ttm{a}{b}_A}\right)^n}=\left(\frac{d_j^2}{\mathcal{D}}\right)^{1-n}.
\end{align}
Then 
\begin{align}
    S\left(\rho^{a|b}_A\right)=-\log\mathcal{D}+2\log d_j.
\end{align}
This is the same value as the ordinary entanglement entropy for any $|\psi_a\rangle$ \cite{Dong:2008ft}.
Since \eqref{toptr} is independent of $n$, the generalized versions $S^{(n,m)}_A$ also take the same value. 
Thus, in this topological example, SVD entropy reduces to the ordinary entanglement entropy for $\ket{\psi_a}$ (or equally $\ket{\psi_b}$). We can interpret this result similarly to the qubit case as follows. In topological theories, a diffeomorphism of the spacetime manifold can be regarded as a unitary operation acting on the Hilbert space \cite{Witten:1988hf}. Applying suitable twisting diffeomorphisms to $\ket{\psi_a}$ and $\ket{\psi_b}$ independently, we can get the same state, e.g. $\ket{\psi_0}$. Therefore, the SVD entropy reduces to the ordinary entanglement entropy with two excitations.

\section{SVD entanglement entropy in quantum spin systems}\label{sec:spin_systems}

In this section, we consider SVD entanglement entropy in quantum spin systems with numerical methods. Since it is difficult to analytically compute the SVD entropy in quantum field theories, numerical results in quantum spin systems turn out to be important to understand the behavior of SVD entropy in quantum many-body systems.

Let us consider a one-dimensional transverse-field Ising model with Hamiltonian given by 
\begin{align}
    H = -J \sum_{i = 0}^{N-1} \sigma_i^z\sigma_{i+1}^z - h \sum_{i=0}^{N-1} \sigma_i^x. 
\end{align}
This Hamiltonian represents a spin chain with $N$ spins labeled by $i$, and the periodic boundary condition is imposed on it. We would like to consider the ground state $\ket{\Omega_{(J,h)}}$ of the above Hamiltonian. In this case, there is a second order quantum phase transition occurring at $J=h$ \cite{sachdev_2000}. The ground state $\ket{\Omega_{(J,h)}}$ is in the ferromagnetic phase when $J>h$ and in the paramagnetic phase when $J<h$. 

We consider two ground states determined from different $(J,h)$. 
\begin{align}
    \ket{\psi_1} = |\Omega_{(J_1,h_1)}\rangle, \\
    \ket{\psi_2} = |\Omega_{(J_2,h_2)}\rangle.
\end{align}
The corresponding transition matrix is given by
\begin{align}
    \tau^{1|2} = \frac{|\Omega_{(J_1,h_1)}\rangle\langle \Omega_{(J_2,h_2)} |}{\langle\Omega_{(J_2,h_2)}|\Omega_{(J_1,h_1)}\rangle}
\end{align}
The pseudo entropy in this setup has been analyzed in \cite{Mollabashi:2020yie,Mollabashi:2021xsd}, where it was pointed out that the pseudo entropy is sensitive when one of the $\ket{\Omega_{(J,h)}}$ experiences a quantum phase transition. 

In the following, we will first study some basic behavior of the SVD entanglement entropy in this setup, and then explore its relationship with this quantum phase transition. We will use the QuSpin package developed in \cite{quspin} in our numerical simulation. The built-in functions in the QuSpin package enables one to perform the exact diagonalization of the many-body Hamiltonian with the locality and symmetries taken into account so that the diagonalization can be more effcient. In our numerical simulation, we will simply use the exact diagonalization to get the ground states and construct the corresponding transition matrices. Then we will numerically compute the singular values of the reduced transition matrices to get the SVD entanglement entropy. 

\subsection{Basic behavior: comparison with entanglement entropy and pseudo entropy}

Before considering the SVD entanglement entropy, let us first review the behavior of the standard entanglement entropy in the transverse Ising model. 

\begin{figure}[H]
    \centering
    \includegraphics[width=0.32\textwidth,trim={1.0cm 0cm 1.0cm 0cm},clip]{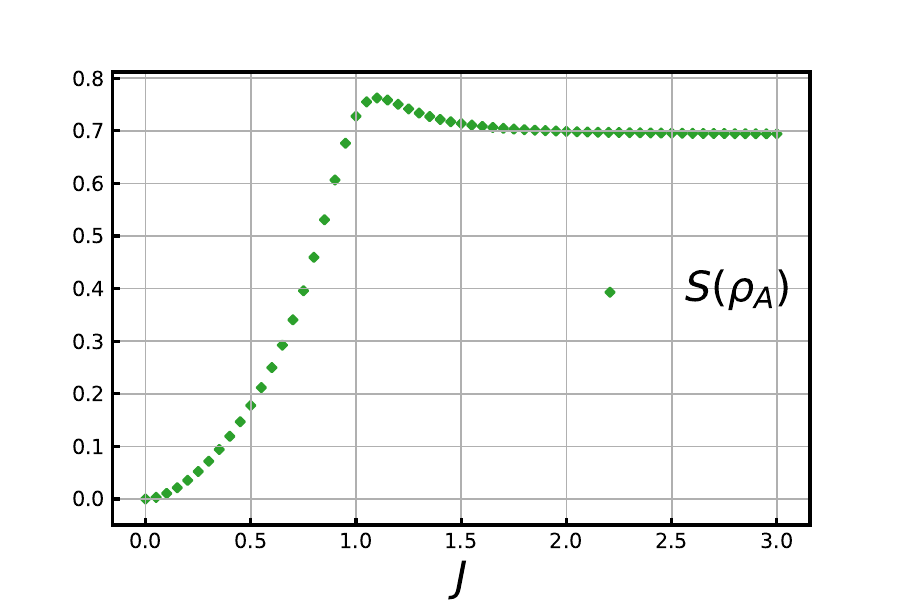}\,
    \includegraphics[width=0.32\textwidth,trim={1.0cm 0cm 1.0cm 0.5cm},clip]{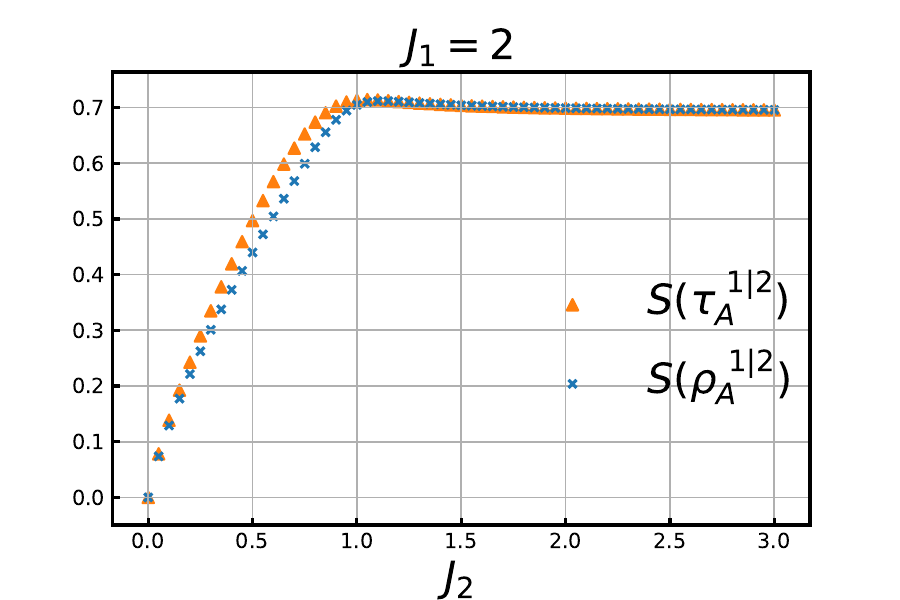}\,
    \includegraphics[width=0.32\textwidth,trim={1.0cm 0cm 1.0cm 0.5cm},clip]{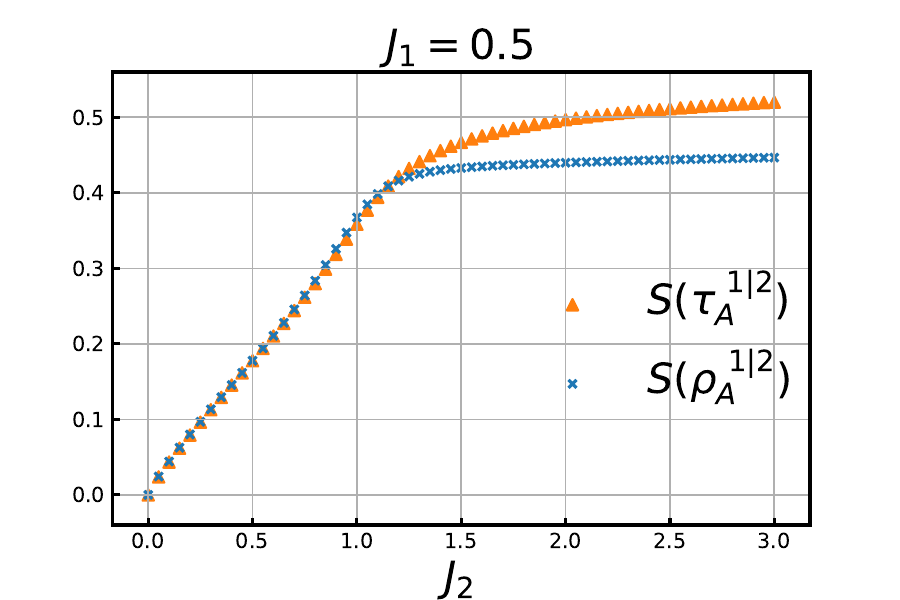}

    \caption{Ground state entanglement entropy, SVD entropy and pseudo entropy in a periodic transverse spin chain with $N = 14$. The subsystem $A$ is taken to be an interval with 7 spins. The parameter $h$ is set to be $h=1$. (Left) Entanglement entropy of $A$ with varying $J$. (Middle) SVD entropy and pseudo entropy of $A$ with fixed $J_1 = 2 $ and varying $J_2$. (Right) SVD entropy and pseudo entropy of $A$ with fixed $J_1 = 0.5 $ and varying $J_2$.}
    \label{fig:TFI_EE_PE}
\end{figure}

Let us consider an even $N$ and take subsystem $A$ to be exactly a half of the whole system. Without loss of generality, we can fix $h=1$. The left panel of Figure \ref{fig:TFI_EE_PE} shows how the ground state entanglement entropy of subsystem $A$ changes when $J$ changes, where $N = 14$. When $J = 0$, the ground state is given by 
\begin{align}
    \ket{\Omega_{(J,h)}} = \left(\frac{1}{\sqrt{2}}\left( \ket{\uparrow} + \ket{\downarrow} \right) \right)^{\otimes N},
\end{align}
which turns out to be a product state. Therefore, the entanglement entropy is $S(\rho_A) = 0$. On the other hand, when $J \gg h = 1$, the ground state is approximately given by 
\begin{align}
    \ket{\Omega_{(J,h)}} \approx \frac{1}{\sqrt{2}}\left( \ket{\uparrow}^{\otimes N} + \ket{\downarrow}^{\otimes N} \right) ,
\end{align}
and the entanglement entropy is approximately $S(\rho_A) = \log 2$. These behaviors can be read out from the left panel of Figure \ref{fig:TFI_EE_PE}. At the thermodynamic limit $N\rightarrow \infty$, there exits a second order phase transition at $J = h = 1$ separating the ferromagnetic phase and the paramagnetic phase. Since we are looking at a finite size system, the phase transition point is not as clear as in the thermodynamic limit. However, we can still observe that the entanglement entropy is enhanced near $J = h = 1$. 

\paragraph{Symmetric bipartition}~\par
Let us then consider two different ground states labeled by $(J_1, h_1)$ and $(J_2, h_2)$. Again, let us consider a spin chain with $N=14$ and $A$ is exactly a half of it. The SVD entanglement entropy $S(\rho_A^{1|2})$ and the pseudo entanglement entropy $S(\tau_A^{1|2})$ are shown in the middle panel and the right panel of Figure \ref{fig:TFI_EE_PE}. One can observe that the SVD entanglement entropy has a qualitatively similar behavior to the pseudo entropy. 

\begin{figure}[H]
    \centering
    \includegraphics[width=0.32\textwidth,trim={1.0cm 0cm 1.0cm 0cm},clip]{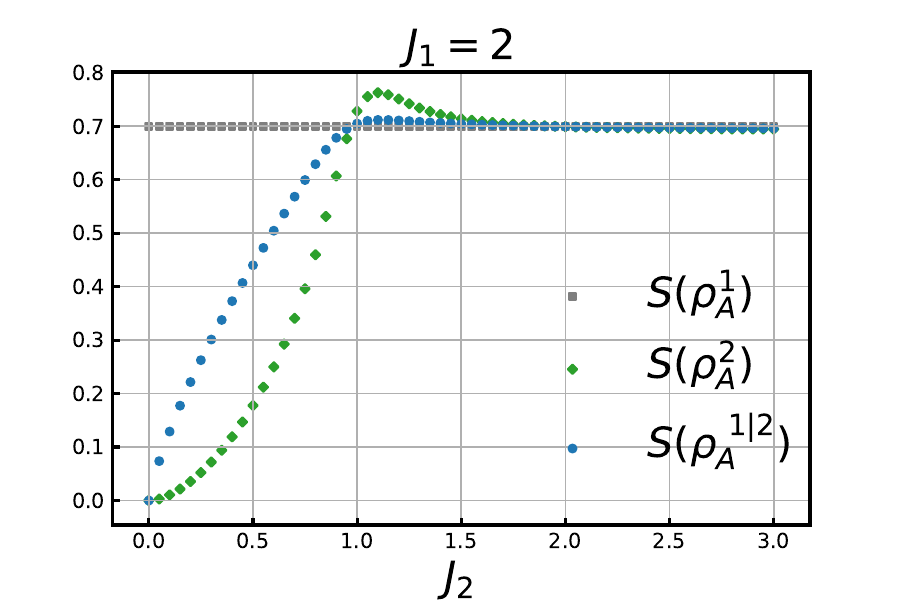}\,
    \includegraphics[width=0.32\textwidth,trim={1.0cm 0cm 1.0cm 0cm},clip]{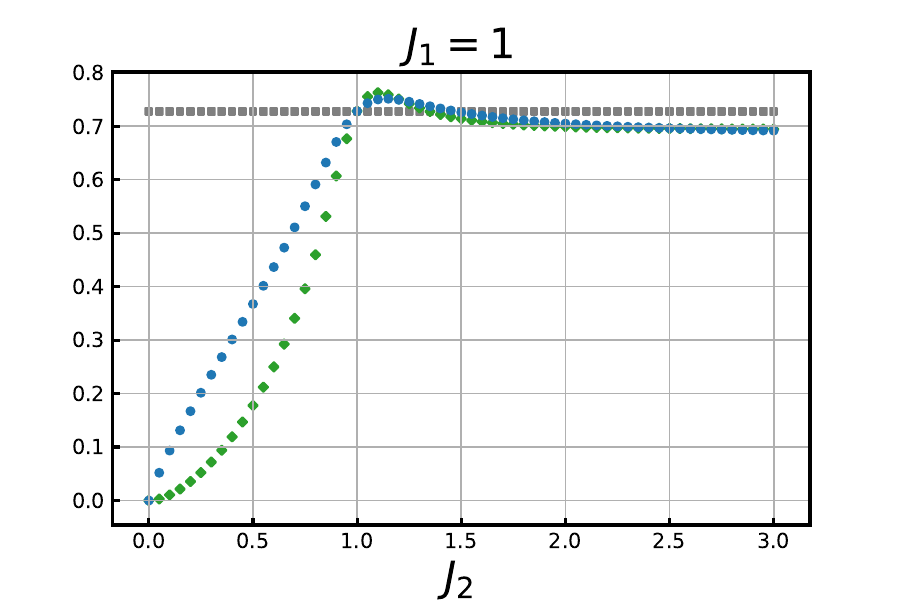}\,
    \includegraphics[width=0.32\textwidth,trim={1.0cm 0cm 1.0cm 0cm},clip]{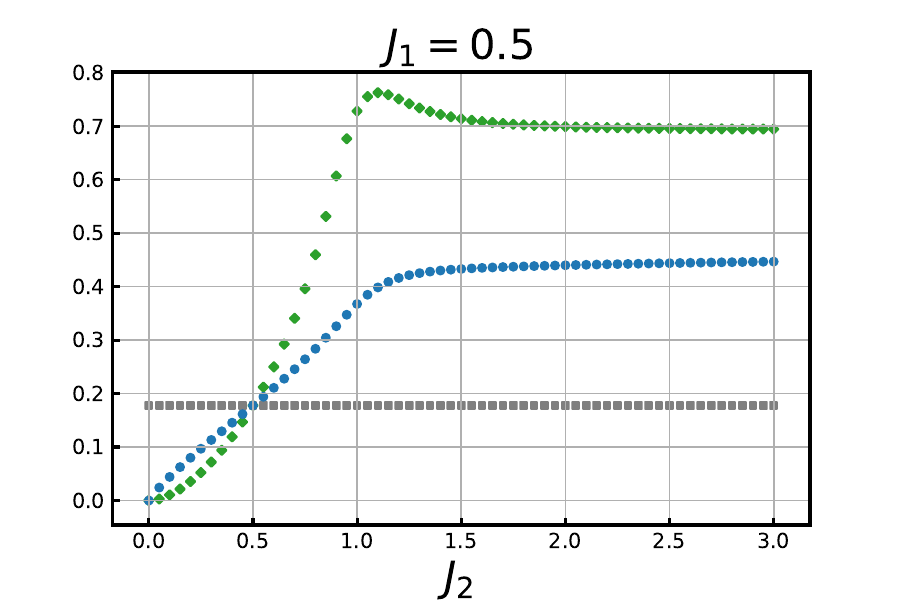}
    \caption{SVD entropy and entanglement entropy in transverse Ising models. Here, we have $N = 14$, $N_A = 7$ and $h_1 = h_2 = 1$. The parameter $J_1$ is fixed as $2$, $1$ and $0.5$ in each figure. The parameter $J_2$ is changed and shown on the horizontal axis. $S(\rho^{1|2}_A)$ is bounded by $\min\{S(\rho_A^1), S(\rho_A^2)\}$ and $\max\{S(\rho_A^1), S(\rho_A^2)\}$ from two sides in most cases.}
    \label{fig:TFI_EE_SVD}
\end{figure}

It is intriguing to compare $S(\rho_A^{1|2})$ with $S(\rho_A^1)$ and $S(\rho_A^2)$, as shown in Figure \ref{fig:TFI_EE_SVD}. From these plots, we can see that $S(\rho^{1|2}_A)$ is bounded by $\min\{S(\rho_A^1), S(\rho_A^2)\}$ and $\max\{S(\rho_A^1), S(\rho_A^2)\}$ from two sides in most cases, while this is not true in general. It is worth noting that the violation only happens when at least one of $J_1$ and $J_2$ is close to $1$ in our numerical results. Therefore, the violation might be a finite size effect, and it is an interesting future problem to test if the bounds hold in infinite spin systems. 

\paragraph{Asymmetric bipartition}~\par

\begin{figure}[H]
    \centering
    \includegraphics[width=0.32\textwidth,trim={1.0cm 0cm 1.0cm 0cm},clip]{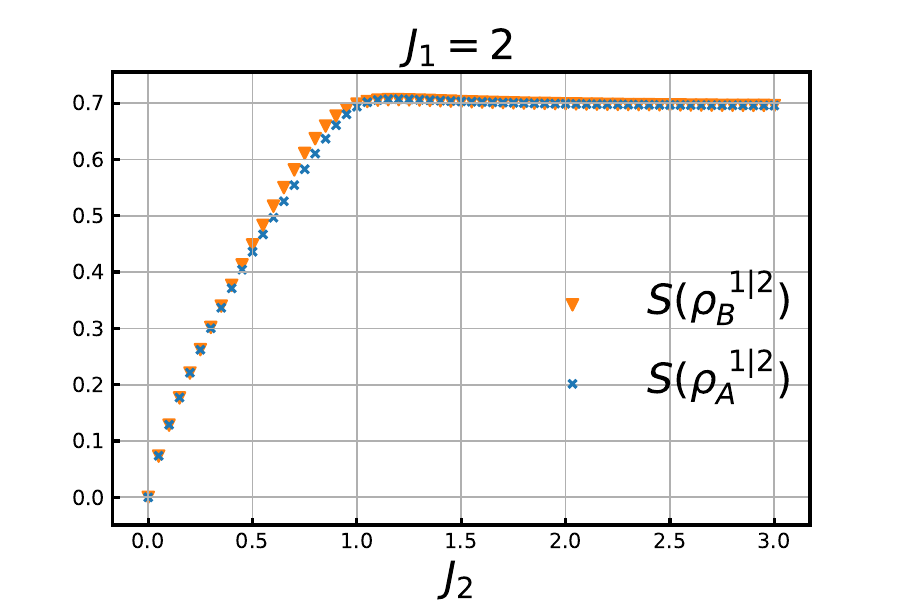}\,
    \includegraphics[width=0.32\textwidth,trim={1.0cm 0cm 1.0cm 0cm},clip]{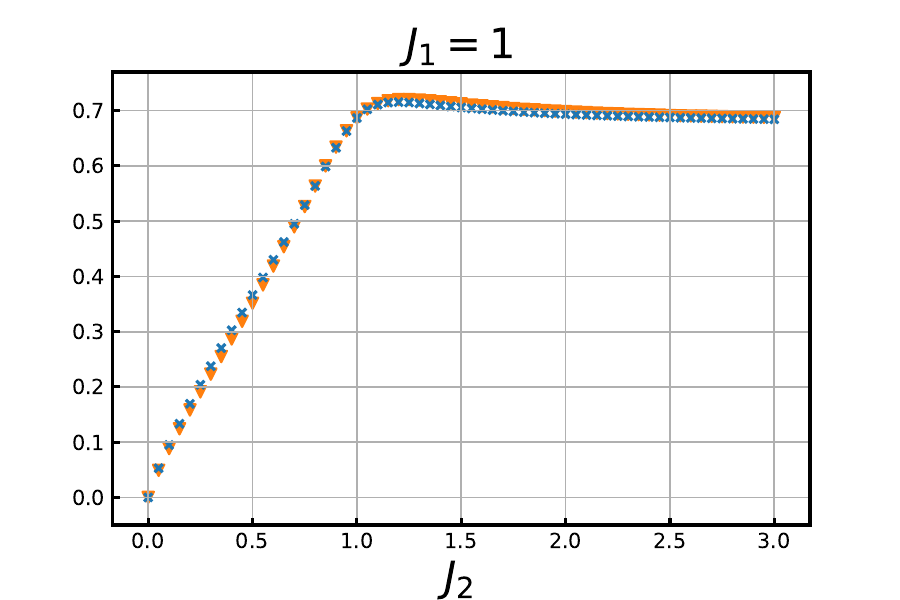}\,
    \includegraphics[width=0.32\textwidth,trim={1.0cm 0cm 1.0cm 0cm},clip]{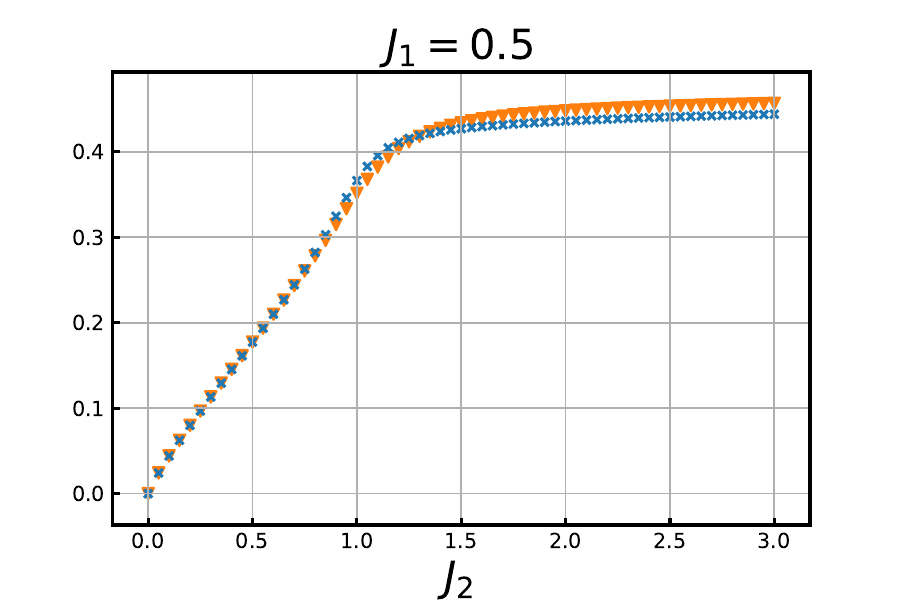}
    \caption{SVD entropy in transverse Ising models with asymmetric bipartition. Here, we have $N=14$, $N_A = 10$, $N_B=4$ and $h_1=h_2=1$. The parameter $J_1$ is fixed as 2, 1 and 0.5 in each figure. As expected, $S(\rho^{1|2}_A)$ and $S(\rho^{1|2}_B)$ take different values, while the qualitative behaviors are similar.}
    \label{fig:TFI_EEAandEEB}
\end{figure}

Let us then change our setup and consider dividing the whole system into $A$ and $B$ where $A$ is an interval with 10 spins and $B$ is an interval with 4 spins. Figure \ref{fig:TFI_EEAandEEB} shows the SVD entropy for subsystem $A$ and $B$ in several different cases. It is clear that the SVD entropy for subsystem $A$ and that for subsystem $B$ take different values. However, they show similar behavior in a qualitative way. 

\paragraph{Multipartition}~\par

\begin{figure}[h]
    \centering
    \includegraphics[width=0.32\textwidth,trim={1.0cm 0cm 1.0cm 0cm},clip]{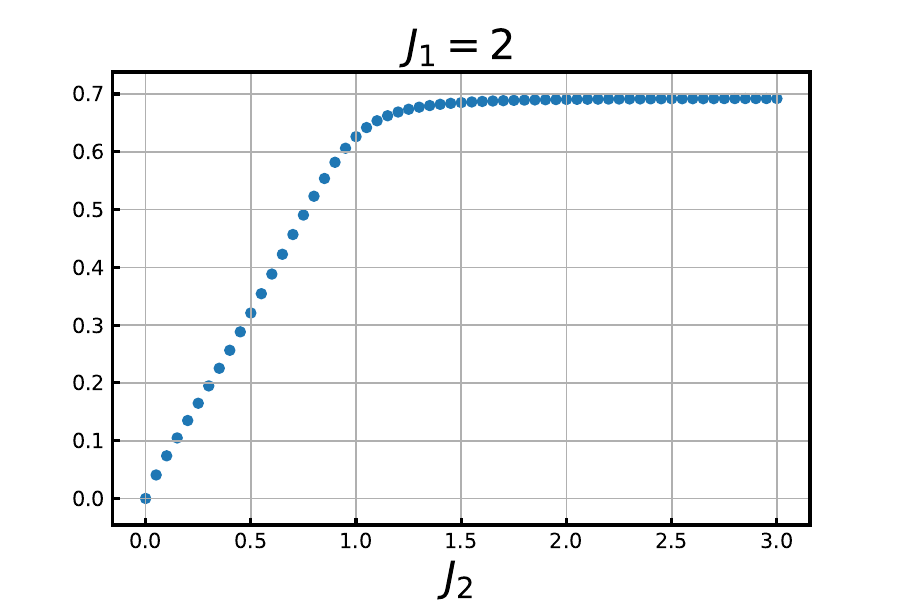}\,
    \includegraphics[width=0.32\textwidth,trim={1.0cm 0cm 1.0cm 0cm},clip]{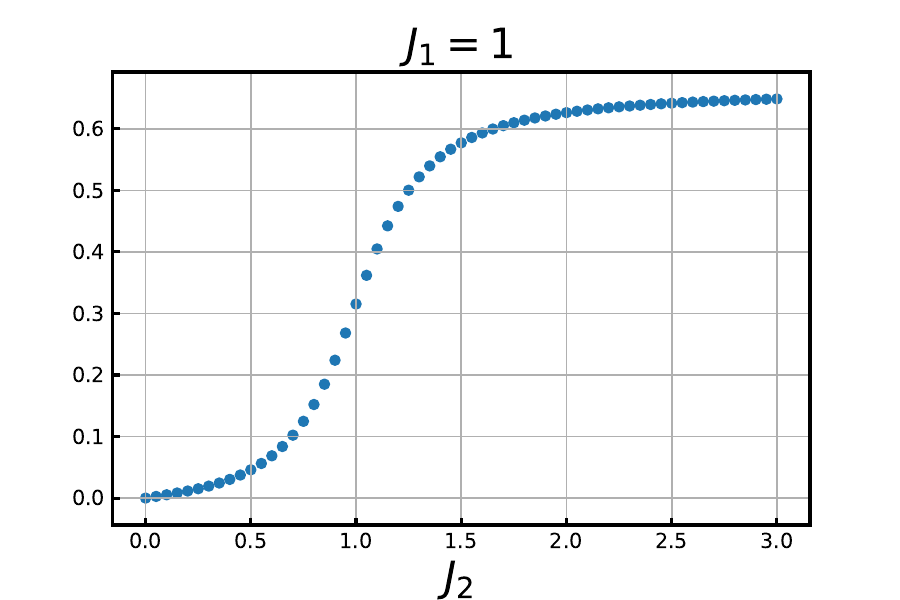}\,
    \includegraphics[width=0.32\textwidth,trim={1.0cm 0cm 1.0cm 0cm},clip]{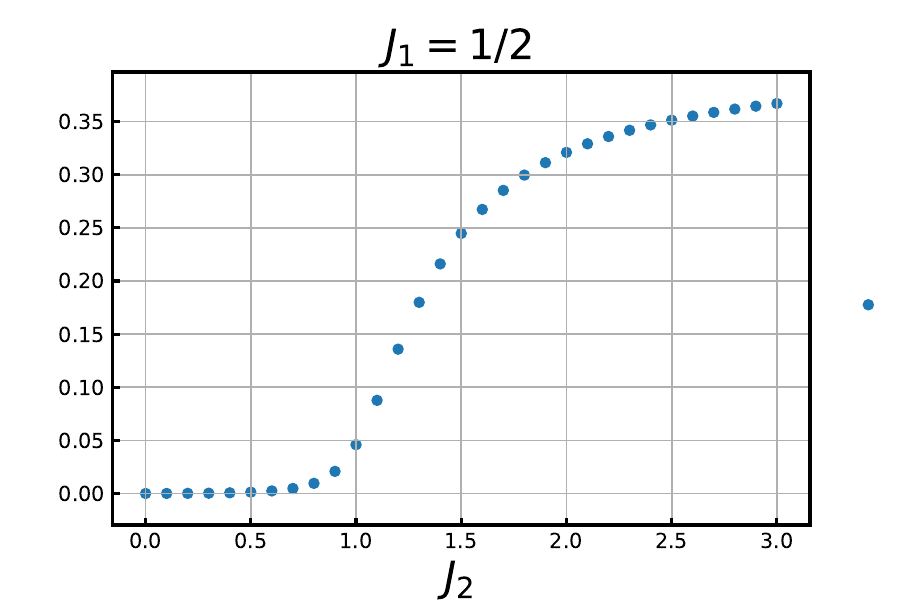}
    \caption{Testing the subadditivity-like inequality. $S(\rho_A^{1|2}) + S(\rho_B^{1|2}) -S(\rho_{AB}^{1|2})$ is shown in several different cases. Here, we have $N=14$, $A=\{i|i=0,1,2\}$, $B = \{i|i=6,7,8\}$ and $h_1=h_2=1$. The parameter $J_1$ is fixed as 2, 1, 0.5 in each figure. $S(\rho_A^{1|2}) + S(\rho_B^{1|2}) -S(\rho_{AB}^{1|2}) \geq 0$ holds in all the examples we have studied.}
    \label{fig:SA}
\end{figure}

We would like to consider dividing the whole system into more parts and test if there is any numerical evidence for other inequalities. We would like to first test if a subadditivity-like inequality, $S(\rho_A^{1|2}) + S(\rho_B^{1|2}) -S(\rho_{AB}^{1|2}) \stackrel{?}{\geq} 0$, holds in our setups. To this end, let us again consider a periodic spin chain with 14 spins, and divide it into three parts: $A$, $B$ and their complement. Let $A$ be an interval composed of three spins labeled by $i=0,1,2$, and let $B$ be another interval composed of three spins labeled by $i=6,7,8$. Figure \ref{fig:SA} shows plots of $S(\rho_A^{1|2}) + S(\rho_B^{1|2}) -S(\rho_{AB}^{1|2})$ in several different cases. We can see that the subadditivity-like inequality holds in all the cases we studied.

\begin{figure}[h]
    \centering
    \includegraphics[width=0.32\textwidth,trim={1.0cm 0cm 1.0cm 0cm},clip]{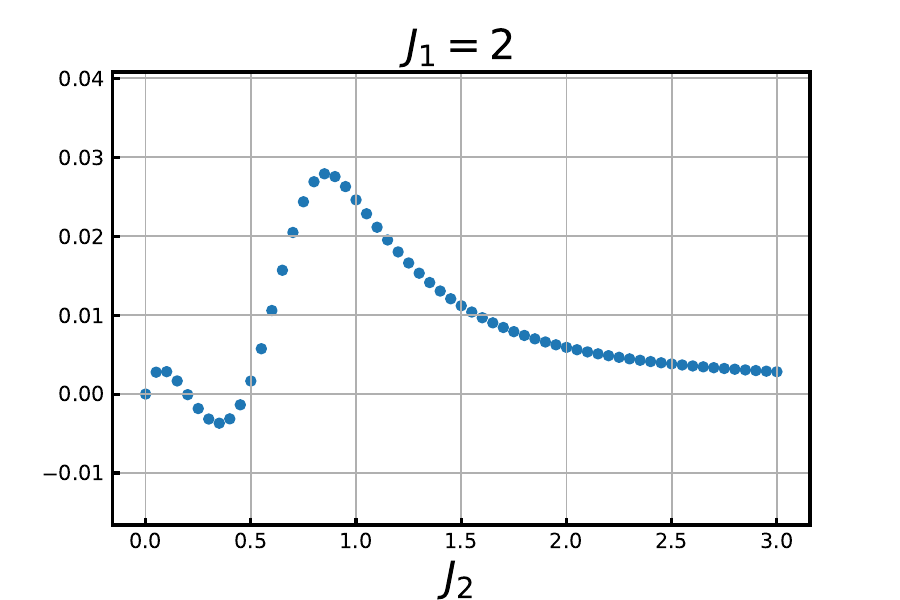}\,
    \includegraphics[width=0.32\textwidth,trim={1.0cm 0cm 1.0cm 0cm},clip]{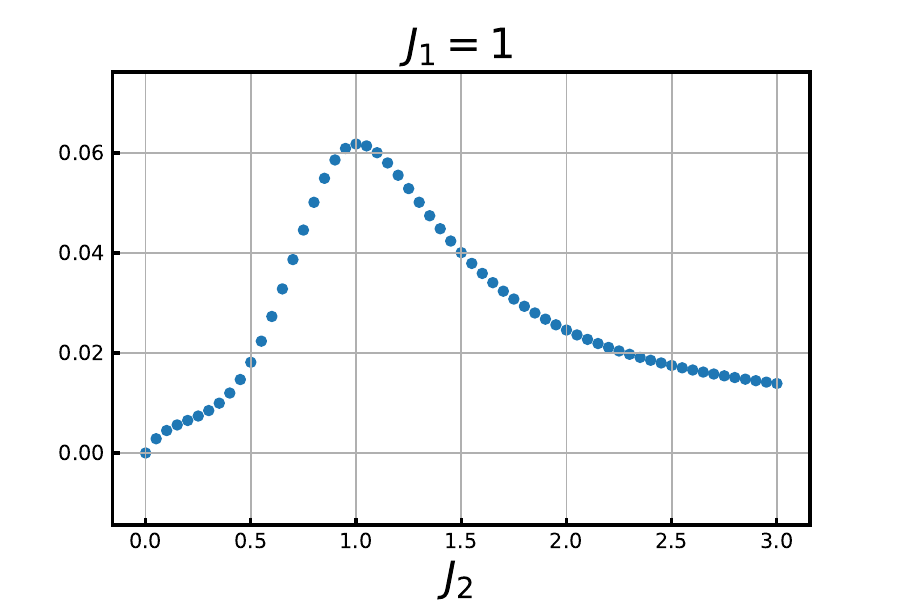}\,
    \includegraphics[width=0.32\textwidth,trim={1.0cm 0cm 1.0cm 0cm},clip]{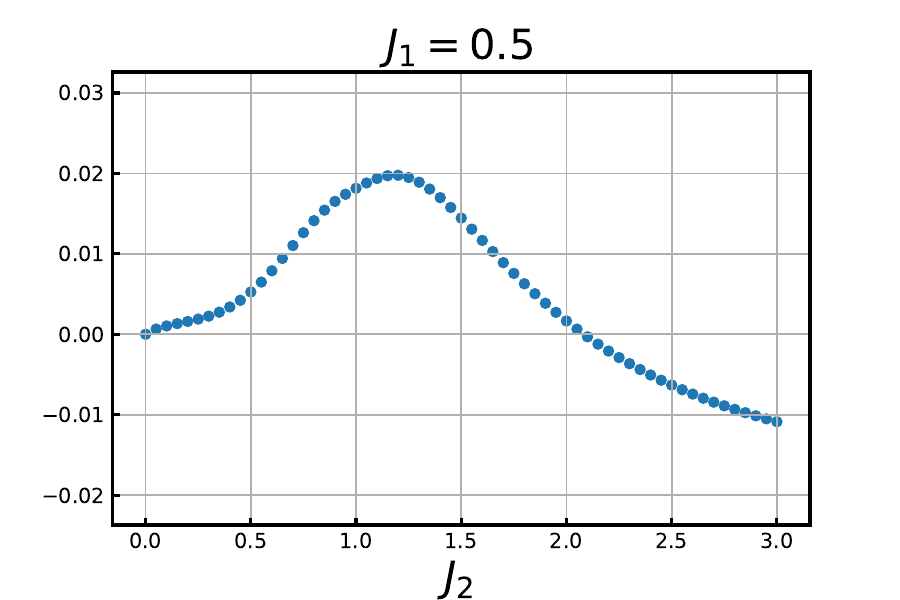}
    \caption{Testing the strong subadditivity-like inequality. $S(\rho_{AB}^{1|2}) + S(\rho_{BC}^{1|2}) - S(\rho_B^{1|2}) - S(\rho_{ABC}^{1|2}) $ is shown in several different cases. Here, we have $N=14$, $A=\{i|i=0,1\}$, $B = \{i|i=2,3\}$, $C = \{i|i=4,5\}$ and $h_1=h_2=1$. The parameter $J_1$ is fixed as 2, 1, 0.5 in each figure. The strong subadditivity-like inequality does not hold.}
    \label{fig:SSA}
\end{figure}

Let us then move on and divide the whole system into four parts: $A$, $B$, $C$ and their complement. We would like to test the strong subadditivity-like inequality 
$S(\rho_{AB}^{1|2}) + S(\rho_{BC}^{1|2}) - S(\rho_B^{1|2}) - S(\rho_{ABC}^{1|2}) \stackrel{?}{\geq} 0$. 
Here, we let $A$ be an interval composed of 2 spins labeled by $0,1$, $B$ be an interval composed of 2 spins labeled by $2,3$ and $C$ be an interval composed of 2 spins labeled by $4,5$. Figure \ref{fig:SSA} shows 
$S(\rho_{AB}^{1|2}) + S(\rho_{BC}^{1|2}) - S(\rho_B^{1|2}) - S(\rho_{ABC}^{1|2}) $ in several cases, and the strong subadditivity-like inequality does not hold. 

\subsection{SVD entanglement entropy as a probe of quantum phase transition}

In \cite{Mollabashi:2020yie,Mollabashi:2021xsd}, the authors point out that the pseudo entropy can be used as a probe of quantum phase transitions. More specifically, the authors find that 
\begin{align}
    S(\tau_A^{1|2}) - \frac{S(\rho_A^1) + S(\rho_A^2)}{2} \leq 0, 
\end{align}
always holds as long as $\ket{\Omega_{(J_1,h_1)}}$ and $\ket{\Omega_{(J_2,h_2)}}$ are in the same quantum phase. On the other hand, however, when $\ket{\Omega_{(J_1,h_1)}}$ and $\ket{\Omega_{(J_2,h_2)}}$ are in different quantum phases, the inequality can be violated. This behavior was confirmed in the transverse Ising model \cite{Mollabashi:2020yie} and the quantum XY model \cite{Mollabashi:2021xsd}, with both direct diagonalization in small systems and the correlation method in large systems. 

\begin{figure}[h]
    \centering
    \includegraphics[width=0.32\textwidth,trim={1.0cm 0cm 1.0cm 0cm},clip]{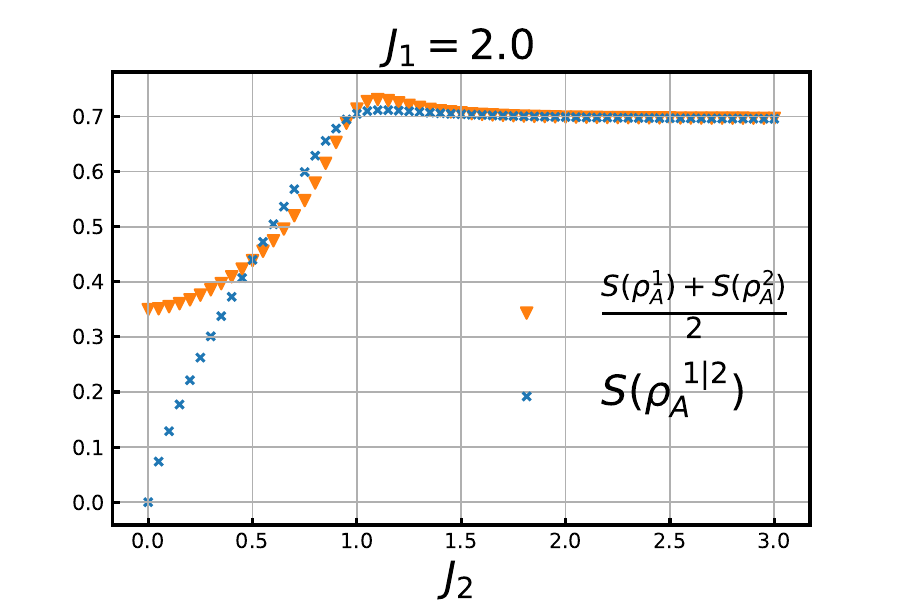}\,
    \includegraphics[width=0.32\textwidth,trim={1.0cm 0cm 1.0cm 0cm},clip]{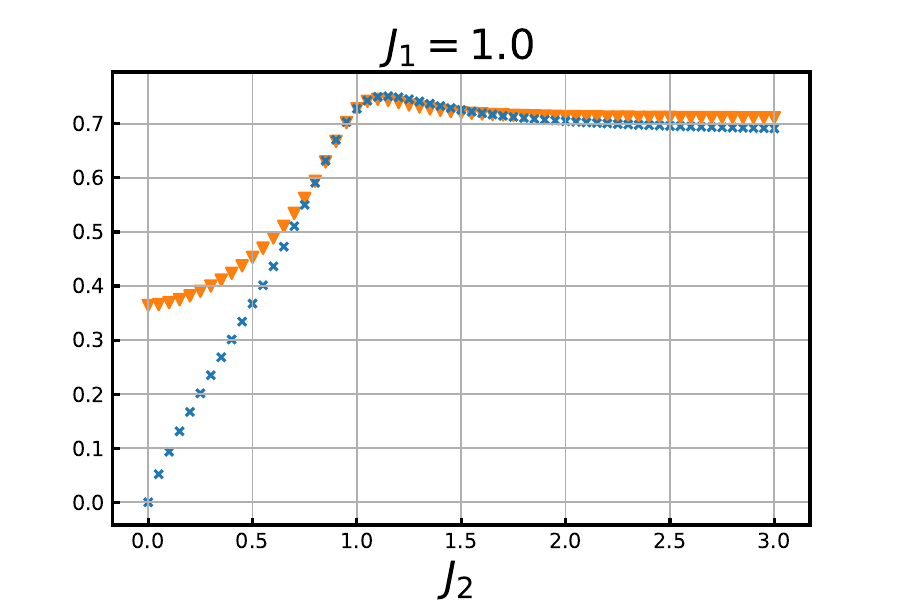}\,
    \includegraphics[width=0.32\textwidth,trim={1.0cm 0cm 1.0cm 0cm},clip]{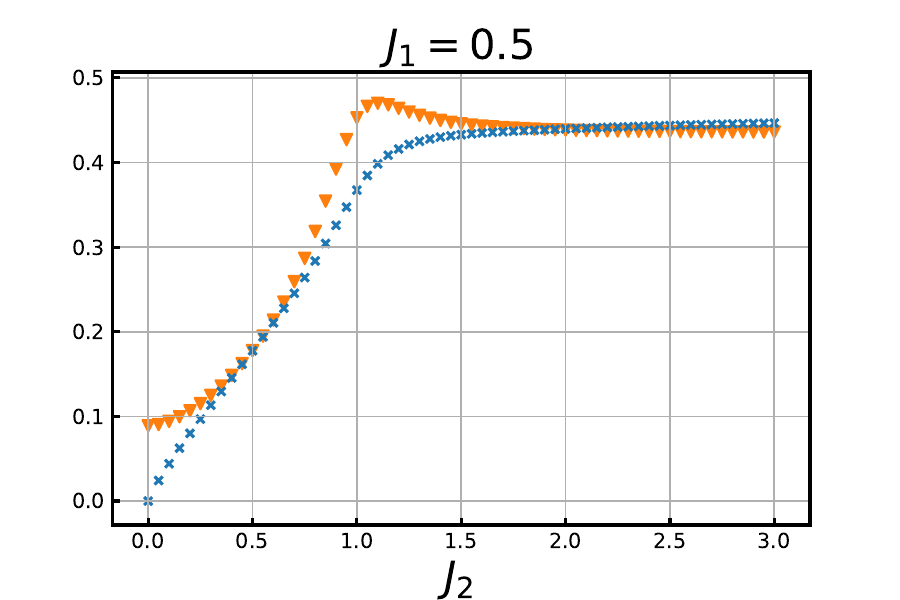}
    \caption{SVD entropy and averaged entanglement entropy in transverse Ising models. Here, we have $N = 14$, $N_A = 7$ and $h_1 = h_2 = 1$. The parameter $J_1$ is fixed as $2$, $1$ and $0.5$ in each figure. The parameter $J_2$ is changed and shown in the horizontal axis. One can observe $ S(\rho^{1|2}_A) - \left(S(\rho^1_A) + S(\rho^2_A)\right)/2 \leq 0$ when the two states are in the same quantum phase. }
    \label{fig:avEETFI}
\end{figure}

\begin{figure}[h]
    \centering
    \includegraphics[width=0.32\textwidth,trim={1.0cm 0cm 1.0cm 0cm},clip]{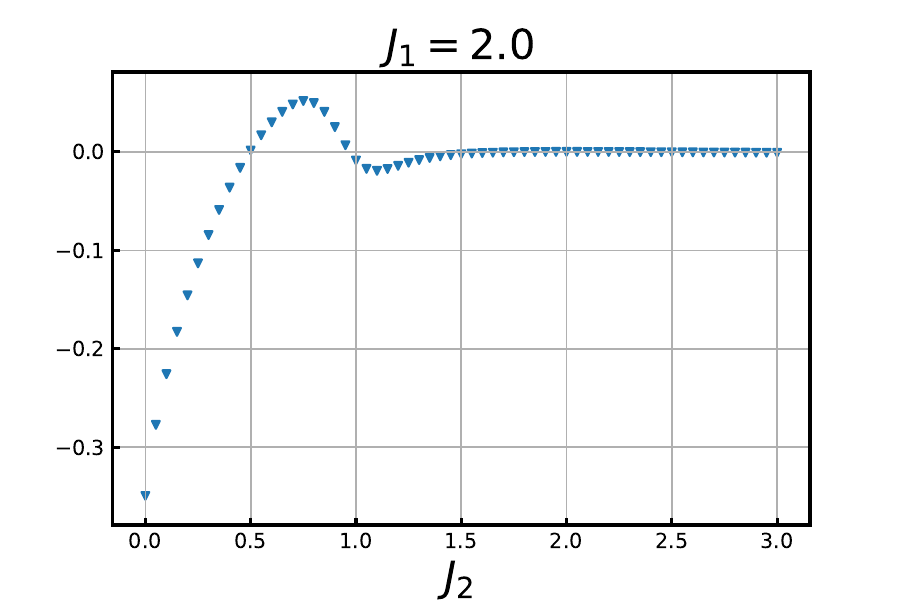}\,
    \includegraphics[width=0.32\textwidth,trim={1.0cm 0cm 1.0cm 0cm},clip]{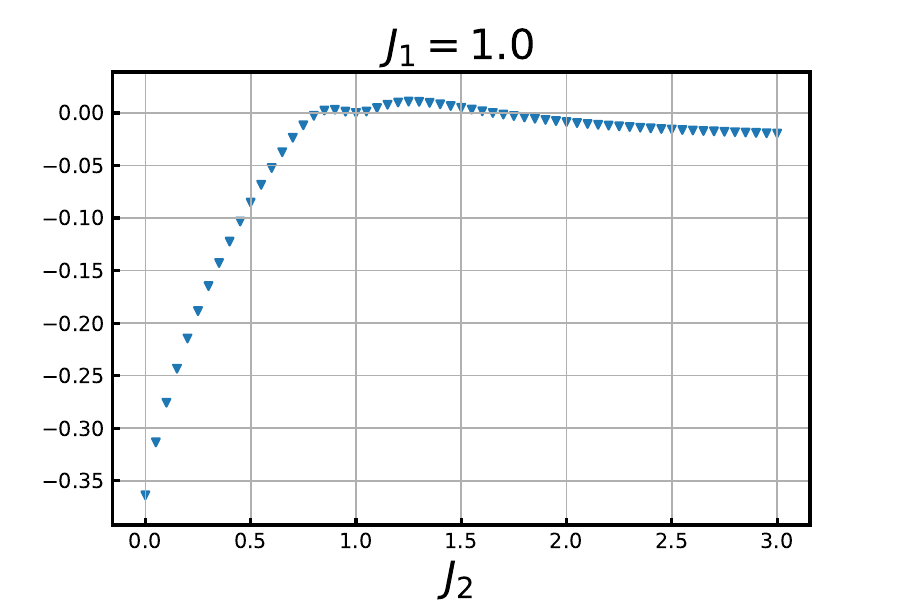}\,
    \includegraphics[width=0.32\textwidth,trim={1.0cm 0cm 1.0cm 0cm},clip]{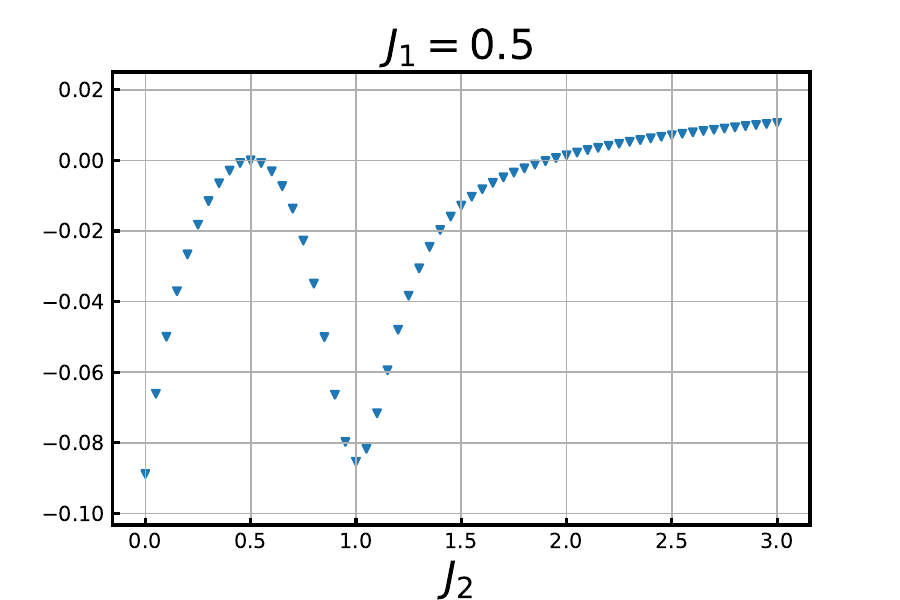}
    \caption{Plots of $S(\rho^{1|2}_A) - \left(S(\rho^1_A) + S(\rho^2_A)\right)/2$ in different cases. $ S(\rho^{1|2}_A) - \left(S(\rho^1_A) + S(\rho^2_A)\right)/2 \leq 0$ when the two states are in the same quantum phase, and can be violated when the two states are in different quantum phases.}
    \label{fig:avEETFI2}
\end{figure}

Since the SVD entanglement entropy is another extension of the entanglement entropy to post-selection setups, it is worth checking if the SVD entropy has similar features. Indeed, we find that 
\begin{align}
    S(\rho_A^{1|2}) - \frac{S(\rho_A^1) + S(\rho_A^2)}{2} \leq 0, 
\end{align}
holds when $\ket{\Omega_{(J_1,h_1)}}$ and $\ket{\Omega_{(J_2,h_2)}}$ are in the same quantum phase, and can be violated when they are in different quantum phases. Plots comparing the SVD entropy and the averaged entanglement entropy are shown in Figure \ref{fig:avEETFI}. We also show plots of $S(\rho_A^{1|2}) - \frac{S(\rho_A^1) + S(\rho_A^2)}{2}$ in Figure \ref{fig:avEETFI2}.

\section{Conclusion and Discussion}\label{sec:conclusions}

In this paper, we introduced a new quantity, called SVD entanglement entropy $S(\rho^{1|2}_A)$. This is a generalization of entanglement entropy depending on two different states $|\psi_1\lb$ and $|\psi_2\lb$, as in pre- and post-selection processes, regarding $|\psi_1\lb$ as the initial state and $|\psi_2\lb$ as the final state. As opposed to the pseudo entropy \cite{Nakata:2021ubr}, this SVD entanglement entropy takes non-negative real values only and is bounded by the log of the Hilbert space dimensions. By analyzing two qubit states, we showed that this SVD entanglement entropy can be interpreted as the average number of Bell pairs distillable from intermediate states that appear between $|\psi_1\lb$ and $|\psi_2\lb$. Furthermore, this quantity is well-defined even if 
$|\psi_1\lb$ and $|\psi_2\lb$ are orthogonal to each other, unlike for pseudo entropy. We may think that these properties provide a characteristic advantage over the pseudo entropy, which constitute motivations for why we introduced this new quantity. Moreover, we observed that the SVD entanglement entropy gets enhanced when the two states are in the different quantum phases in explicit examples of transverse-field Ising model. Therefore we can employ the SVD entanglement entropy as a quantum order parameter to distinguish different quantum phases. This property looks analogous to the behavior of the real part of pseudo entropy. 

The calculations of SVD entanglement entropy, however, turn out to be more complicated than those of the pseudo entropy.
This is mainly because we need to take the square root of the product of the reduced transition matrix associated with the subregion and its complex conjugate, which is difficult to perform in a field-theoretic analysis, especially in the path-integral formalism. To avoid this problem, we introduced R{\'e}nyi extensions $S^{(n,m)}_A=S^{(n)}(\rho^{(m)1|2}_A)$ of the SVD entanglement entropy, and we calculated this quantity for $m=2$ in various two-dimensional conformal field theories (CFTs). For field-theoretic calculations, we focused on the difference $\Delta S^{(2,2)}_A$ between $S^{(2,2)}_A$ for excited states and its value when two states are both the vacuum state. We mainly analyzed locally excited states created by a primary operator. First we focused on the case where $|\psi_1\lb$ is a locally excited state and $|\psi_2\lb$ is the vacuum in free scalar CFT, then we found $\Delta S^{(2,2)}_A=0$ at $t=0$. However, under time evolution, we found that $\Delta S^{(2,2)}_A$  approaches a positive constant. On the other hand, we also analyzed the Ising CFT in the same setup and we found that even at $t=0$,  $\Delta S^{(2,2)}_A$ takes positive values. Moreover, we observed that when the local operator is acted on the subsystem $A$, $\Delta S^{(2,2)}_A$ vanishes and $\Delta S^{(2,2)}_B$ becomes enhanced, and vice versa. We also calculated $\Delta S^{(2,2)}_A$ in the free scalar CFT when both $|\psi_1\lb$ and $|\psi_2\lb$ are locally excited states. This setup has also been considered for pseudo entropy in \cite{Nakata:2021ubr}. We find that SVD entropy somewhat inherits from pseudo entropy the dependence on the points of insertions, with the absolute value $|\Delta S^{(2,2)}|$ being suppressed compared to the pseudo entropy case. One difference from pseudo entropy is $\Delta S^{(2,2)}_A\neq\Delta S^{(2,2)}_B$. The behaviors of $\Delta S^{(2,2)}_A$ and $\Delta S^{(2,2)}_B$ can be interpreted in analogy with the discussion of local unitary operation given in Section~\ref{subsec:localunitary}.

Similarly, we can provide a holographic calculation of the R{\'e}nyi SVD entropy $S^{(1,2)}_A$. As an example, we calculated this geometrically when $|\psi_1\lb$ and $|\psi_2\lb$ are vacuum states in two-dimensional CFTs that are related by an exactly marginal transformation. We also calculated $\Delta S^{(1,2)}_A$ for locally excited states in holographic CFTs, which has qualitative features similar to the results for integrable CFTs.

As for Chern-Simons theory, the SVD entropy can be calculated analytically, thus allowing us to capture some characteristic properties of SVD entropy. We considered the four-excited states realized by inserting two Wilson lines, which is a case where the pseudo entropy is known to admit a rather complicated form. On the other hand, we found that SVD entropy takes a simple form and is equal to the entanglement entropy. This result is understood by the fact that the singular value decomposition is done by acting with two different unitary operators, which correspond to acting via diffeomorphisms on the spacetime. 

In terms of the mathematical properties, we  formulated SVD entropy as an extension of the von~Neumann entropy from density matrices to arbitrary square matrices. We showed that the SVD entropy satisfies some properties in common with the von~Neumann entropy, including additivity, two-sided independent unitary invariance, and a weakened form of concavity. In addition, we numerically illustrated that the Page curve seems to take on a similar form to the Page curve for the usual von~Neumann entropy, albeit with a few minor differences due to the fact that the subregion SVD entanglement entropies are in general different. Although these provide the similarities between SVD entropy and von~Neumann entropy, some properties do not extend, such as subadditivity and the Araki-Lieb inequality, for example. We provided some numerical evidence, and sometimes explicit counter-examples, showing how these properties can fail in general. Nevertheless, the SVD entropy has played an important role in other fields, such as data analysis, which has seen applications in understanding the complexity of ecological networks and the stock market index, to name a few. Our work has therefore illustrated its additional role in a field-theoretic context, holography, and lattice systems. 

It would be intriguing to explore more properties of SVD entropy, including more refined inequalities satisfied by this quantity and by its R{\'e}nyi extensions that have not been addressed here.
In particular, although subadditivity of the SVD entropy is known to fail, is there a suitable correction term that provides a weaker version of subadditivity?
At the same time, it is helpful to study more examples of calculations of this quantity. For example, it is interesting to analyze the SVD entropy for various quantum phase transitions in quantum many-body systems and their field-theoretic counterparts. Finally, it is a fascinating direction to understand its relevance to quantum gravity, such as the black hole information problem. One additional potential justification for this connection is due to the fact that the SVD entropy has been used to study the complexity of data and networks, both of which have provided us with new insights into the black hole information problem in recent years~\cite{Brown:2016PRL,Brown:2016PRD}.

\section*{Acknowledgements}

We are grateful to James Fullwood, Dongsheng Ge, Jonathan Harper, Zhian Jia, Ali Mollabashi, Yoshifumi Nakata, Pratik Nandy, Shinsei Ryu, and Aephraim Steinberg for discussions.

This work is supported by the Simons Foundation through the ``It from Qubit'' and by MEXT KAKENHI Grant-in-Aid for Transformative Research Areas (A) through the ``Extreme Universe'' collaboration: Grant Number 21H05182, 21H05183 and 21H05187.
This work is also supported by Inamori Research Institute for Science and by JSPS Grant-in-Aid for Scientific Research (A) No.~21H04469.
The work of YT is supported by Grant-in-Aid for JSPS Fellows No.~22J21950, No.~22KJ1971. ZW was supported by Grant-in-Aid for JSPS Fellows No. 20J23116.


\bibliographystyle{JHEP}
\bibliography{NewPE}


\end{document}